\documentclass[useAMS,usenatbib]{mn2e}
\usepackage{times}
\usepackage{graphicx}
\usepackage{amssymb}
\usepackage{longtable}

\title[Quenching of Centrals and Satellites]{The impact of galactic properties and environment on the quenching of central and satellite galaxies: A comparison between SDSS, Illustris and L-Galaxies}
\author[Asa F. L. Bluck et al.]{Asa F. L. Bluck$^{1,2,*}$, J. Trevor Mendel$^{3}$, Sara L. Ellison$^{2}$, David R. Patton$^{4}$, Luc Simard$^{5}$, \newauthor Bruno M. B. Henriques$^{1}$, Paul Torrey$^{6,7}$, Hossen Teimoorinia$^{2}$, Jorge Moreno$^{8,7,9}$,\newauthor Else Starkenburg$^{10}$
\\$^1$ Institute for Astronomy, Department of Physics, ETH Zurich, Wolfgang-Pauli-Strasse 27, Zurich, 8093, Switzerland
\\$^2$ Department of Physics and Astronomy, University of Victoria, 3800 Finnerty Road, Victoria, British Columbia, V8P 1A1, Canada
\\$^3$ Max-Planck-Institut f\"ur Extraterrestrische Physik (MPE), Giessenbachstrasse, D-85748 Garching, Germany 
\\$^4$ Department of Physics and Astronomy, Trent University, 1600 West Bank Drive, Peterborough, Ontario, K9J 7B8, Canada 
\\$^5$ National Research Council of Canada, Herzberg Institute of Astrophysics, 5071 West Saanich Road, Victoria, British Columbia, V9E 2E7, Canada
\\$^6$ MIT Kavli Institute for Astrophysics \& Space Research, Cambridge, MA, 02139, USA
\\$^7$ TAPIR, Mailcode 350-17, California Institute of Technology, Pasadena, CA 91125, USA
\\$^8$ Department of Physics and Astronomy, California State Polytechnic University Pomona, Pomona, CA 91768, USA
\\$^9$ Harvard-Smithsonian Center for Astrophysics, 60 Garden Street, Cambridge, MA, 02138, USA
\\$^{10}$ Leibniz-Institut f\"ur Astrophysik Potsdam (AIP), An der Sternwarte 16, 14482 Potsdam, Germany
\\$*$ Email: asa.bluck@phys.ethz.ch}

\begin{document}

\maketitle

\begin{abstract}

\noindent We quantify the impact that a variety of galactic and environmental properties have on the quenching of star formation. We collate a sample of $\sim$ 400,000 central and $\sim$ 100,000 satellite galaxies from the Sloan Digital Sky Survey Data Release 7 (SDSS DR7). Specifically, we consider central velocity dispersion ($\sigma_{c}$), stellar, halo, bulge and disk mass, local density, bulge-to-total ratio, group-centric distance and galaxy-halo mass ratio. We develop and apply a new statistical technique to quantify the impact on the quenched fraction ($f_{\rm Quench}$) of varying one parameter, while keeping the remaining parameters fixed. For centrals, we find that the $f_{\rm Quench} - \sigma_{c}$ relationship is tighter and steeper than for any other variable considered. We compare to the Illustris hydrodynamical simulation and the Munich semi-analytic model (L-Galaxies), finding that our results for centrals are qualitatively consistent with their predictions for quenching via radio-mode AGN feedback, hinting at the viability of this process in explaining our observational trends. However, we also find evidence that quenching in L-Galaxies is too efficient and quenching in Illustris is not efficient enough, compared to observations. For satellites, we find strong evidence that environment affects their quenched fraction at fixed central velocity dispersion, particularly at lower masses. At higher masses, satellites behave identically to centrals in their quenching. Of the environmental parameters considered, local density affects the quenched fraction of satellites the most at fixed central velocity dispersion.

\end{abstract}

\section{Introduction}

Understanding why galaxies stop forming stars is an important unresolved question in the field of galaxy formation and evolution. Only $\sim$10\% of baryons reside within galaxies (e.g., Fukugita \& Peebles 2004; Shull et al. 2012), yet since galaxies lie at nodes in the cosmic web corresponding to local minima in the gravitational potential, naively one would expect far more baryons to collate in galaxies, ultimately forming more stars. Theoretical models offer a wide range of solutions to this problem, relying on the physics of gas, stars, and black hole accretion disks as so called `baryonic feedback' (e.g., Cole et al. 2000; Croton et al. 2006, Bower et al. 2006, 2008; Somerville et al. 2008; Guo et al. 2011; Vogelsberger et al. 2014a,b; Henriques et al. 2015; Schaye et al. 2015; Somerville et al. 2015). However, observational studies are required to test these models and provide evidence for their range and applicability. 

Observationally, the fraction of quenched (passive/non-star forming) galaxies in a given population has been shown to have a strong dependence on galaxy stellar mass (e.g., Baldry et al. 2006; Peng et al. 2010, 2012) and galaxy structure, e.g. bulge-to-total light/ mass ratio, $B/T$, or S\'{e}rsic index, $n_{S}$ (e.g., Driver et al. 2006; Cameron et al. 2009; Cameron \& Driver 2009; Wuyts et al. 2011; Mendel et al. 2013; Bluck et al. 2014; Lang et al. 2014; Omand et al. 2014). Additionally, the quenched fraction depends on environment, particularly the surface density of galaxies in a given region of space, the halo mass of the group or cluster, and the distance a galaxy resides at from the centre of its group (e.g., Balogh et al. 2004; van den Bosch et al. 2007, 2008; Peng et al. 2012; Woo et al. 2013; Bluck et al. 2014).

It has become evident that understanding quenching processes in galaxies requires separate consideration of central and satellite galaxies, since the mechanisms for quenching star formation in these systems most likely differ (e.g., Peng et al. 2012; Woo et al. 2013; Bluck et al. 2014; Knobel et al. 2015). Central galaxies are most commonly defined as the most massive galaxy in their group or cluster, with satellites being any other group member (e.g., Yang et al. 2007, 2009). The dominant galaxy in any given dark matter halo is taken to be the central, so isolated galaxies are considered to be the central galaxy of their group of one. Observationally, satellites in general depend on both intrinsic and environmental parameters for their quenching, whereas centrals depend primarily only on intrinsic properties (e.g., Peng et al. 2012). In many simulations and models, the quenching of central galaxies is governed primarily by AGN feedback and the quenching of satellite galaxies is governed primarily by environmental processes, such as, e.g., strangulation or stripping (e.g., Guo et al. 2011; Vogelsberger et al. 2014a,b; Henriques et al. 2015; Schaye et al. 2015; Peng et al. 2015; Somerville et al. 2015). 

More recent work has linked the quenched (or red) fraction of large populations of galaxies to the central density within 1 kpc (Cheung et al. 2012, Fang et al. 2013, Woo et al. 2015), the central velocity dispersion (Wake et al. 2012), and to the mass of the galactic bulge (Bluck et al. 2014, Lang et al. 2014, Omand et al. 2014). An artificial neural network (ANN) analysis performed by Teimoorinia, Bluck \& Ellison (2016) established that for central galaxies the most accurate predictions for whether a galaxy will be star forming or not are given by central velocity dispersion, which outperforms all other variables considered, including bulge mass, stellar mass and halo mass. All of these inner-region galaxy properties are expected to correlate strongly with the mass of the central black hole (e.g., Magorrian et al. 1998; Gebhardt et al. 2000; Ferrarese \& Merritt 2000, Haring \& Rix 2004, McConnell et al. 2011; McConnell \& Ma 2013; Saglia et al. 2016) and hence may provide qualitative support for the AGN feedback driven quenching paradigm. However, it is certainly conceivable that other quenching processes could give rise to these trends without AGN feedback.

Since the idea that most galaxies contain a supermassive black hole was first suggested (e.g., Kormendy \& Richstone 1995), the energy released from forming these objects has become a popular mechanism for regulating gas flows and star formation in simulations, particularly for massive galaxies (e.g., Croton et al. 2006; Bower et al. 2006, 2008; Somerville et al. 2008; Guo et al. 2011; Vogelsberger et al. 2014a,b; Henriques et al. 2015; Schaye et al. 2015). In fact, substantial feedback from accretion around supermassive black holes is required in cosmological semi-analytic models, semi-empirical models, and hydrodynamical simulations to achieve the steep slope of the high-mass end of the galaxy stellar mass function (e.g., Vogelsberger et al. 2014a,b;  Henriques et al. 2015; Schaye et al. 2015). Observationally, direct measurements of AGN driven winds in galaxies and radio jet induced bubbles in galaxy haloes have provided evidence for the mechanisms by which AGN feedback can affect galaxies, but typically only for a very small number of galaxies (e.g., McNamara et al. 2000; Nulsen et al. 2005; McNamara et al. 2007; Dunn et al. 2010; Fabian 2012; Cicone et al. 2013; Liu et al. 2013; Harrison et al. 2014, 2016). Hence, whether or not AGN feedback actually quenches galaxies in statistically significant numbers remains an open question.

Alternatives to AGN feedback driven quenching of central galaxies do exist in the theoretical literature, and there is some observational support for these as well. Virial shock heating of gas in haloes above some critical dark matter halo mass ($M_{\rm crit} \geq 10^{12} M_{\odot}$) can lead to a stifling of gas supply and hence an eventual shutting off of star formation in galaxies (e.g., Dekel \& Birnboim 2006; Dekel et al. 2009; Dekel et al. 2014). Recent observations suggest that halo mass is more constraining of the quenched fraction of centrals than stellar mass, qualitatively in line with this view (e.g., Woo et al. 2013). However, the stronger dependence of central galaxy quenching on bulge mass and central density (e.g., Bluck et al. 2014, Woo et al. 2015) imply that this cannot be the sole, or dominant, route to quenching centrals. Further to this, elevated gas depletion and supernovae feedback in galaxy mergers, and the growth of the central potential and its stabilizing influence on giant molecular cloud collapse, have both been evoked as potential alternatives to the more commonly utilised AGN feedback (e.g., Martig et al. 2009; Darg et al. 2010; Moreno et al. 2013). To fully distinguish between these various processes careful comparison of observational data to simulations and models must be made.

Satellites are potentially subject to a wide range of additional physical processes for quenching than centrals, resulting from their relative motion across the hot gas halo, and their increased group potential, and galaxy - galaxy, tidal interactions. Processes such as ram pressure stripping, harassment, strangulation from removal of the satellites' hot gas halo, and pre-processing in groups prior to the cluster environment can all lead to a removal of gas or gas supply and hence a reduction and eventual cessation of star formation (e.g., Balogh et al. 2004; Cortese et al. 2006; Font et al. 2008; Tasca et al. 2009; Peng et al. 2012; Hirschmann et al. 2013; Wetzel et al. 2013). Additionally, if a central galaxy enters a group or cluster environment for the first time, transitioning to becoming a satellite, it will no longer reside at a local gravitational minimum in the cosmic web. Thus, cold gas streams will no longer feed the new satellite galaxy and hence this will also contribute to its star formation quenching (e.g., Guo et al. 2011; Henriques et al. 2015). It is important to stress that all of these environmentally dependent quenching processes work in addition to the mass-correlating quenching associated with centrals, and thus that we might expect to see evidence for two distinct regimes in satellite quenching, one where environment dominates and one where internal properties dominate.

In Bluck et al. (2014) we conclude that `bulge mass is king' in the sense that bulge mass is a tighter and steeper correlator to the quenched fraction for centrals than stellar mass, halo mass, disk mass, local galaxy density, and galactic structure ($B/T$). For a smaller list of variables (not including bulge or halo mass) Wake et al. (2012) established that central velocity dispersion outperforms stellar mass, morphology and environment in constraining the quenching of a general population of local galaxies. Recently, Teimoorinia, Bluck \& Ellison (2016) found strong evidence from an ANN technique that central velocity dispersion is the best single variable for parameterizing the quenching of centrals, improving upon even bulge mass. Additionally, Cheung et al. (2012), Fang et al. (2013) and Woo et al. (2015) find strong evidence for the central stellar mass density within 1 kpc being a particularly tight correlator to the quenched fraction. This quantity is also demonstrated to scale tightly with both bulge mass and central velocity dispersion. Taken together, it is clear that a high central mass concentration and hence central velocity dispersion is a prerequisite for quenching central galaxies.

The primary motivation for this paper is to expand on the work of Wake et al. (2012), Bluck et al. (2014) and Teimoorinia et al. (2016) by investigating the impact on the quenched fraction of central and satellites galaxies from varying galaxy and environmental properties at fixed central velocity dispersion. For centrals, this allows us to look for additional dependencies of quenching, whilst controlling for the parameter which matters most statistically. For satellites, fixing the central velocity dispersion allows us to effectively control for the most important intrinsic parameter before studying the impact of environment on these systems. We then compare these results to a cosmological hydrodynamical simulation (Illustris, Vogelsberger et al. 2014a,b) and a semi-analytic model (the Munich model of galaxy formation: L-Galaxies, Henriques et al. 2015), to gain insight into the possible physical processes responsible for our observed results. 

The paper is structured as follows. We give a review of our data sources and measurements in Section 2, and define our quenched fraction method in Section 3. In Section 4 we give a brief overview of our results. Section 5 presents our results for central galaxies, including a new method for ascertaining the statistical influence on the quenched fraction of varying a given galaxy property at fixed other galaxy properties. We discuss the possible interpretations of our results for centrals in Section 6, and make a detailed comparison to a cosmological simulation and a semi-analytic model. In Section 7 we present our results for satellites and compare them to the centrals. We conclude in Section 8. We also include two appendices, the first giving an example of our area statistics approach (Appendix A) and the second showing the stability of our results to different scaling laws (Appendix B). Throughout the paper we assume a $\Lambda$CDM cosmology with H$_{0}$ = 70 km s$^{-1}$ Mpc$^{-1}$, $\Omega_{m}$ = 0.3, $\Omega_{\Lambda}$ = 0.7, and adopt AB magnitude units.

\section{Data Overview \& Parameter Measurements}

\begin{figure*}
\includegraphics[width=0.49\textwidth]{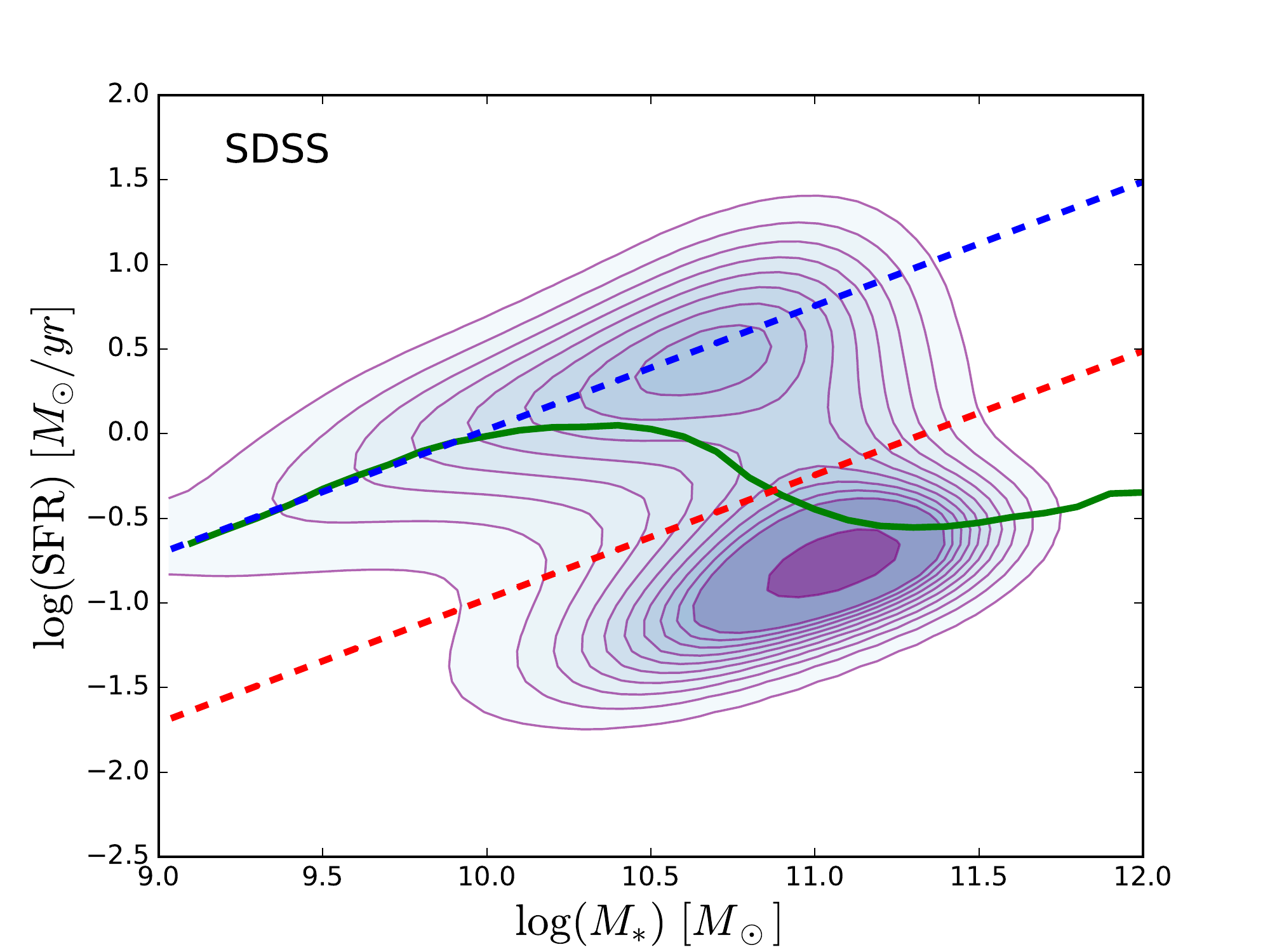}
\includegraphics[width=0.49\textwidth]{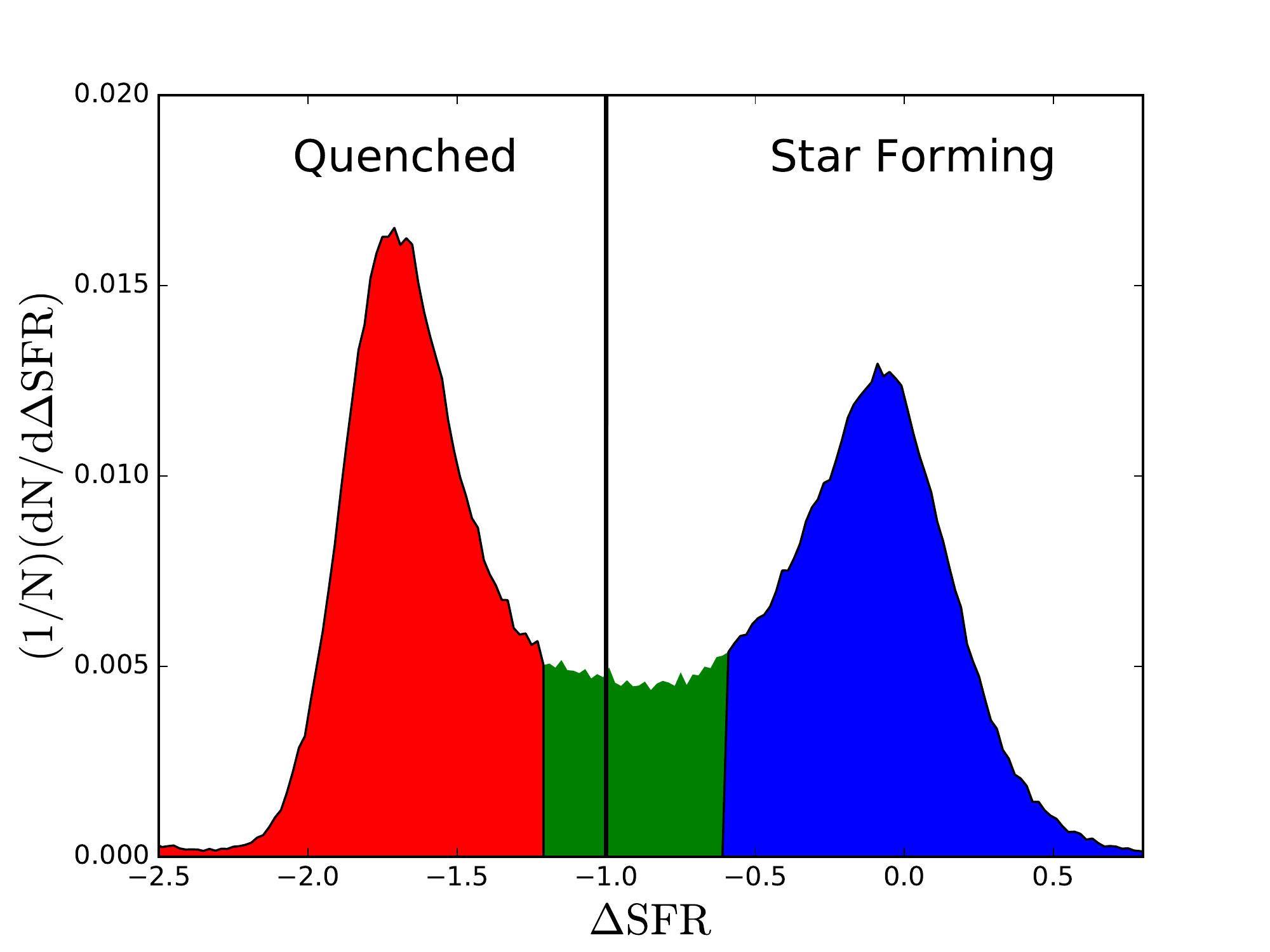}
\caption{{\it Left panel}: the star forming main sequence for SDSS galaxies at z $<$ 0.2. The green line traces the median SFR relation with stellar mass. The blue dashed line indicates a least squares linear fit to the median relation at $M_{*} < 10^{10} M_{\odot}$, which approximates the star forming main sequence (with the average relationship for the redshift range given by: $\log ({\rm SFR} [M_{\odot}/yr]) = (0.73\pm0.05) \times \log(M_{*}[M_{\odot}]) - (7.3 \pm 0.3)$). The red dashed line shows a schematic rendering of our quenched fraction cut, at an order of magnitude below the main sequence relation. {\it Right panel}: Distribution of $\Delta$SFR, which is the logarithmic distance from the star forming main sequence, defined as a function of stellar mass and redshift. Here the main sequence is more precisely defined via star forming emission line galaxies only, which are not AGN, as in Bluck et al. (2014). We define quenched galaxies to have $\Delta {\rm SFR} < -1$, which separates the two peaks effectively. We also consider for some analyses the green valley region, where $-1.2 < \Delta {\rm SFR} < -0.6$. The shaded regions indicate star forming (blue), green valley (green) and quenched, or passive, galaxies (red). }
\end{figure*}

We use the Sloan Digital Sky Survey Data Release 7 (SDSS DR7, Abazajian et al. 2009) spectroscopic sample as our data source. From this we collate a sample of 538046 galaxies (423480 centrals and 114566 satellites) with  $8 < {\rm log}(M_{*}/M_{\odot}) < 12$ at z $<$ 0.2. In this paper we investigate the star forming properties of central and satellite galaxies, as a function of various galaxy and environmental properties. The essential details of these parameters are outlined in this section (but see Bluck et al. 2014 for a more detailed account).

Star formation rates (SFR) are calculated from extinction corrected emission lines (H$\alpha$, H$\beta$, [NII], [OIII]) for non-AGN star forming galaxies and from the strength of the 4000 \AA \hspace{0.05cm} break for non-emission line galaxies and AGN (Brinchmann et al. 2004). To use the emission line method, the strength of each of the BPT (Baldwin, Phillips \&Terlevich 1981) lines must have an S/N $>$ 3 and additionally galaxies must not be identified as AGN via the Kauffmann et al. (2003) line ratio cut. A fibre correction is applied based on galaxy colour and magnitude outside the aperture. This is the same sample of SFRs used in many recent quenching papers (e.g., Woo et al. 2013; Bluck et al. 2014; Woo et al. 2015; Teimoorinia et al. 2016). All of the results and conclusions of this work are recovered qualitatively even if we use photometric SFRs from SED fitting, or construct the analogous red fraction instead of the quenched fraction from star formation rates. This implies that the aperture corrections are not unduly biasing our results on quenching for centrals and satellites, since the photometric techniques do not depend on them.  

The stellar masses for the galaxies, and their component disks and spheroids, are derived in Mendel et al. (2014), based on SED fitting to a dual S\'{e}rsic fit of the $ugriz$ wavebands (Simard et al. 2011). An $n_{s}$ = 4 bulge and $n_{s}$ = 1 disk model is used, and we test the reliability of this approach in Mendel et al. (2014) and Bluck et al. (2014) via model data. We define the galaxy structure (or morphology) to be the bulge-to-total stellar mass ratio: $B/T = M_{\rm bulge} / M_{*}$, where $M_{*}$ is the total stellar mass of the galaxy (defined as $M_{*} = M_{\rm bulge} + M_{\rm disk}$). Similarly, disk-to-total stellar mass ratio is defined as: $D/T = 1-B/T = M_{\rm disk} / M_{*}$.

Velocity dispersions are derived from broadened template fits to the widths of absorption lines taken from Bernardi et al. (2003) with an updated method implemented as in Bernardi et al. (2007) to the later data releases. We use the Princeton velocity dispersion measurements as opposed to the SDSS pipeline (e.g., Bolton et al. 2012) because the latter restricts the sample to early-type spectra and the former does not. Velocity dispersions from absorption lines with a S/N $<$ 3.5 are discarded from our sample, and those with $\sigma$ $<$ 70 km/s are removed from most analyses, due to the instrumental resolution of the SDSS spectra. We also restrict our final sample to galaxies with an error on the velocity dispersion of $\sigma_{\rm err}$ $<$ 35 km/s. $\sim$ 80 \% of our parent sample pass these data quality cuts. To avoid biasing the sample by removal of galaxies without substantial bulge components, we allow the low velocity dispersions to re-enter some of our analyses as `low' values, where we incorporate only the minimal information that there is a low velocity dispersion and do not utilise their specific values for any purpose. For these analyses, we set all velocity dispersions less than 70 km/s equal to 50 km/s, allowing us to retain the information that they are low without inferring anything about their specific values.

For the measured velocity dispersions, we first make an aperture correction, so that all measurements are made at the same effective aperture. We use the formula in Jorgensen et al. (1995), specifically calculating the centralised velocity dispersion as:

\begin{equation}
\sigma_{c}  \equiv  \sigma_{Re/8} = \big(\frac{R_{e}/8}{R_{\rm ap}}\big)^{-0.04} \sigma_{\rm ap}
\end{equation}

\noindent where $\sigma_{\rm ap}$ is the velocity dispersion measured within the aperture, $R_{\rm ap}$ is the aperture radius, and $R_{e}$ is the effective radius of the bulge or spheroid (taken from the morphological catalogues of Simard et al. 2011). The factor of 1/8 is chosen to be in line with the measurements made in the literature. We note that the aperture correction only affects the final estimate of the central velocity dispersion by typically $<$10\%. 

To combat the effect of kinematic contamination on velocity dispersion measurements from disk rotation into the plane of the sky, we restrict all late type galaxies (LTGs, $B/T \leq 0.5$) to be face-on (b/a $>$ 0.9) for the remainder of the paper (although see Appendix B where we relax this criterion). However, this introduces a bias whereby there are far fewer LTGs in our sample relative to early type galaxies (ETGs, $B/T > 0.5$), which will affect the ratio of star forming to passive systems. To counter this, we weight each galaxy by the inverse of the probability of its inclusion in the sample. Specifically, we calculate a weight:

\begin{equation}
w_{i} = \frac{1}{1 - f_{\rm rem}(B/T)_{i}} 
\end{equation}

\noindent where $f_{\rm rem}$ is the fraction of galaxies removed due to our axis ratio cut at the $B/T$ value of each galaxy. For LTGs this varies as a function of morphology, but for ETGs it is equal to one due to the fact that we do not cull ETGs from our sample. This weight is then multiplied by $1/V_{max}$ and used as a new weight for computing each statistic in our analysis, e.g. for the quenched fractions (see Section 3). None of our results or conclusions are strongly affected by restricting to face-on LTGs and weighting (see Appendix B). However, we use this technique as a conservative approach for incorporating velocity dispersions into our analysis, and using these to estimate black hole masses for disk-dominated galaxies, when comparing to models in Section 6. The mean bulge fraction in the fibre for ETGs is $\langle (B/T)_{\rm fib} \rangle \sim 0.9$, indicating that no restriction in their orientation is needed to first order, which also aids our analysis by leaving a substantial number of galaxies to perform our statistics on. 

We consider several measurements of environment in this paper, including halo mass, group/ cluster-centric distance, satellite-halo mass ratio and local (over)-density. The halo masses are derived from an abundance matching technique applied to the total stellar mass of the group or cluster (from Yang et al. 2007, 2008, 2009). Testing of the group finding algorithm on model galaxies from the Millennium Simulation (Springel et al. 2005) showed that over 90\% of galaxies are correctly assigned to groups at $M_{\rm halo} > 10^{12} M_{\odot}$. Using these group catalogues, centrals are defined as the most massive galaxy in the group, with satellites being any other group member. The projected distance of each satellite to its central galaxy, in units of the virial radius, is used as another environmental metric in this work (defined as: $D_{cc} = R/R_{vir}$). Where $R$ is the projected distance to the central, and $R_{vir}$ is the virial radius. We define the mass ratio:

\begin{equation}
\mu_{*} = \frac{M_{*,{\rm sat}}}{M_{\rm halo}}
\end{equation} 

\noindent which indicates the relative mass of the satellite to the halo, and hence is a measure of how major or minor a component of the group or cluster the satellite is. We also use measurements of the the normalised surface galaxy density (based on measurements in Baldry et al. 2006). Over-densities are computed as:

\begin{equation}
\delta_{n} = \frac{\Sigma_{n}}{\langle \Sigma_{n}(\pm \delta z) \rangle}
\end{equation}

\noindent where

\begin{equation}
\Sigma_{n} = \frac{n}{\pi r^{2}_{p,m}}
\end{equation}

\noindent where $r_{p,n}$ is the projected distance to the nth nearest neighbour, and $\langle \Sigma_{n}(\pm \delta z) \rangle$ is the mean value of the density parameter at each 0.01 redshift slice, which normalises the the density parameter accounting for the flux limit of the SDSS.

Full details on the observational data and measurements used in this paper are given in the prior works of this series, Bluck et al. (2014) and Teimoorinia et al. (2016). In addition to the SDSS observations, we also compare to the Illustris simulation (Vogelsberger et al. 2014a,b) and to the latest version of the Munich model of galaxy formation (L-Galaxies, Henriques et al. 2015). We select the output galaxy catalogues at z = 0.1 (equivalent to the median redshift in our observations) and take all measurements (e.g., stellar, halo and black hole mass and star formation rate) from these public catalogues. More details on the simulations are provided in Section 6.1.

\section{Defining the Quenched Fraction}

We follow the prescription for defining the quenched (or passive) fraction in Bluck et al. (2014) and Teimoorinia, Bluck \& Ellison (2016). A galaxy is defined to be passive if it is forming stars at a rate at least a factor of ten times lower than (emission line, non-AGN) star forming galaxies, matched at the same stellar mass and redshift. We start by defining the main sequence as the ${\rm SFR} - M_{*}$ relation for star forming galaxies. Star forming galaxies are defined observationally as emission line galaxies (with S/N $>$ 5), which are furthermore not identified as AGN by the Kauffmann et al. (2003) line cut on the Baldwin, Phillips \& Terlevich (1981, BPT) emission line diagnostic plot. This relationship is shown for our sample in Fig. 5 of Bluck et al. (2014).  We then determine the logarithmic distance each galaxy in the SDSS resides at from the star forming main sequence, computing:

\begin{equation}
\Delta {\rm SFR} = \log \bigg( \frac{{\rm SFR}(M_{*},z)}{{\rm median}({\rm SFR}_{SF}(M_{*} \pm \delta M_{*}, z \pm \delta z)} \bigg) \hspace{0.3cm} ,
\end{equation}

\noindent where SFR$_{SF}$ indicates the star formation rate of main sequence star forming galaxies matched by stellar mass and redshift for each galaxy in the full SDSS sample. Matching thresholds are set to 0.1 dex for stellar mass and 0.005 for redshift. Typically $>$ 200 star forming `controls' are found for each galaxy, and only a few percent require a broadening of these thresholds to find at least ten matches. The star forming main sequence and the distribution in the $\Delta$SFR statistic are shown in Fig. 1.

A threshold of $\Delta$SFR $<$ -1 cleanly divides the star forming and passive peak (see Fig. 1, right panel). Furthermore, we emphasise here that our results are not sensitive to the exact location of the cut. Varying the $\Delta$SFR threshold anywhere throughout the green valley region ($-1.2 < \Delta {\rm SFR} < -0.6$, indicated in green in Fig. 1 right panel) leads to almost identical results, and no change in the conclusions of this work.

The quenched fraction is then defined as the ratio of quenched-to-total galaxies in each binning of galaxy or environmental parameters. We correct for the flux limits of the SDSS by weighting each galaxy in the quenched fraction by the inverse of the volume over which its $ugriz$ magnitudes would pass all of the selection criteria ($1/V_{\rm max}$), which varies as a function of stellar mass and colour (see Mendel et al. 2014). Specifically we calculate:

\begin{equation}
f_{{\rm Quench}, j} = \sum_{i} \bigg( \frac{(w_{i}/V_{{\rm max}, i}) [\Delta {\rm SFR} < -1]}{(w_{i}/V_{{\rm max}, i}) [{\rm ALL}]} \bigg) \hspace{0.3cm} ,
\end{equation}

\noindent where $w_{i}$ is the weighting from the inclination cut, given by eq. 2. The errors on the quenched fraction are computed in this work via the jack-knife technique, as in Bluck et al. (2014), which we find to give comparable results to a full Monte Carlo implementation taking into account the errors on all galaxy properties. In general, both of these more sophisticated techniques lead to a larger total error on average than in the simple Poisson statistics case. See Bluck et al. (2014) \S 2 \& 3 for full details on these data and techniques.

For comparison to simulations and models later in the paper, we define a simplified version of our quenched fraction criterion. In general, the models do not have reliable enough information on emission lines to construct the main sequence identically to how we proceed with the observational data (outlined above). Thus, we must construct an alternative method. It is common in the literature for such comparisons to be made at fixed sSFR (= ${\rm SFR}/M_{*}$). However, given that the normalisation of the main sequence varies from model to model, this is not an ideal way to define the main sequence and hence quenched fraction, and can lead to systematic error in the quenched fraction dependence on galaxy properties.

In Fig. 1 (left panel) we show the median relationship of SFR with stellar mass (green line), and note that this is very close to a straight line at $M_{*} < 10^{10} M_{\odot}$. As such, we construct a linear fit to the median main sequence relation at low masses (shown as a blue dashed line). This method relies on the fact that the median galaxy at low masses resides on the star forming main sequence, which is reasonable. The linear fit goes cleanly through the centre of the density contours of the star forming main sequence, indicating that it is indeed a successful approach for defining the main sequence, in lieu of more sophisticated emission line diagnostics. We then define galaxies to be quenched exactly as before, i.e. if they lie one order of magnitude or greater below the main sequence ($\Delta$SFR $<$ -1, indicated by a red dashed line in Fig. 1 left panel). All of our observational results are identical if we use either method to define quenched galaxies, once care is taken to perform this at each redshift slice separately. Thus, the rendering in Fig. 1 shows a schematic only of the method. We use this simplified approach for the simulated and model data (which are taken at a fixed redshift slice), avoiding complicated issues of emission line diagnostics in the models.

\begin{figure}
\includegraphics[width=0.49\textwidth]{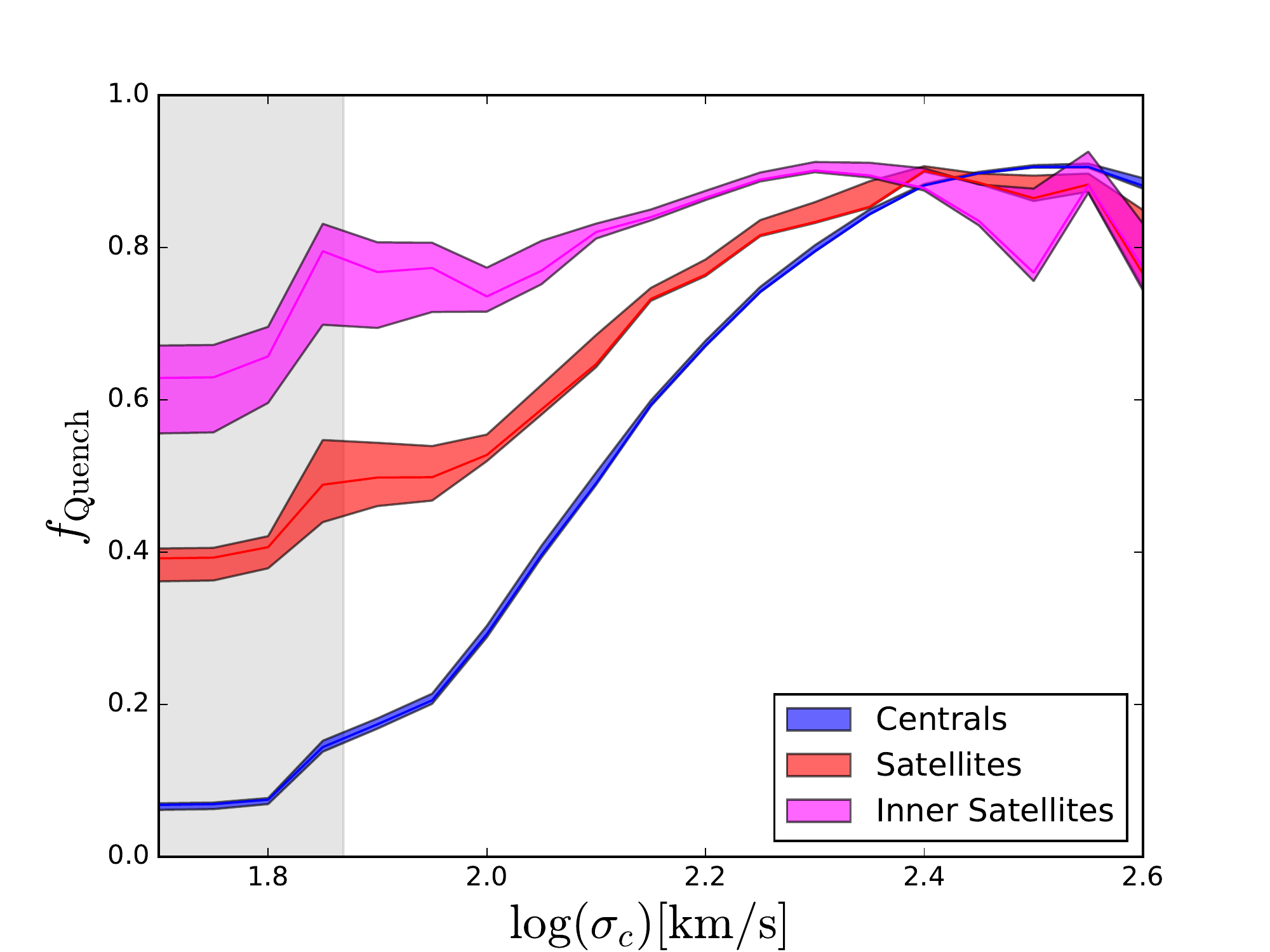}
\caption{The quenched fraction -- central velocity dispersion relationship for central, satellite, and inner-satellite galaxies. The 1 $\sigma$ error on the quenched fraction is computed via the jack-knife technique in each binning, and shown as the shaded region for each sub-sample. Central galaxies are taken to be the most massive members of their groups or clusters, with satellites being any other group member. Inner satellites are defined as satellites within 0.1 virial radii (projected) of their central. At a fixed $\sigma_{c}$, satellites are more frequently quenched than centrals, with inner satellites being more frequently quenched than the general satellite population. This effect is much more pronounced at low $\sigma_{c}$, and disappears entirely at $\sigma_{c} > 250$ km/s.}
\end{figure}

\section{Results Overview}

Recent observations have established that the quenched (or red) fraction of central galaxies is most tightly correlated with the inner regions of these galaxies, e.g., surface mass density within 1 kpc, bulge mass or central velocity dispersion (Cheung et al. 2012; Wake et al. 2012; Fang et al. 2013; Bluck et al. 2014; Lang et al. 2014; Omand et al. 2014; Woo et al. 2015). Teimoorinia et al. (2016) found strong evidence from an ANN analysis that central velocity dispersion is the most predictive, and hence most tightly constraining, observable for central galaxy quenching out of the following list of variables: stellar, halo, bulge and disk mass; local galaxy density and galactic structure ($B/T$). Moreover central velocity dispersion is found to be tightly correlated with surface mass density within 1 kpc. In this work we concentrate on central velocity dispersion because it is more responsive to differences in the structural properties of host galaxies and furthermore has better calibrated empirical relationships with dynamically measured black hole mass (e.g., Saglia et al. 2016). Throughout the results sections we explore the quenched fraction dependence on various galaxy and environmental parameters, at a fixed central velocity dispersion, for central and satellite galaxies. The aim of this approach is to establish to what extent other parameters affect central and satellite galaxy quenching, in what manner (i.e. do they lead to positive or negative correlations at fixed $\sigma_{c}$?), and ultimately compare the observational results to predictions from contemporary simulations and models (in Section 6).

\subsection{Comparison of the Quenched Fractions of Central and Satellite Galaxies at Fixed Central Velocity Dispersion}

In Fig. 2 we show the quenched fraction relationship with central velocity dispersion, for central, satellite and inner satellite galaxies. Centrals are defined as the most massive galaxy in the group, according to the SDSS group catalogues of Yang et al. (2007, 2009). Satellites are any other group members, with inner satellites being satellites found within 0.1 virial radii (projected). This plot may be compared to the differences between central and satellite galaxies at fixed $M_{*}$, $B/T$ and $M_{\rm halo}$ shown in Bluck et al. (2014). The grey region indicates $\sigma_{c} < 70$ km/s, which is approximately the instrumental resolution of the SDSS spectrograph. It is clear that there is a strong dependence of central galaxy quenching on central velocity dispersion, with a progressively weaker dependence for satellites and inner satellites. At a fixed $\sigma_{c}$, satellites are in general more frequently passive than centrals, and inner satellites are more frequently passive than the general satellite population. This effect is significantly more pronounced at low central velocity dispersions, and disappears entirely by $\sigma_{c} > 250$ km/s.

Central galaxies have a 50\% chance of being quenched at an average central velocity dispersion of $\sigma_{c} = 140 \pm 5 {\rm km/s}$, with satellites achieving a 50\% quenched fraction at a significantly lower central velocity dispersion of $\sigma_{c} =  90 \pm 5 {\rm km/s}$. Interestingly, inner satellites are always more than 50\% quenched in every central velocity dispersion range we consider, down to at least the spectroscopic resolution of $\sim$ 70 km/s. 

The higher frequency of quenched satellite and inner satellite galaxies at low central velocity dispersions, relative to centrals, suggests that environmental processes are important in the quenching of these systems (as argued for in, e.g., van den Bosch et al. 2008; Baldry et al. 2008; Peng et al. 2010, 2012). For centrals, the very low fraction of quenched systems at low central velocity dispersion, and steep rise in probability of being quenched out to higher central velocity dispersions, is qualitatively consistent with quenching from AGN feedback (in either the radio or quasar mode, e.g. Croton et al. 2006; Hopkins et a. 2008). This is due to the observed $M_{BH} - \sigma$ relation (e.g., McConnell \& Ma 2013; Saglia et al. 2016), and the strong dependence of AGN driven quenching on supermassive black hole mass in most models (e.g., Henriques et al. 2015; Schaye et al. 2015). However, given the many inter-correlations between galaxy properties, it is not yet established which, if any, galaxy property is truly fundamental to central galaxy quenching, and hence which physical mechanism(s) are the most probable cause.

Due to the observed differences in the quenched fraction relation with central velocity dispersion between central and satellite galaxies, we consider each of these populations separately throughout our analyses in the following results sections.

\begin{figure*}
\includegraphics[width=0.49\textwidth]{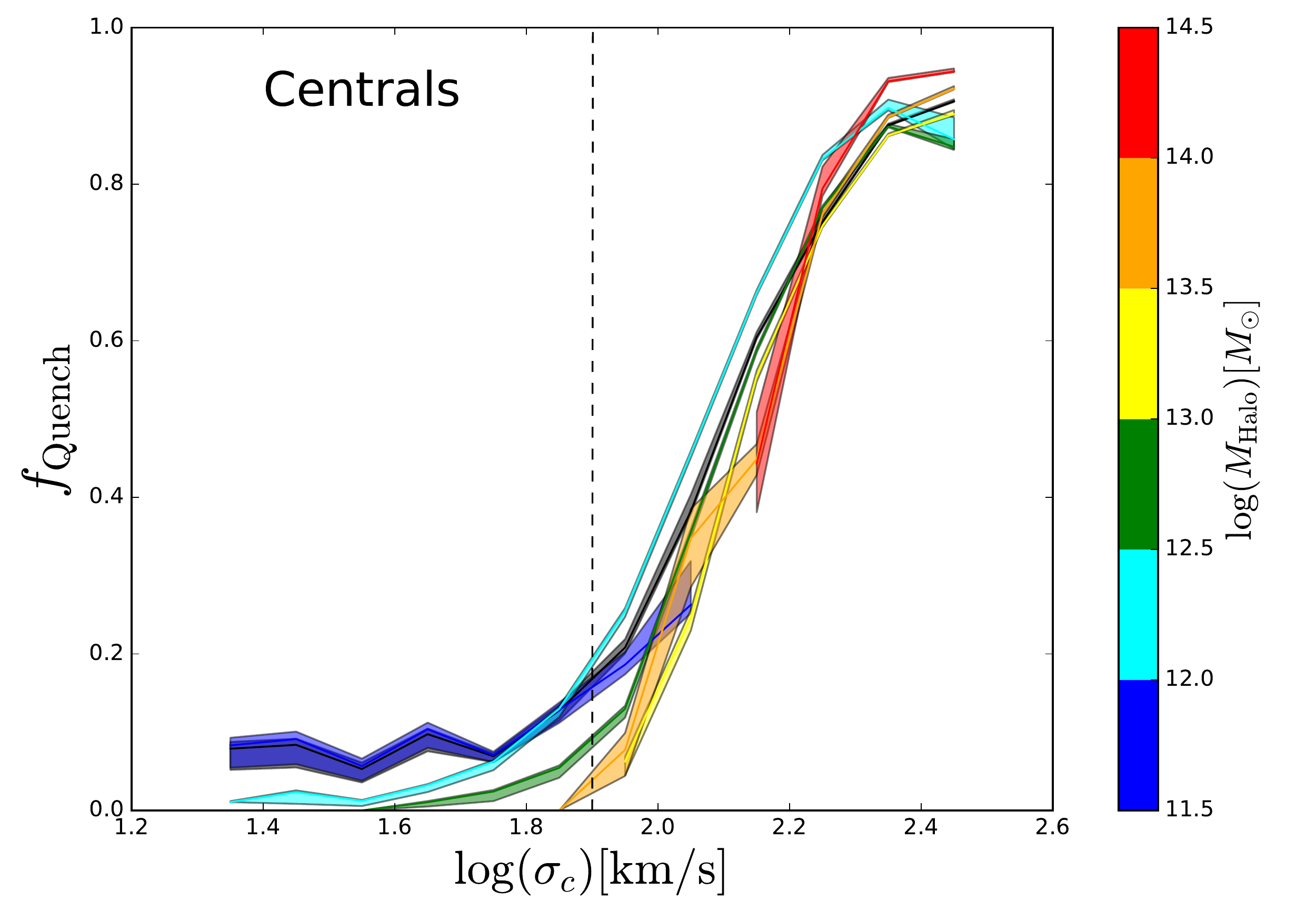}
\includegraphics[width=0.49\textwidth]{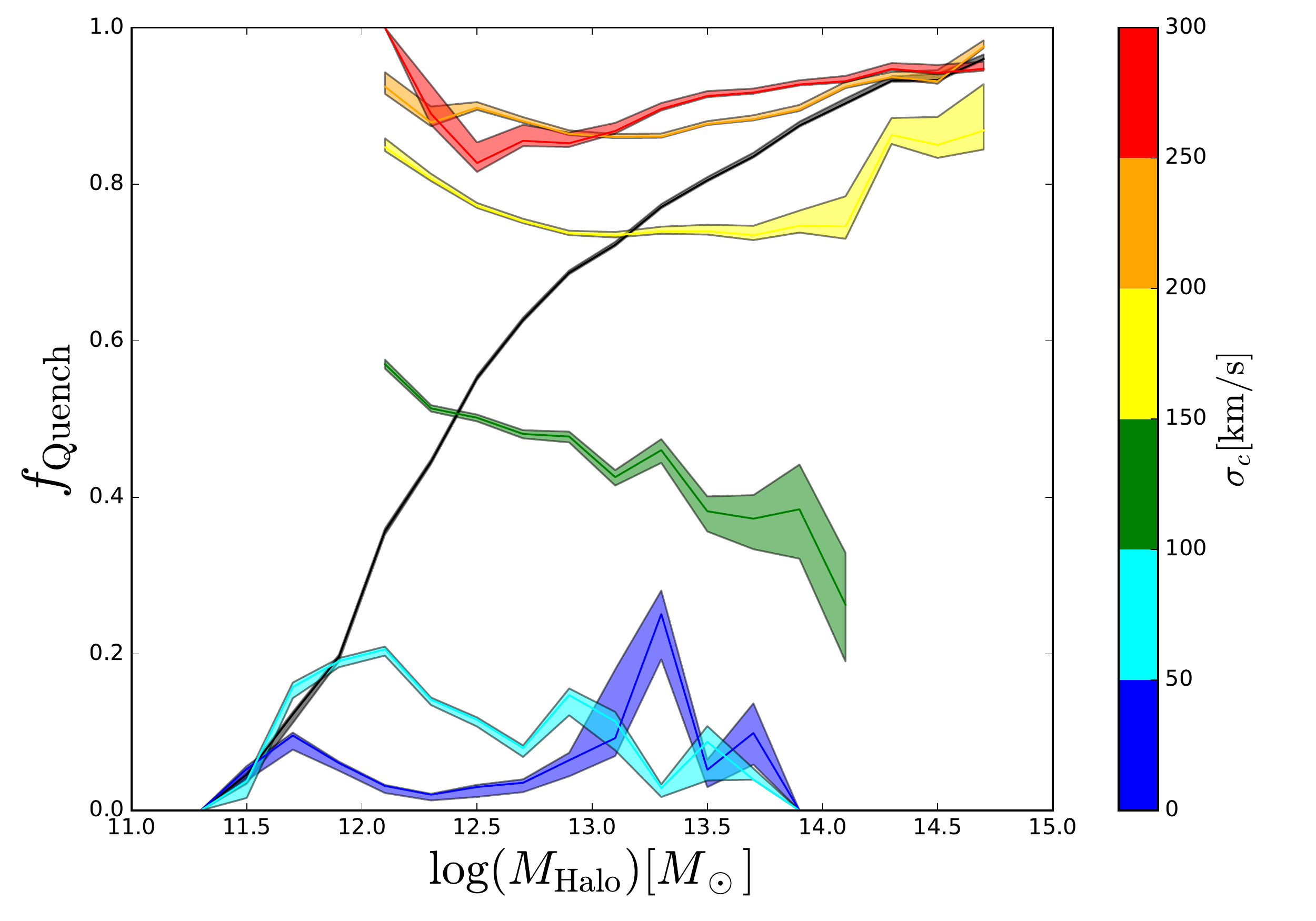}
\includegraphics[width=0.49\textwidth]{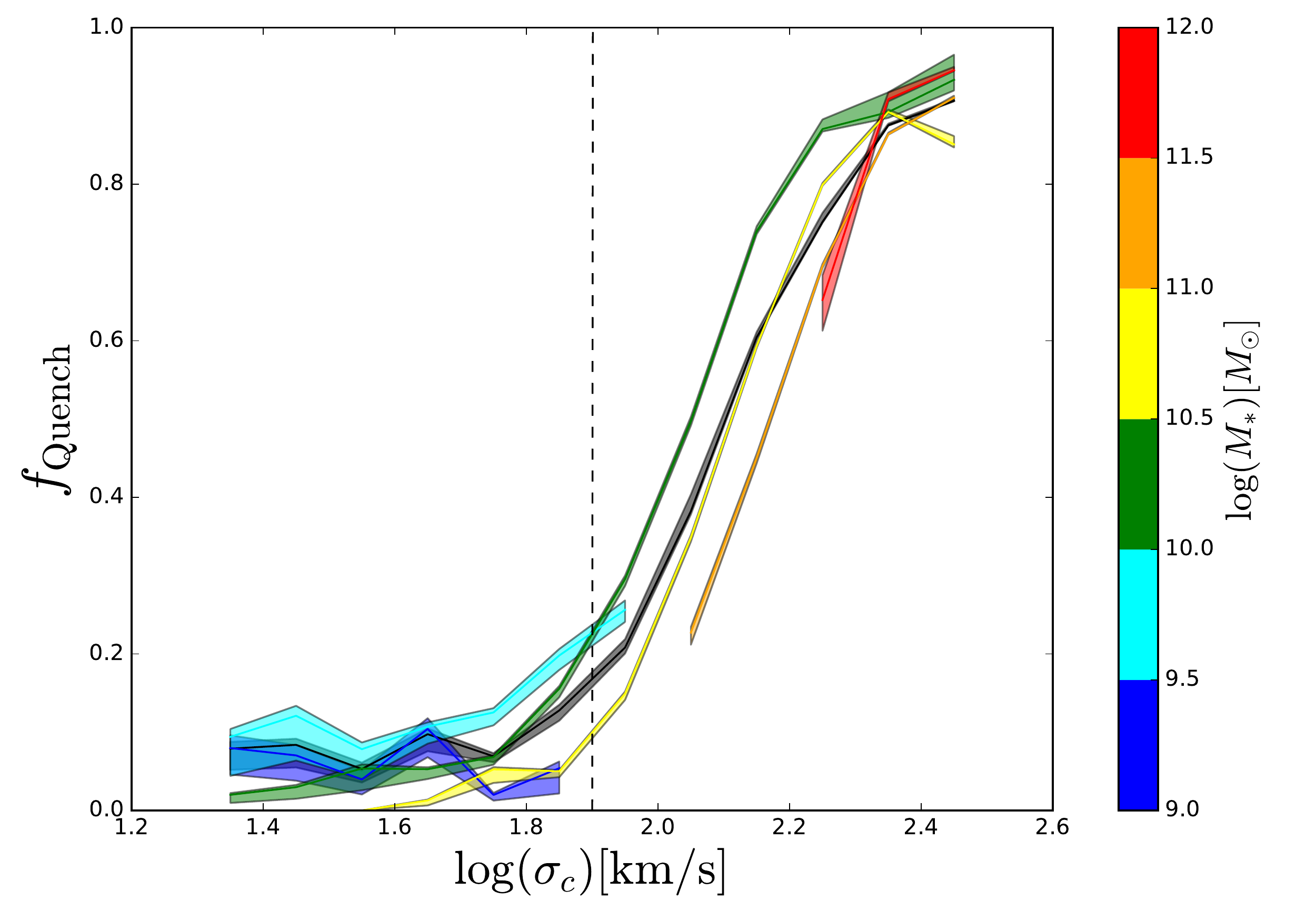}
\includegraphics[width=0.49\textwidth]{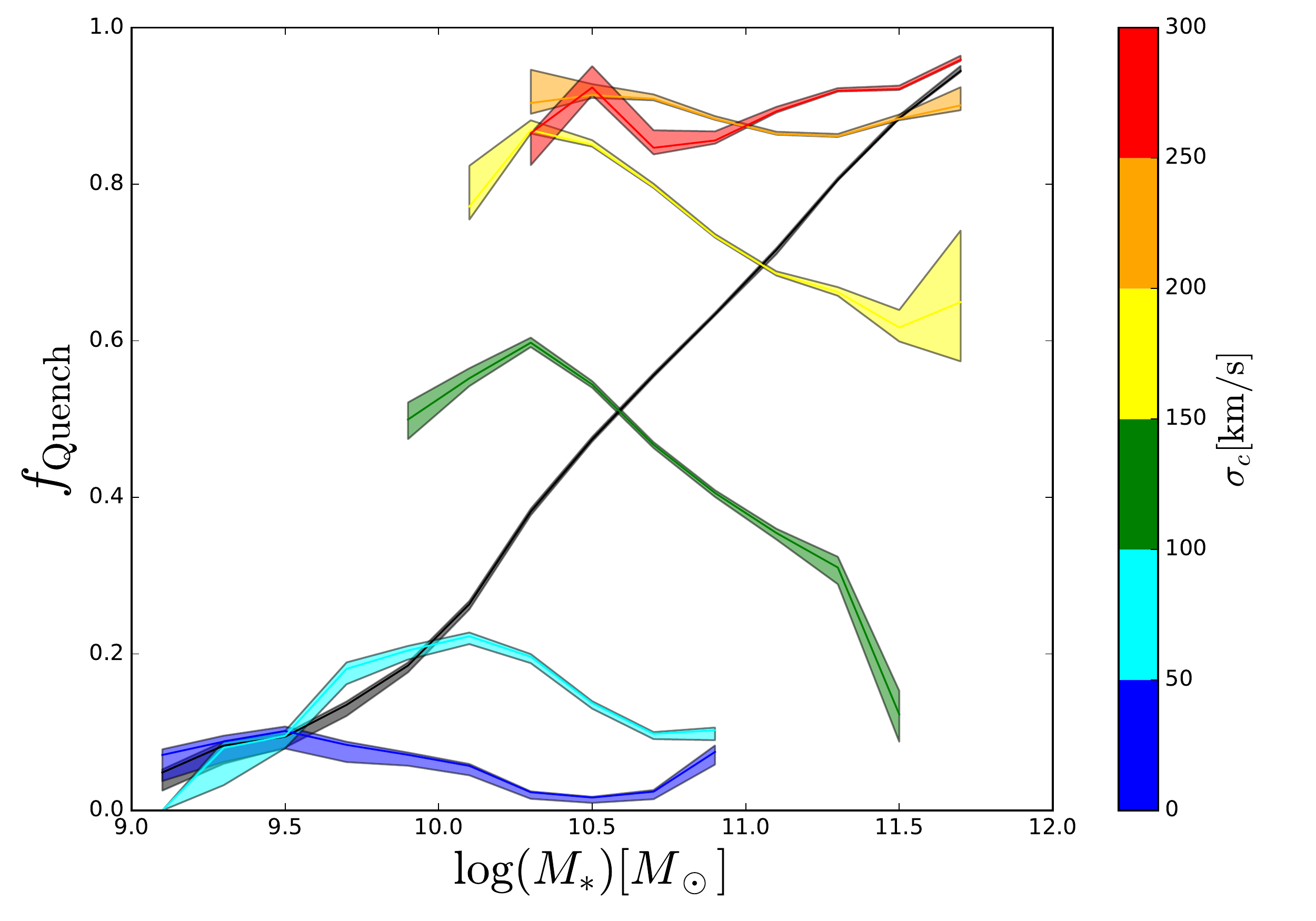}
\includegraphics[width=0.49\textwidth]{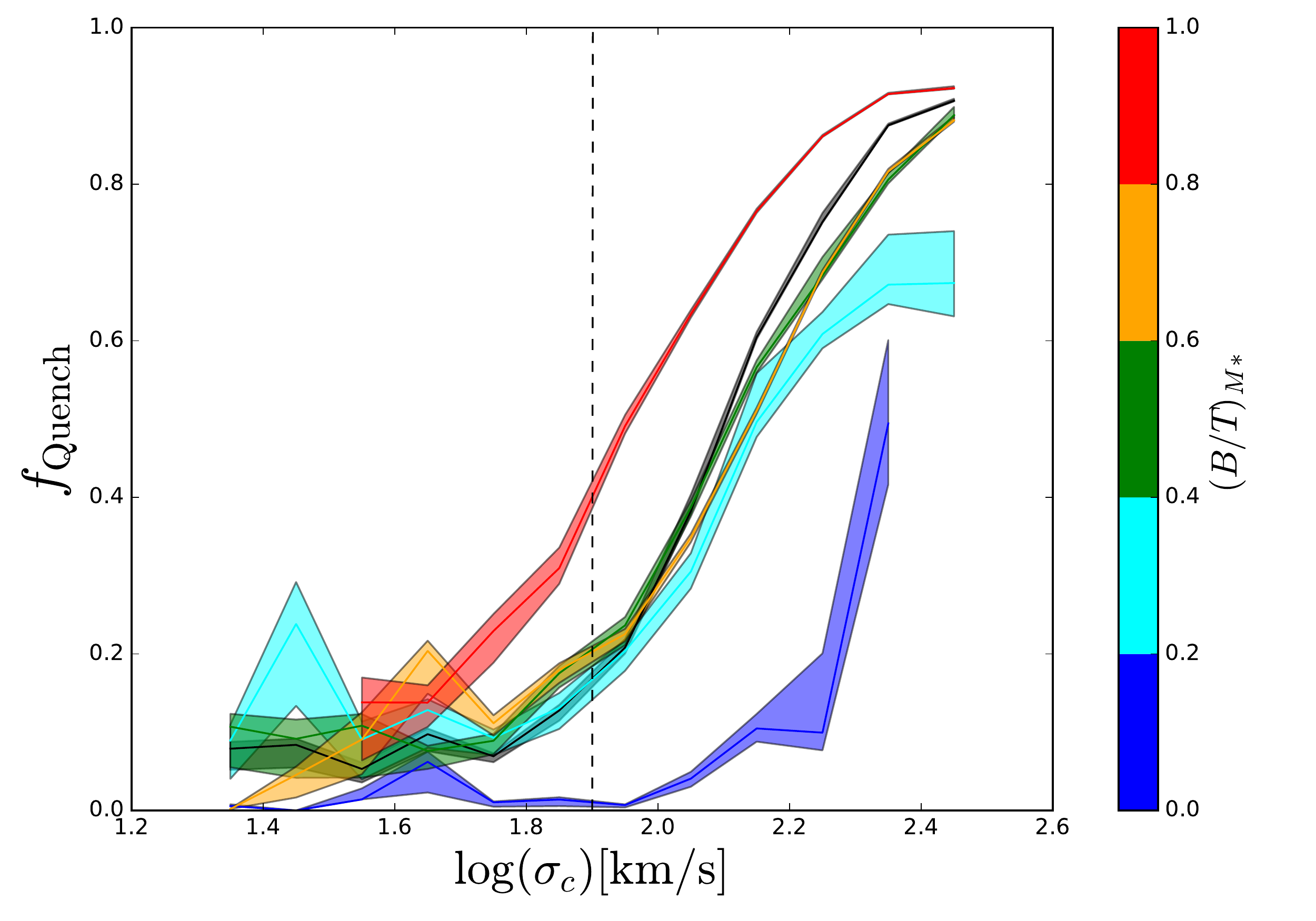}
\includegraphics[width=0.49\textwidth]{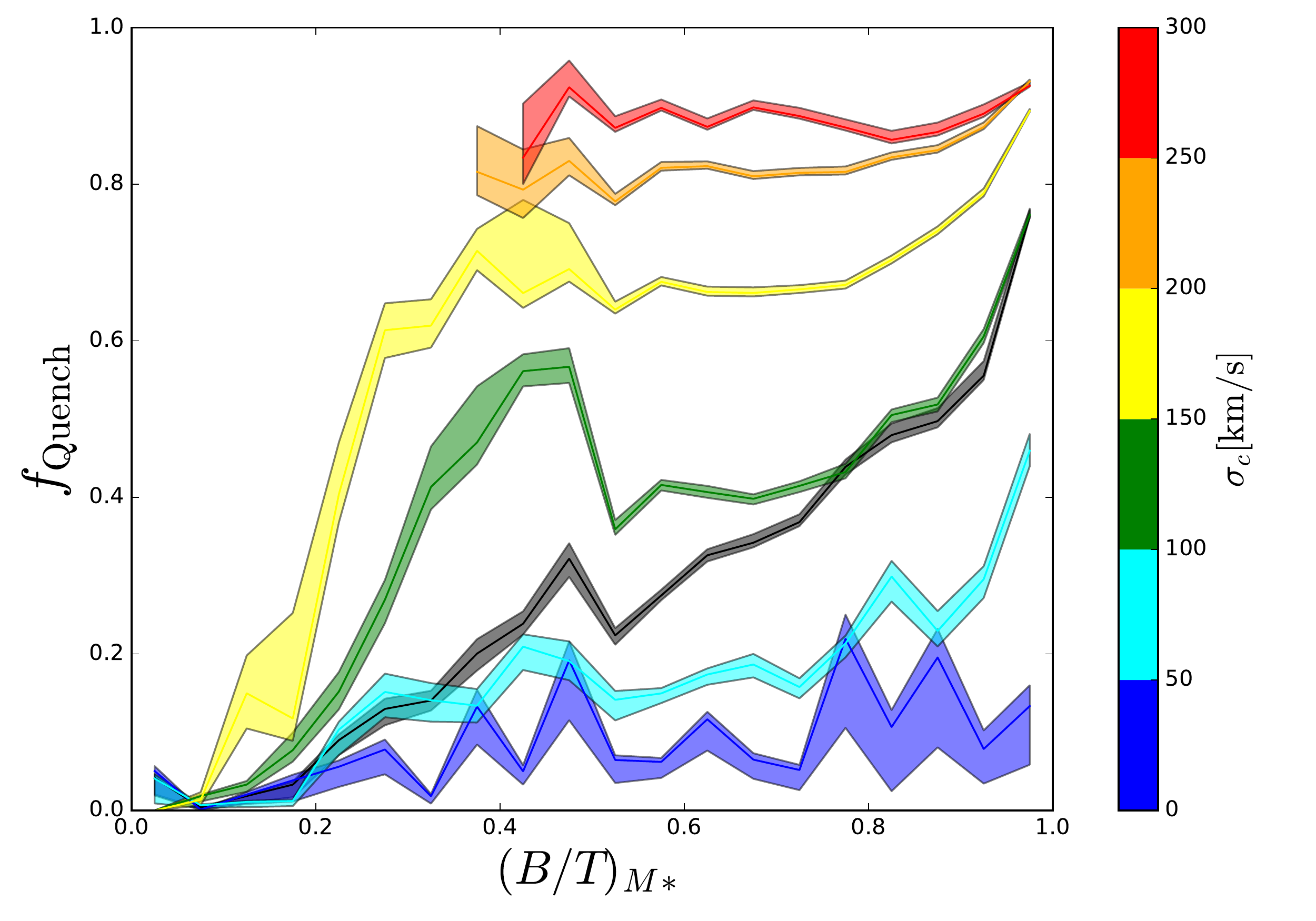}
\caption{The quenched fraction dependence for central galaxies on: {\it left column} - central velocity dispersion, subdivided (from top to bottom) by group halo mass, total stellar mass, and bulge-to-total stellar mass ratio ($B/T$); {\it right column} - group halo mass, total stellar mass, and $B/T$, each subdivided by central velocity dispersion. The black lines shows the unbinned relationships, with the coloured lines showing the relationships at fixed values of the quantities indicated in the colour bar. Varying stellar or halo mass at fixed central velocity dispersion leads to essentially no impact on the quenched fraction (top left panels), whereas varying central velocity dispersion at fixed stellar or halo mass dramatically affects the quenched fraction (top right panels). Both $\sigma_{c}$ and $B/T$ affect the quenched fraction at fixed values of the other parameter; however, the affect on the quenched fraction of varying $\sigma_{c}$ at fixed $B/T$ is larger than the other way around. The shaded colour regions represent the 1 $\sigma$ error on the quenched fraction computed via the jack-knife technique. }
\end{figure*}

\section{Results for Centrals}

\subsection{The Relationship Between Quenched Fraction and Central Velocity Dispersion, at fixed Stellar Mass, Halo Mass and Galaxy Structure}

Correlation does not imply causation; thus, we must be cautious of claiming a physical connection between central velocity dispersion and the quenching of central galaxies. One simple test is to determine whether or not the $f_{\rm Quench} - \sigma_{c}$ relation is still evident when other galaxy properties are held constant, and additionally to explore the corollary, of whether or not, e.g., the $f_{\rm Quench} - M_{*}$, $M_{\rm halo}$ and $B/T$ relations are still evident when $\sigma_{c}$ is held constant. The left column of Fig. 3 shows the $f_{\rm Quench} - \sigma_{c}$ relation for central galaxies, in fixed ranges of (from top to bottom): $M_{\rm halo}$, $M_{*}$ and $B/T$. Varying the halo mass or stellar mass of galaxies at constant central velocity dispersion (by even three orders of magnitude) has very little impact of the fraction of quenched galaxies. Furthermore, the $f_{\rm Quench} - \sigma_{c}$ relationships at fixed ranges in stellar or halo mass (shown as coloured lines, labelled by the colour bar) are almost identical to the unbinned relationship (shown in black). $B/T$, on the other hand, does lead to a significant impact on the fraction of quenched galaxies at a fixed $\sigma_{c}$ (bottom left panel). This notwithstanding, the $f_{\rm Quench} - \sigma_{c}$ relation does remain evident and steep, even for a constant range in galaxy structure ($B/T$).

The right column of Fig. 3 shows (from top to bottom) the relationship between quenched fraction and halo mass, stellar mass and $B/T$ structure. The unbinned relations are shown in black and the relations at constant central velocity dispersion are shown in varying colours (labelled by the colour bar). For both halo and stellar mass, the positive relationship with quenched fraction in the unbinned case is entirely transformed when binned by central velocity dispersion. There is in fact no evidence for a positive correlation between the fraction of quenched centrals and their total stellar mass or the mass of their dark matter haloes, at constant central velocity dispersion. Moreover, in some $\sigma_{c}$ ranges there is even evidence for an anti-correlation between quenched fraction and mass in stars or halo. Thus, it is highly unlikely that either halo mass or stellar mass can be causally related to the quenching of central galaxies, given that their correlations with the quenched fraction are entirely dependent on a third quantity, namely central velocity dispersion. 

In the bottom right panel of Fig. 3, we find a residual dependence of quenched fraction on $B/T$ structure, even at fixed central velocity dispersion. However, this is mostly evident at low $B/T$ and at low central velocity dispersion, where our measurements of the pressure supported kinematics are most uncertain. Furthermore, the effect on the quenched fraction of varying $\sigma_{c}$ at fixed $B/T$ (bottom right) is significantly larger than the other way around (bottom left). Thus, both $B/T$ and $\sigma_{c}$ affect the quenched fraction of central galaxies at fixed values of the other parameter, but $\sigma_{c}$ has a larger impact on quenching than $B/T$. We discuss the possible meaning of these results in the discussion, including a comparison to simulations (see Section 6).

\subsection{Area Statistics Approach}

\subsubsection{Method}

In the previous sub-section we investigate the $f_{\rm Quench} - \sigma_{c}$ relationship at fixed values of several other galaxy properties, and make some general inferences from the structure of these plots. However, it is desirable to be more quantitative about this process. One potential issue with the fixed variable approach of \S 5.1 is how to choose the range in each parameter to set fixed. We solve this issue here by first binning the data by one variable (e.g., $\sigma_{c}$) and then sorting the data by a secondary variable (e.g., $M_{\rm halo}$). We then construct the weighted quenched fraction for percentile ranges of the second variable at each bin in the first variable. For example, for our fiducial definition (${\rm Area_{50}}$) we compute the area contained within the passive fraction computed for upper and lower 50\% of the data in the secondary variable, for each value of the primary variable. This requires no a priori bin structure, and moreover always utilises 100\% of the data available in assessing the quenched fraction dependence, i.e. it is much less sensitive to outliers than the fixed binning approach where some bins may contain only a few percent of the data. Another weakness of the qualitative approach of \S 5.1 is that we can identify which parameter is more important to quenching, but not by how much or at what confidence level. To combat this we use our new percentile range quenched fraction plots to construct two new statistics. 

First, we define the area contained within the quenched fraction relationship between upper and lower percentile ranges as:

\begin{equation}
{\rm Area = \frac{1}{\Delta \alpha} \int^{\alpha max}_{\alpha min} \big| f_{Q}(\alpha|_{\beta (upp)}) - f_{Q}(\alpha|_{\beta (low)}) \big| \hspace{0.1cm} d\alpha }
\end{equation}

\noindent where $\alpha$ indicates the primary variable (i.e. the x-axis of the quenched fraction plot) and $\beta$ indicates the secondary variable, i.e. the variable we sort by to obtain the percentile ranges of the quenched fraction, $f_{Q}$ (which we have abbreviated from $f_{\rm Quench}$). For example, in the top left plot of Fig. 4, $\alpha = \sigma_{c}$ and $\beta = M_{\rm halo}$. The top right plot swaps these variables around. A larger area for varying $\beta$ at fixed $\alpha$ than the other way around indicates that variable $\beta$ is more constraining of the quenched fraction than variable $\alpha$. 
The error on the area statistic is computed by adding in quadrature the positive and negative errors on the upper and lower percentile range (respectively), which are themselves constructed by convolving the jack-knife 1 $\sigma$ statistical error on each binning over the full range in the primary variable ($\Delta \alpha = \alpha_{\rm max} - \alpha_{\rm min}$). Note that the areas are defined to be positive due to the modulus in the definition. Thus, they give a prescription for ascertaining which parameter out of a set of two is more constraining of the quenched fraction, but they do not determine whether the impact on the quenched fraction is positive or negative. In all plots and tables the area statistics are quoted for upper and lower 50\% binnings, i.e. it is in effect ${\rm Area_{50}}$ which we show. We recover qualitatively similar results for all reasonable definitions, e.g. upper and lower 25\% which is shown as a light shaded region on each area statistic plot (see, e.g., Fig. 4 \& 5).

We define another statistic which is sensitive to the directionality of the dependence (i.e. whether increasing a given parameter at fixed values of another parameter increases or decreases the quenched fraction). This statistic is weighted by the number of data points in each range. Thus, we define the weighted average difference as:

\begin{equation}
{\rm Avg = \langle\Delta f_{Q}\rangle = \frac{\sum_{i} \big( f_{Q}(\alpha_{i}|_{\beta (upp)}) - f_{Q}(\alpha_{i}|_{\beta (low)}) \big) \times N_{i}}{\sum_{i} N_{i}}  }
\end{equation}

\noindent where $\alpha$ and $\beta$ are defined as before, and $N_{i}$ is the number of galaxies in each $\alpha$-binning. Note that this statistic can be positive or negative, depending upon how variable $\beta$ impacts the quenched fraction at fixed values of variable $\alpha$. The errors on $\langle\Delta f_{p}\rangle$ are computed in exact analogy to the errors on the area statistic. The area statistic can be used to determine which variable leads to the tighter quenched fraction relationship for each row in the area plots (e.g., Figs. 4 \& 5), and the weighted average difference gives the directionality of the trend (positive or negative). As with the Area statistic, all average differences are quoted for upper and lower 50\% of the secondary variable across the range in the primary variable.

In Appendix A a set of examples are given, demonstrating how the area statistics approach works on simulated data. This is intended to build intuition with the approach, and we recommend reading this appendix before continuing with the results sections. At this point we reintroduce the observational data, and construct areas and average differences for a number of interesting physical parameter pairings for centrals.

\begin{figure*}
\includegraphics[width=0.49\textwidth]{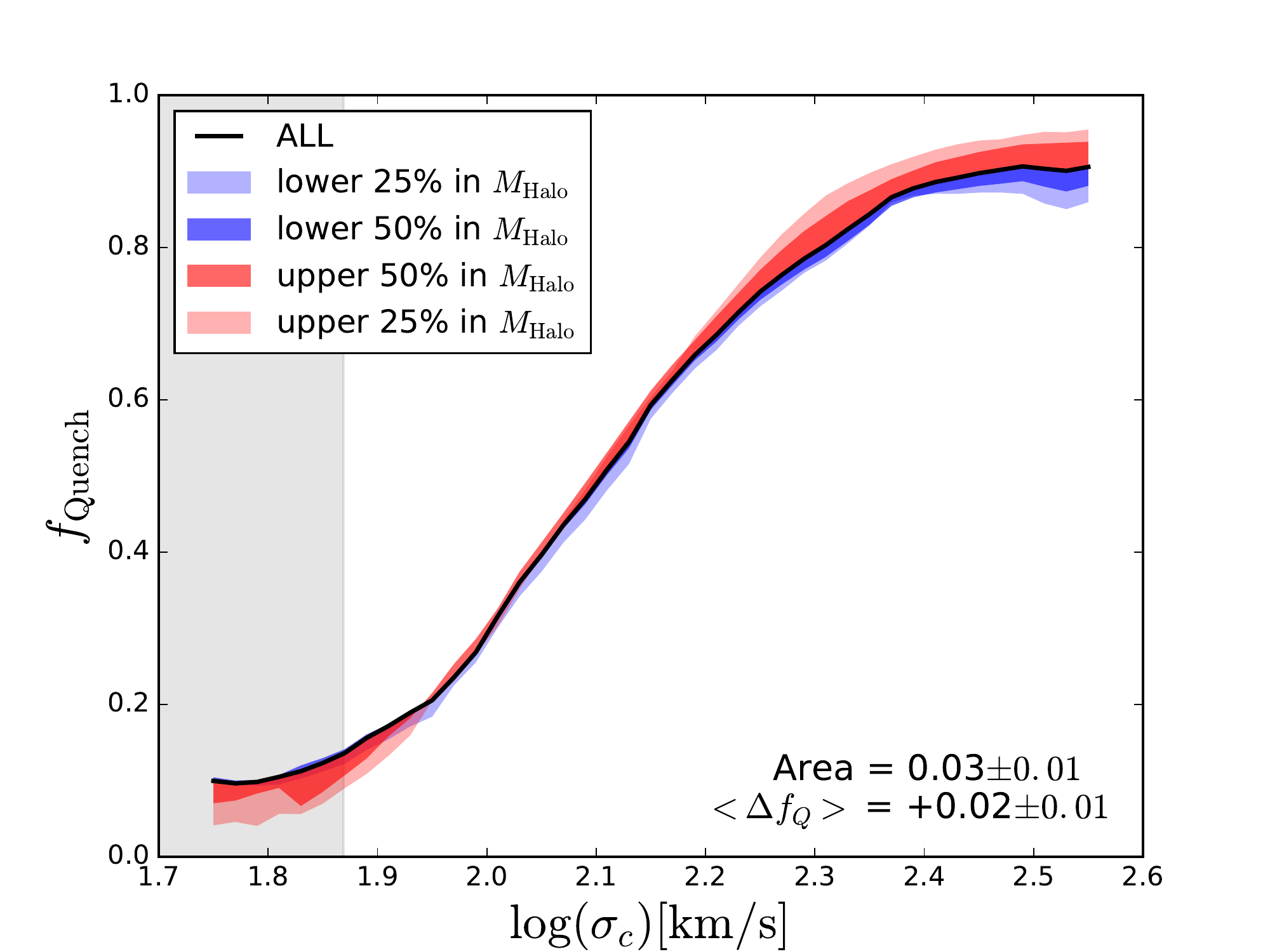}
\includegraphics[width=0.49\textwidth]{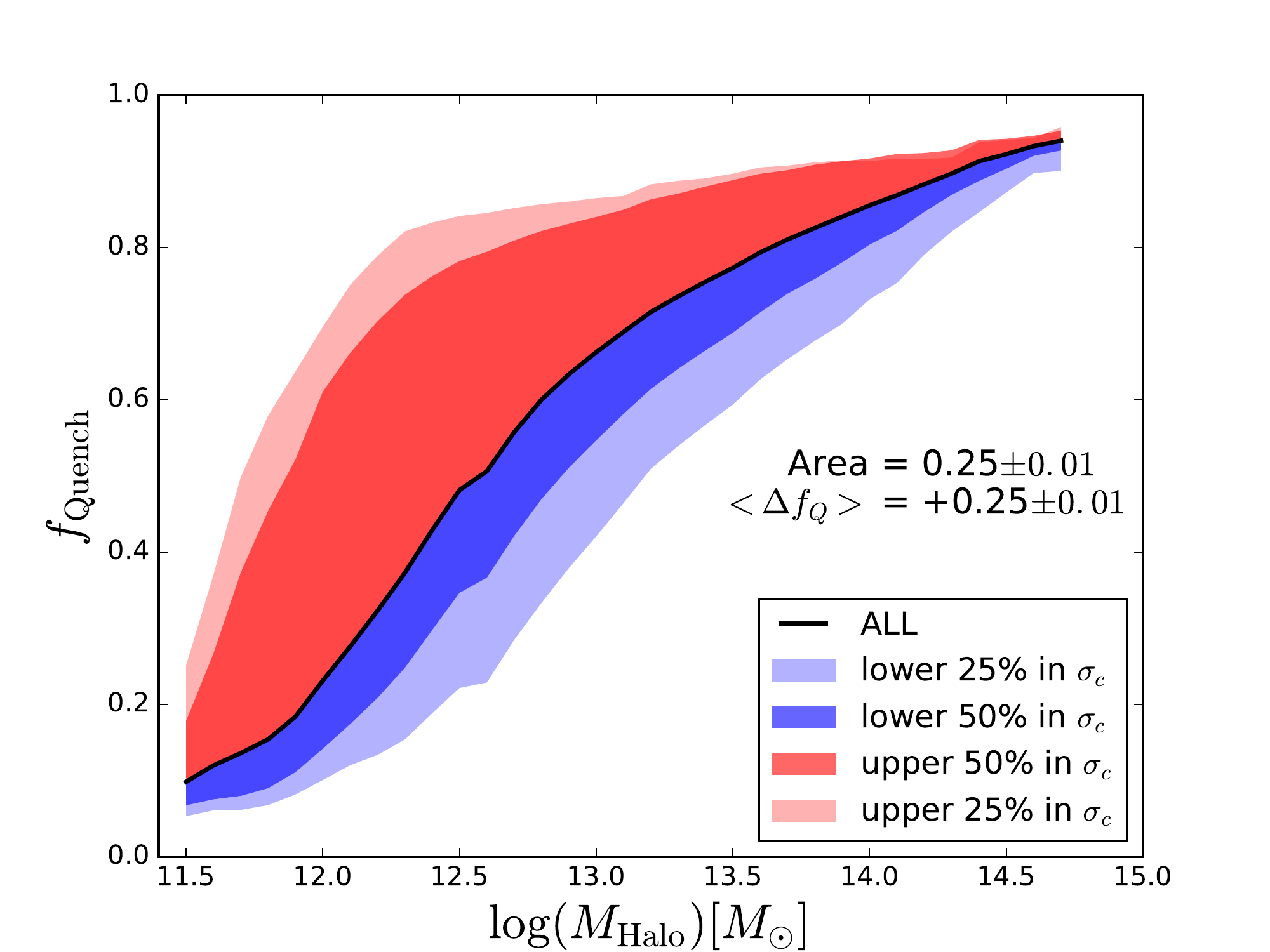}
\includegraphics[width=0.49\textwidth]{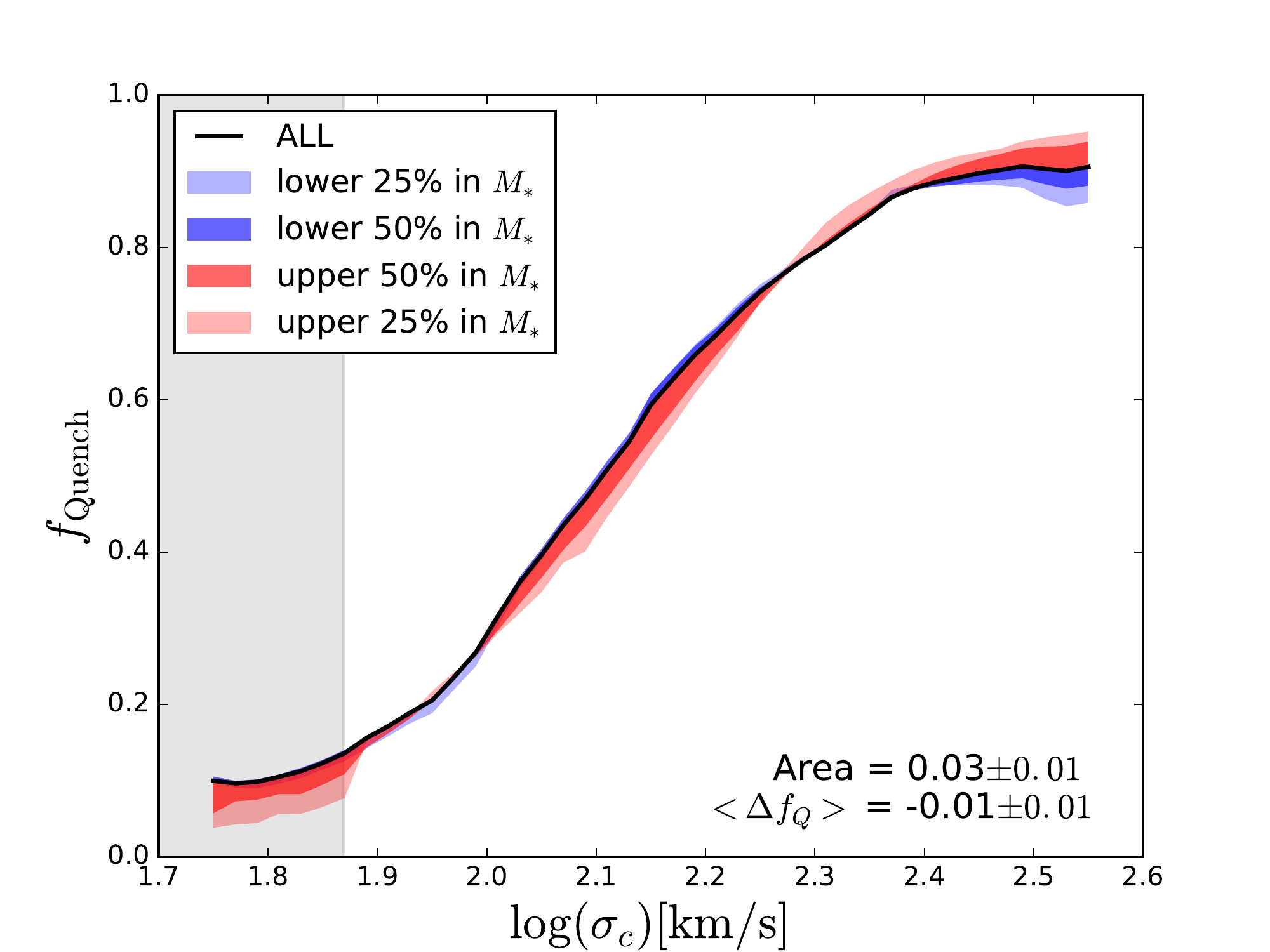}
\includegraphics[width=0.49\textwidth]{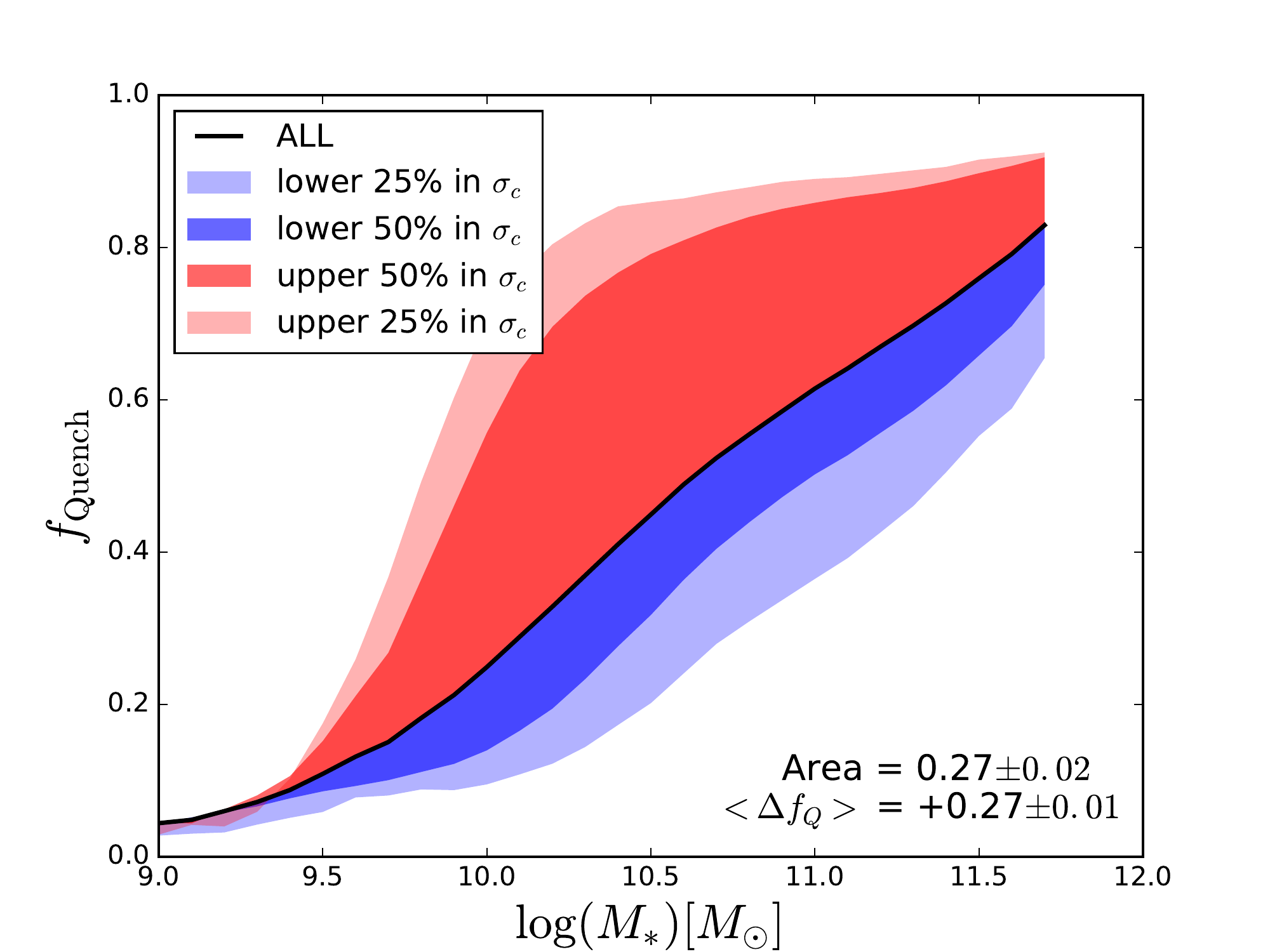}
\includegraphics[width=0.49\textwidth]{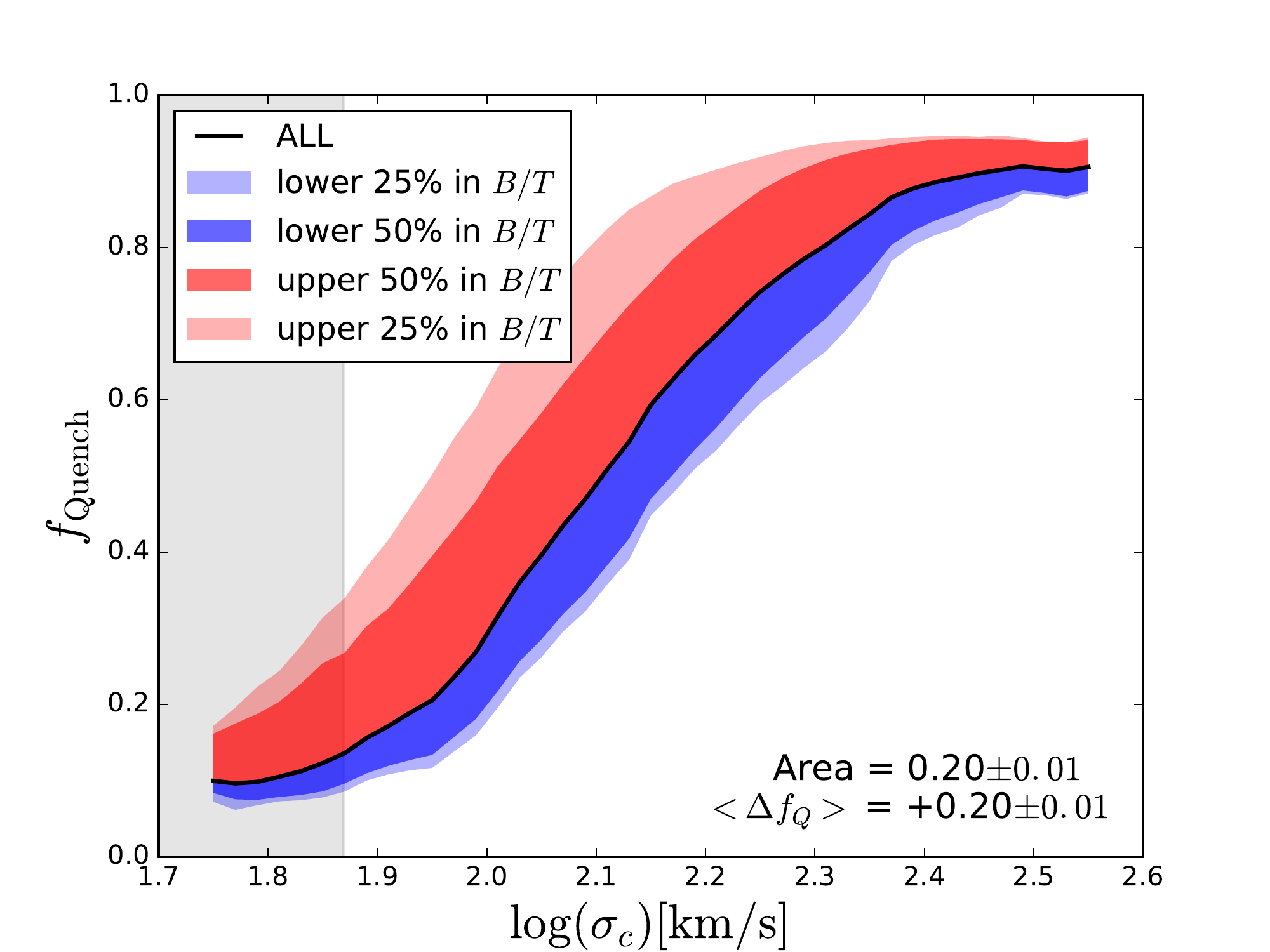}
\includegraphics[width=0.49\textwidth]{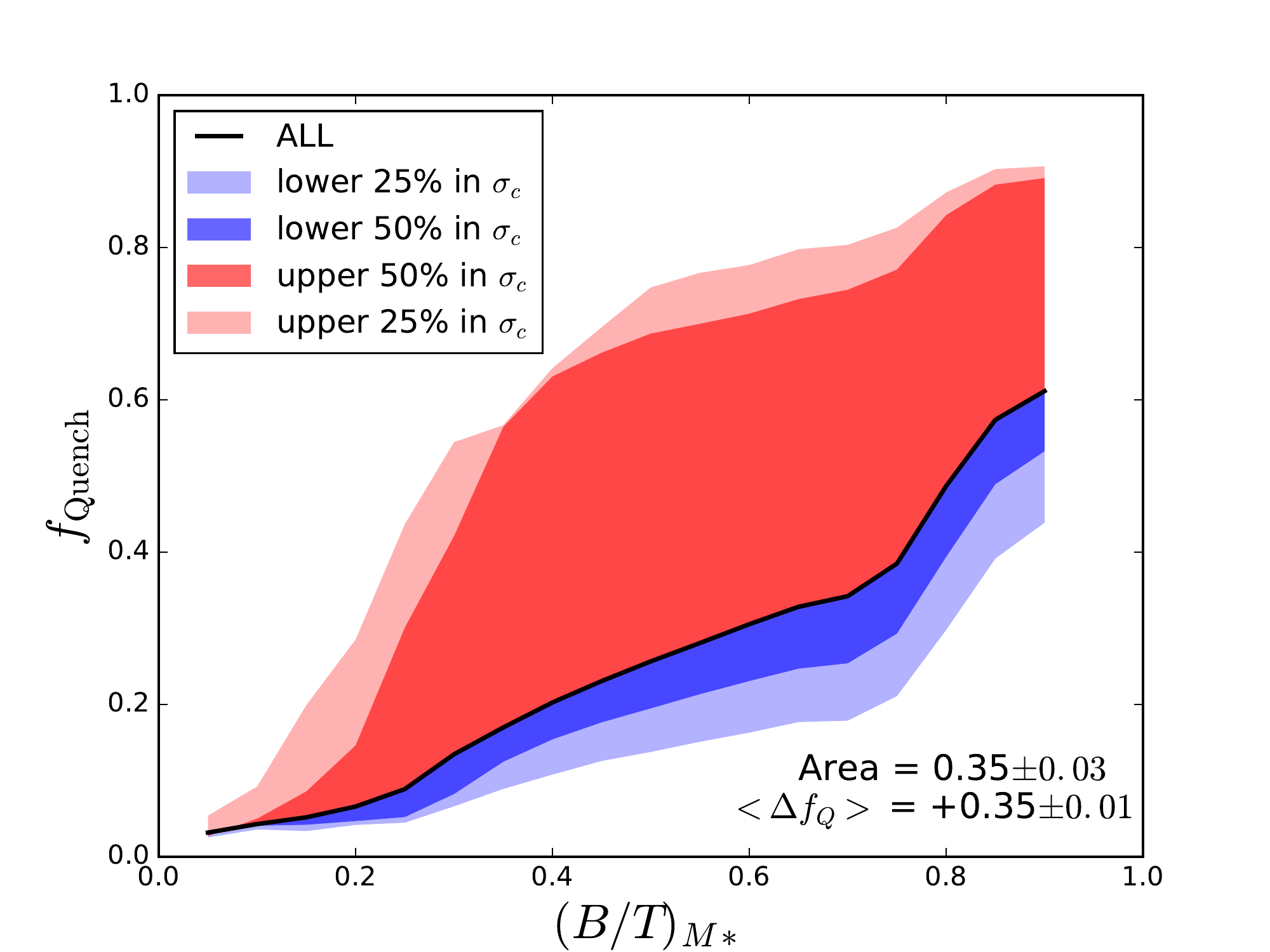}
\caption{Area statistics plots for centrals (1). The left column shows the $f_{\rm Quench} - \sigma_{c}$ relationship, divided from top to bottom by halo mass, stellar mass, and $B/T$ structure. The right column shows the quenched fraction relationships with (from top to bottom): halo mass, stellar mass, and $B/T$, each split by central velocity dispersion range. This is a similar plot to Fig. 3, but instead of splitting by fixed values of each parameter, here we divide the quenched fraction relationship into percentiles of the secondary variable, at fixed values of the primary variable (as indicated on each plot). We find tighter correlations between the quenched fraction and central velocity dispersion than with halo mass, stellar mass or $B/T$ structure. The area contained within, and the mean difference between upper and lower 50th percentiles are shown on each plot, with errors computed via convolving the jack-knife errors from each binning. The statistical improvement of parameterizing the passive fraction with $\sigma_{c}$ (over $M_{\rm halo}$, $M_{*}$ or $B/T$) is highly significant.}
\end{figure*}

\begin{figure*}
\includegraphics[width=0.49\textwidth]{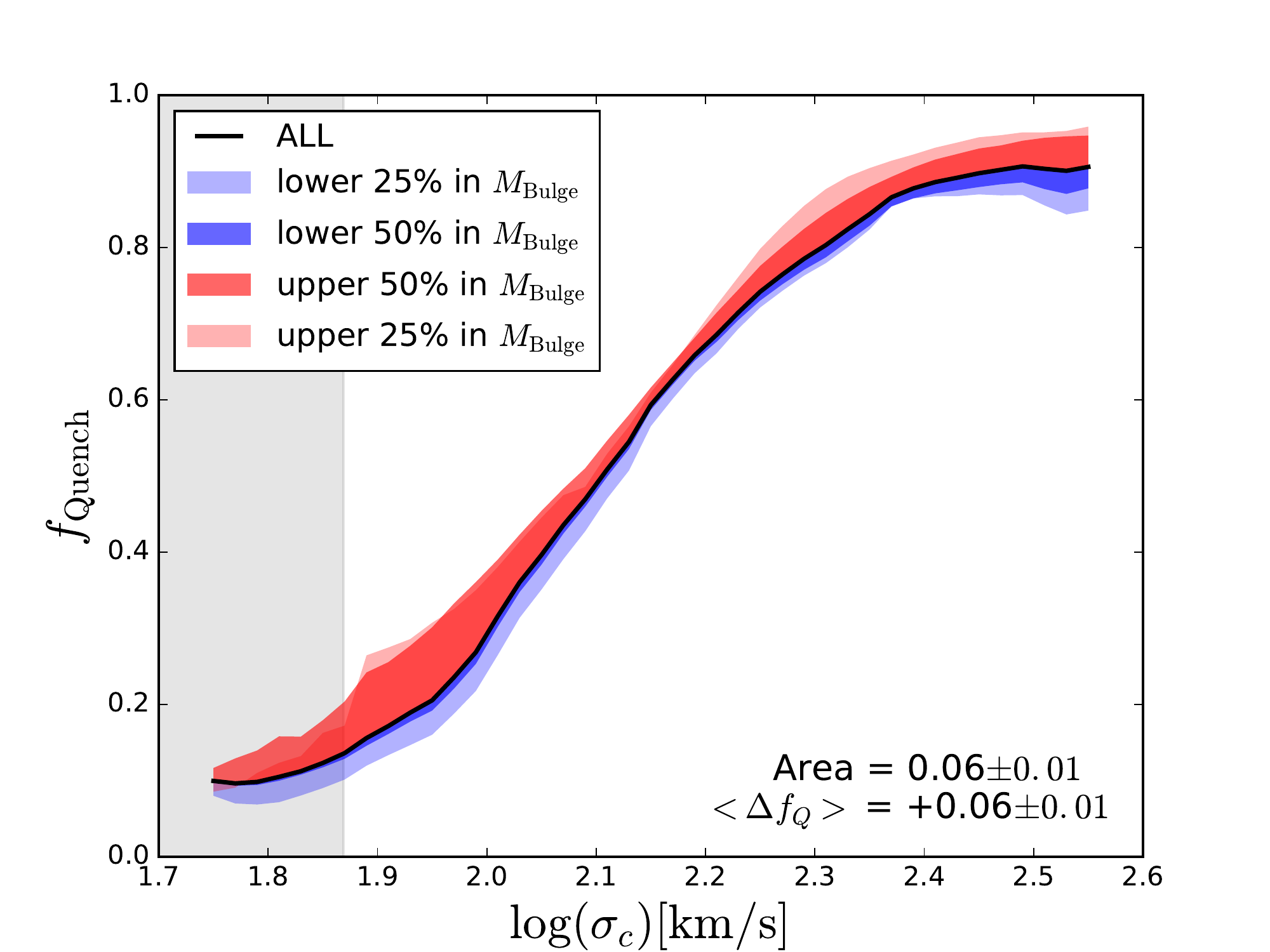}
\includegraphics[width=0.49\textwidth]{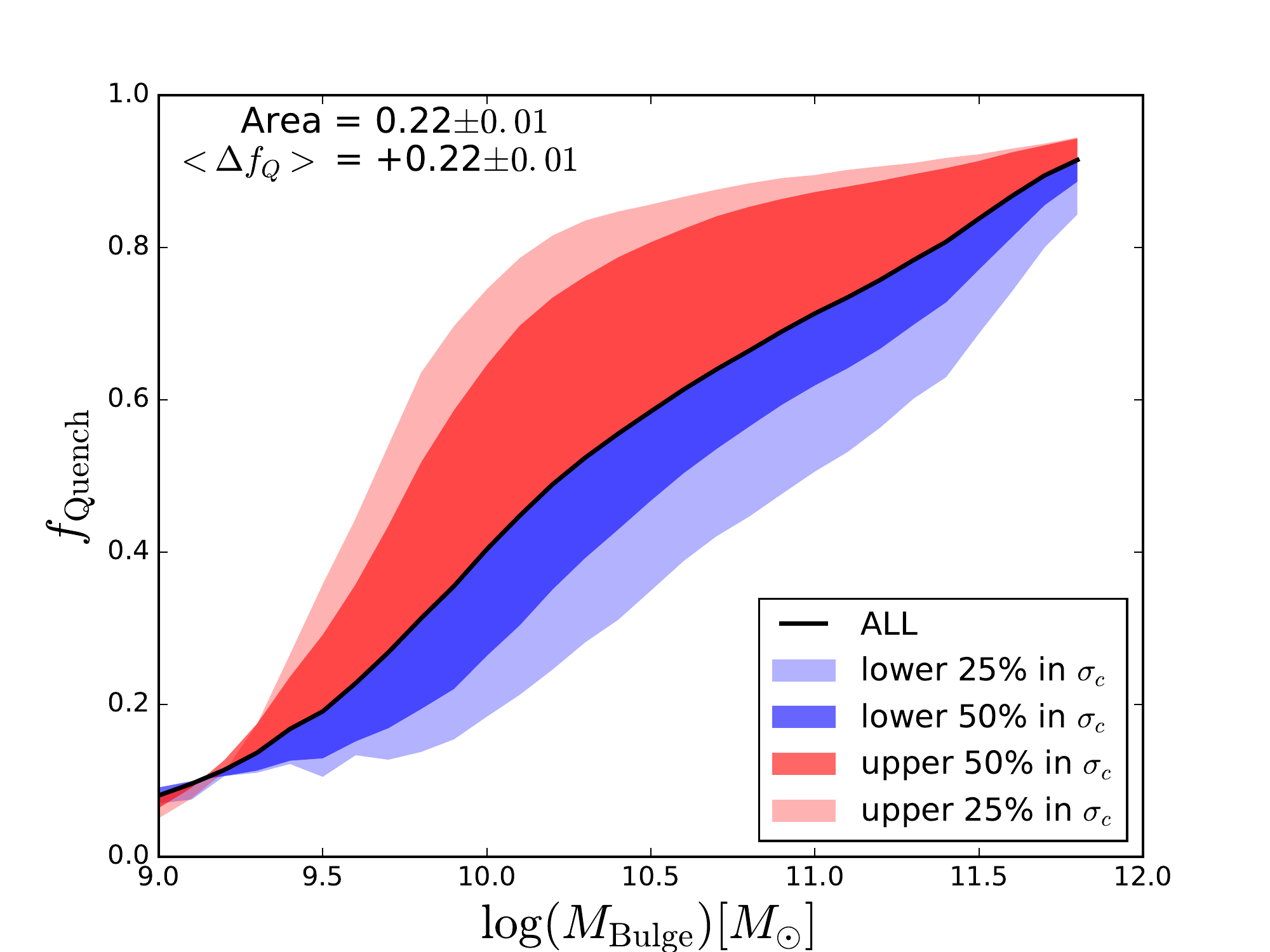}
\includegraphics[width=0.49\textwidth]{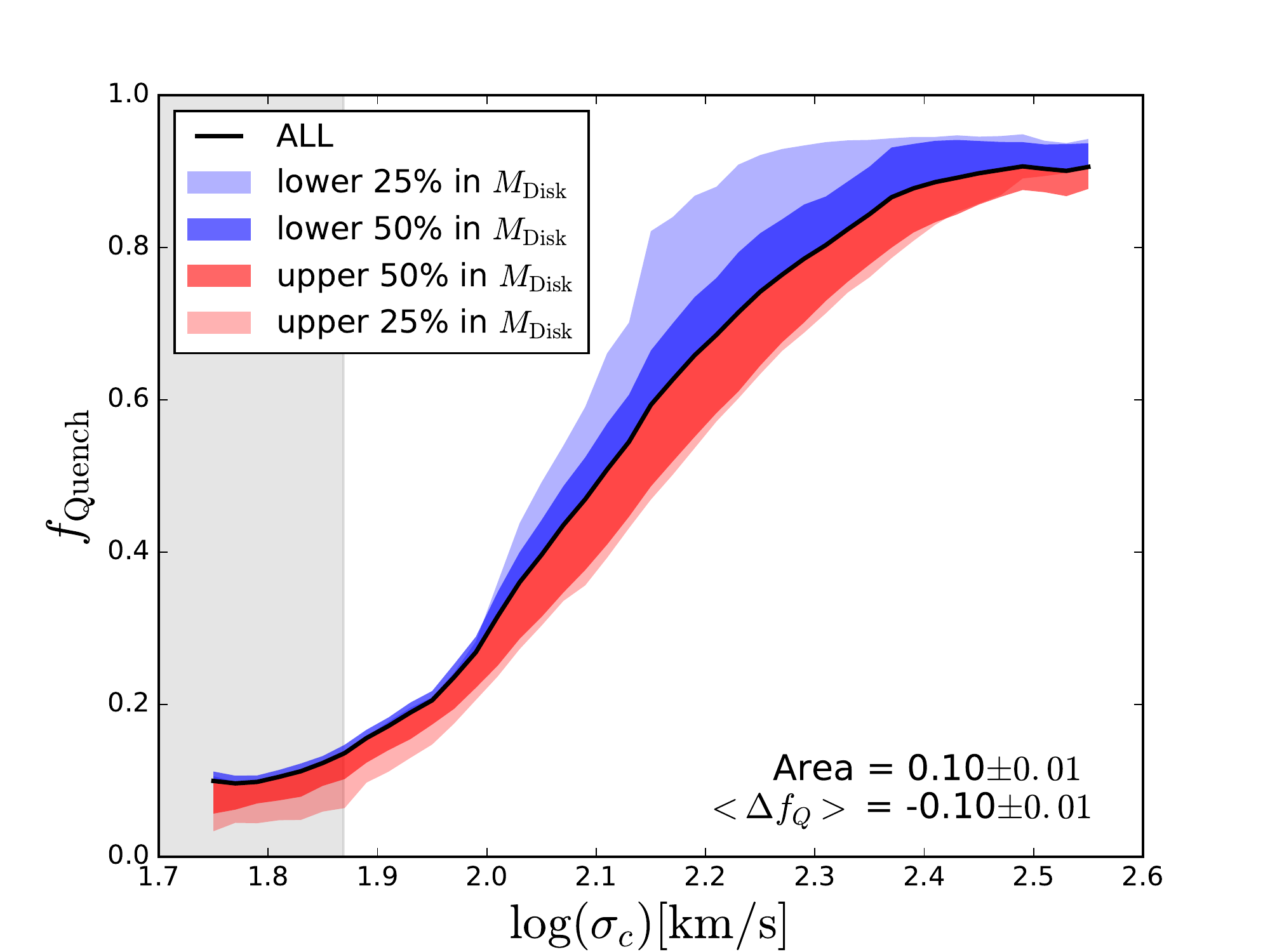}
\includegraphics[width=0.49\textwidth]{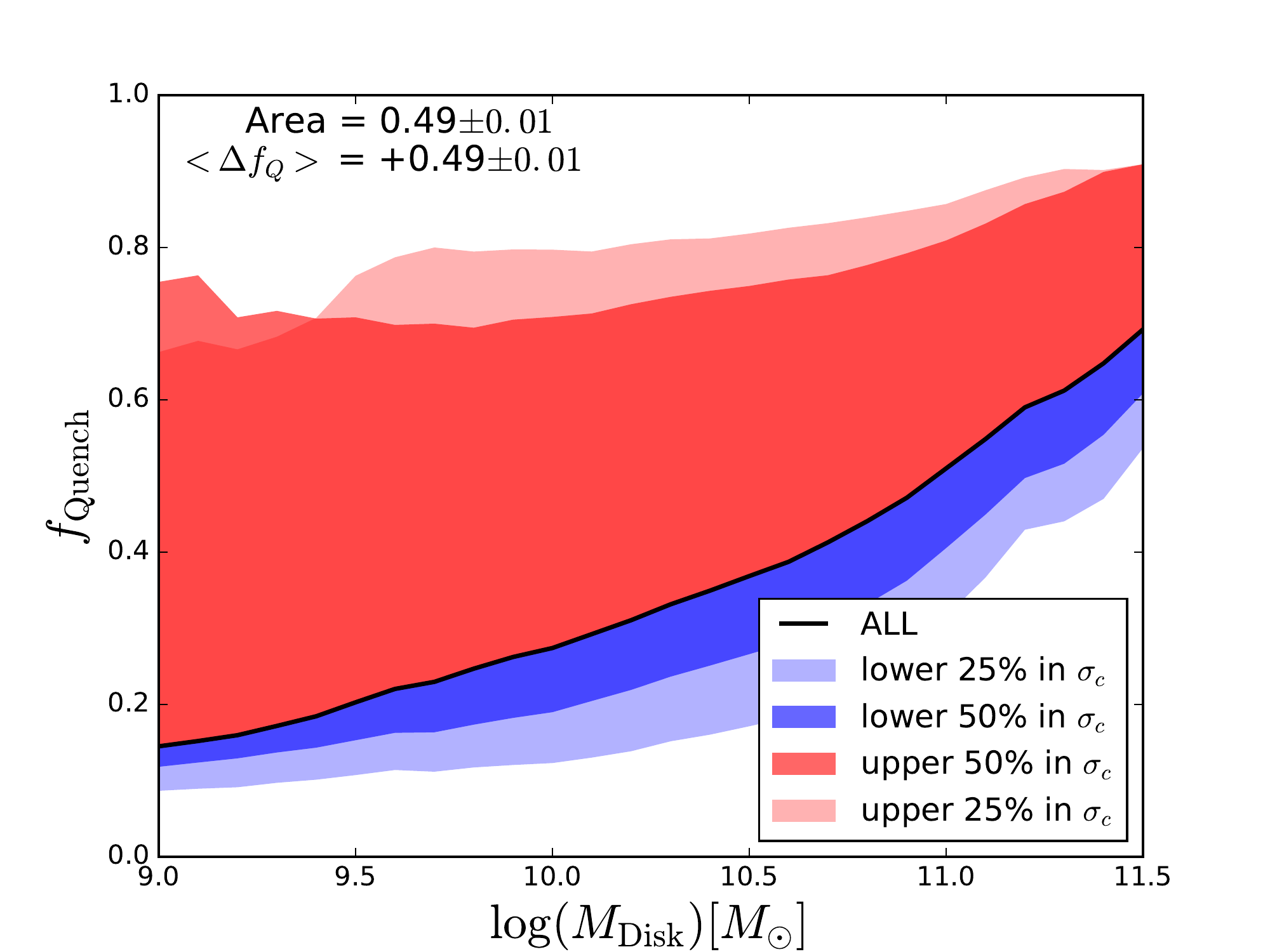}
\includegraphics[width=0.49\textwidth]{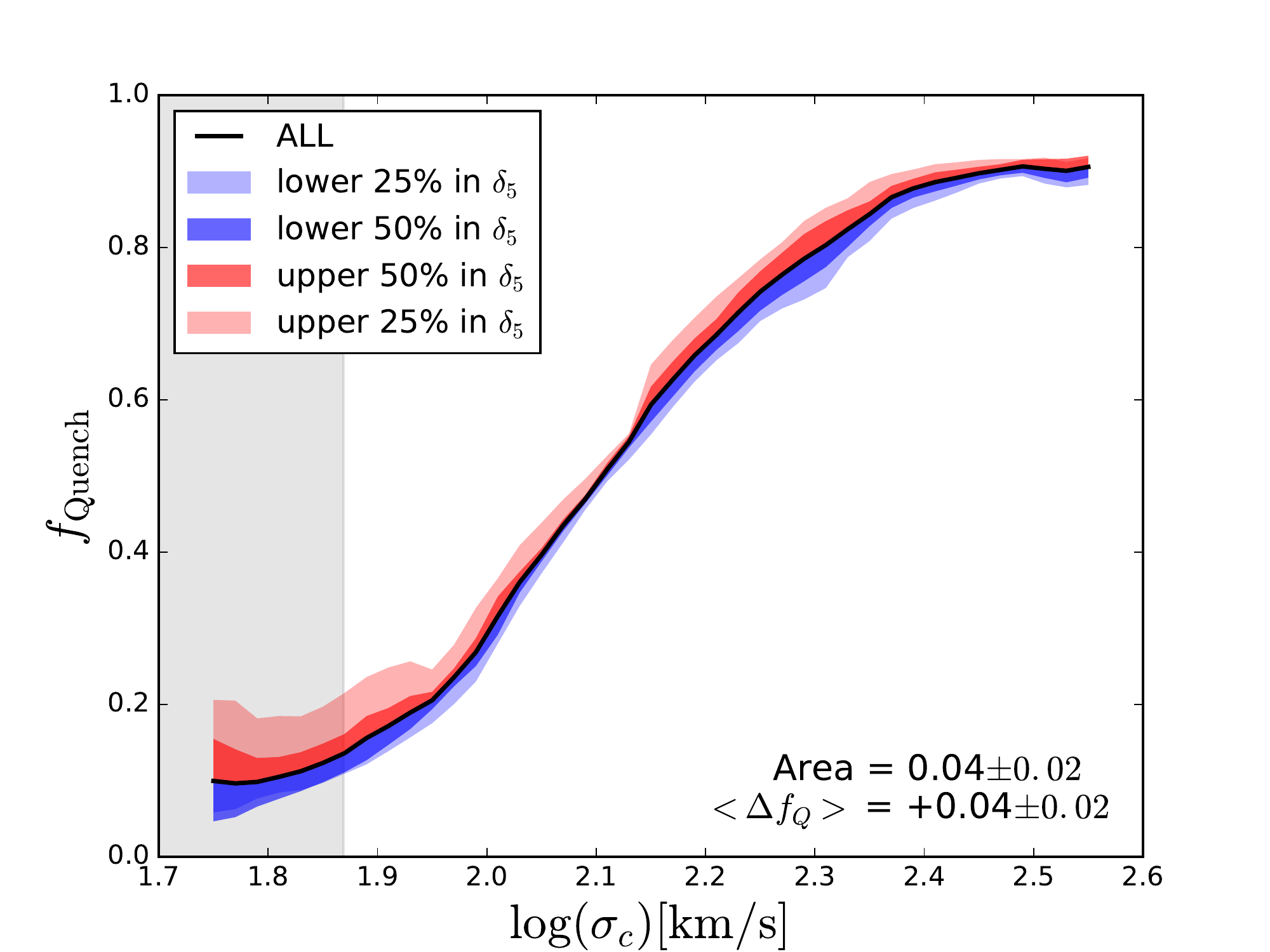}
\includegraphics[width=0.49\textwidth]{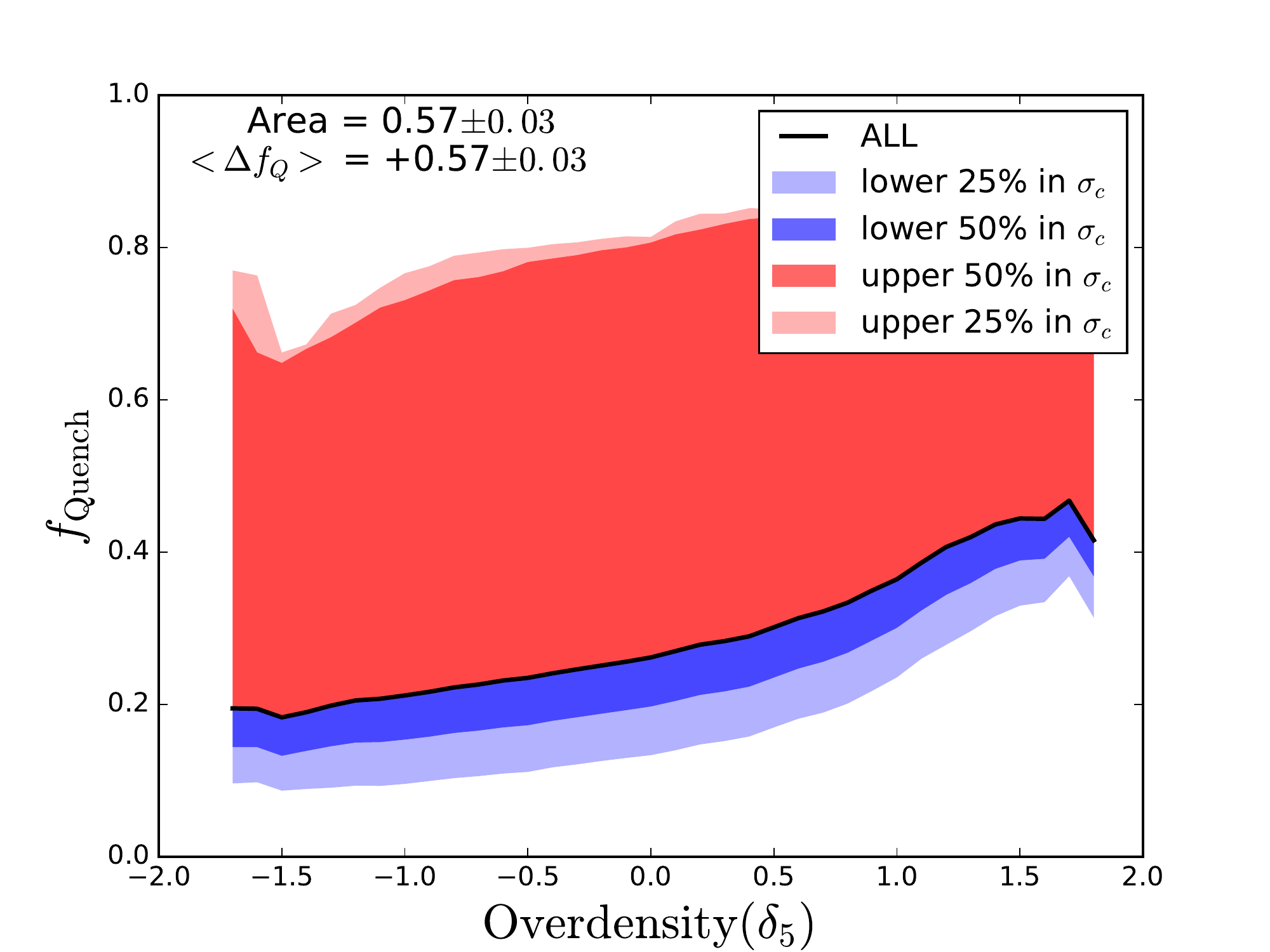}
\caption{Area statistics plots for centrals (2). The left column show the $f_{\rm Quench} - \sigma_{c}$ relationship, divided from top to bottom by bulge mass, disk mass, and overdensity at 5th nearest neighbour ($\delta_{5}$). The right column shows the quenched fraction relationships with (from top to bottom): bulge mass, disk mass, and $\delta_{5}$, each split by central velocity dispersion range. We find significantly tighter correlations between the quenched fraction and central velocity dispersion than with bulge mass, disk mass or $\delta_{5}$. Furthermore, we find that increasing disk mass at fixed values of central velocity dispersion {\it decreases} the quenched fraction (blue regions lying above red regions), whilst increasing central velocity dispersion at fixed values of all of the other parameters always leads to a significant {\it positive} effect on the quenched fraction (red regions lying above blue regions). The area contained within, and the mean difference between upper and lower 50th percentiles are shown on each plot, with errors computed via convolving the jack-knife errors from each binning. The statistical improvement of parameterizing the passive fraction with $\sigma_{c}$ (over $M_{\rm bulge}$, $M_{\rm disk}$ or $\delta_{5}$) is highly significant.}
\end{figure*}

\subsubsection{Results for Centrals}

In Fig. 4 we reproduce our results in Fig. 3 for the area statistics approach (see above and Appendix A). The left column shows the $f_{\rm Quench} - \sigma_{c}$ relation, split by percentile ranges in (from top to bottom): halo mass, stellar mass and $B/T$ structure. The right hand column shows the quenched fraction relationship with (from top to bottom): halo mass, stellar mass and $B/T$ structure, each split by percentile ranges in $\sigma_{c}$. This plot should be read by rows. The solid red and blue shaded regions represent upper and lower 50\% of the data in the $\beta$-variable, respectively (see eqs. 8 \& 9), with the semi-transparent shading indicating the upper and lower 25\% of the data in the $\beta$-variable. The areas are considerably smaller for parameterizing quenching as a function of central velocity dispersion than for stellar mass, halo mass or $B/T$. This indicates that quenching depends more fundamentally on central velocity dispersion than any of these alternatives, for central galaxies. Additionally, the average difference is always positive for varying the central velocity dispersion at fixed other galaxy properties. These results are highly significant, typically $>$ 4 $\sigma$, where the error is constructed by convolving the individual jack-knife error on each bin and adding in quadrature the positive and negative contributions (for red and blue shadings, respectively).

In Fig. 5 we investigate three more cases: bulge mass, disk mass and overdensity at the 5th nearest neighbour ($\delta_{5}$). Here again we find by far the smallest areas for central velocity dispersion acting as the primary variable, than for any of the other cases. Central velocity dispersion performs significantly better than even bulge mass (top row), which was previously found to outperform the rest of the variables considered in this work (Bluck et al. 2014). The case of disk mass is especially interesting, because increasing its value at fixed central velocity dispersion lowers the quenched fraction, and furthermore leads to the highest area at fixed central velocity dispersion from this set of variables. This also explains the slight negative trend seen with stellar mass, and (at least partially) the positive trend seen with $B/T$. Whilst the dominant correlator to the quenched fraction, $\sigma_{c}$, drives quenching (i.e., always leads to positive increases of the quenched fraction at fixed other galaxy properties), information about the disk structure in some sense  `resists' quenching. 

Given that disk mass and $D/T$ (= 1 - $B/T$) correlates with gas mass and gas fraction (e.g., Saintonge et al. 2011; Maddox et al. 2015), it is likely that these variables give information on what remains to be quenched in a given galaxy, and hence how much work must be done to quench it. Alternatively, central velocity dispersion (which is known to correlate strongly with dynamically measured $M_{BH}$) likely gives information regarding the available energy to do work quenching the system. In any case, if the quenching of central galaxies is to be parameterized by a single variable, central velocity dispersion is by far the best choice out of our set of parameters (in agreement with a complementary analysis via artificial neural networks performed in Teimoorinia, Bluck \& Ellison 2016, and also in agreement with a smaller set of comparisons made in Wake et al. 2012). The results for all areas and average differences in the quenched fraction, for each combination considered here, are presented in Table 1. We discuss the possible meaning of these results further in the discussion (Section 6).

\begin{table}
\caption{Summary of area and mean difference statistics for centrals (taken from Figs. 4 \& 5.)}
 \label{tab2}
 \begin{center}
 \begin{tabular}{@{}ccc}
  \hline
\hline
$[\alpha, \beta]$ & Area  & $\langle\Delta f_{Q}\rangle$ \\ 
\hline
$[\sigma_{c}, M_{\rm halo}]$ & 0.03 $\pm$ 0.01 & + 0.02 $\pm$ 0.01 \\ 
$[M_{\rm halo}, \sigma_{c}]$ & 0.25 $\pm$ 0.01 & + 0.25 $\pm$ 0.01 \\ 
\hline
$[\sigma_{c}, M_{*}]$ & 0.03$\pm$ 0.01 & - 0.01 $\pm$ 0.01 \\
$[M_{*}, \sigma_{c}]$ & 0.27 $\pm$ 0.02 & + 0.27 $\pm$ 0.01 \\
\hline
$[\sigma_{c}, B/T]$ & 0.20 $\pm$ 0.01 & + 0.20 $\pm$ 0.01 \\
$[B/T, \sigma_{c}]$ & 0.35 $\pm$ 0.03 & + 0.35 $\pm$ 0.01 \\
\hline
\hline
$[\sigma_{c}, M_{\rm bulge}]$ & 0.06 $\pm$ 0.01 & + 0.06 $\pm$ 0.01 \\
$[M_{\rm bulge}, \sigma_{c}]$ & 0.22 $\pm$ 0.01 & + 0.22 $\pm$ 0.01 \\
\hline
$[\sigma_{c}, M_{\rm disk}]$ & 0.10 $\pm$ 0.01 & - 0.10 $\pm$ 0.01 \\
$[M_{\rm disk}, \sigma_{c}]$ & 0.49 $\pm$ 0.01 & + 0.49 $\pm$ 0.01 \\
\hline
$[\sigma_{c}, \delta_{5}]$ & 0.04 $\pm$ 0.02 & + 0.04 $\pm$ 0.02 \\
$[\delta_{5}, \sigma_{c}]$ & 0.57 $\pm$ 0.03 & + 0.57 $\pm$ 0.03 \\
\hline
\end{tabular}
\end{center}
Note: $\alpha$ and $\beta$ are defined as in eqs. 8 \& 9, they correspond to the x-axis variable and the percentile (colour) variable in Figs. 4 \& 5, respectively.
\end{table}

\section{What quenches central galaxies?}

In Section 5, we have demonstrated that the fraction of quenched galaxies is more accurately constrained by central velocity dispersion, than by halo, stellar, bulge or disk mass, bulge-to-total stellar mass ratio ($B/T$) or overdensity of galaxies evaluated at the 5th nearest neighbour ($\delta_{5}$). These observational findings are in agreement with prior analyses of the role of central velocity dispersion in quenching (e.g., Wake et al. 2012; Teimoorinia, Bluck \& Ellison 2016), and are consistent with the strong dependence of central galaxy quenching on surface mass density within 1 kpc (e.g., Cheung et al. 2012; Fang et al. 2013; Woo et al. 2015; Lilly \& Carollo 2016) and bulge mass (Bluck et al. 2014; Lang et al. 2014; Omand et al. 2014). Furthermore, we find that varying local density, stellar, halo or bulge mass at fixed central velocity dispersion leads to essentially no impact whatsoever on the quenched fraction (even when these parameters are varied by over three orders of magnitude). This fact has profound implications for the mechanism(s) which can be responsible for quenching centrals. 

Given our results, it seems implausible that the quenching of central galaxies can be governed by halo mass quenching, which depends critically on the dark matter gravitational potential and hence halo mass (e.g., Dekel \& Birnboim 2006; Dekel et al. 2009; Woo et al. 2013; Dekel et al. 2014). Additionally, conventional `mass quenching' (Peng et al. 2010, 2012), which is parameterized by stellar mass, is clearly not an optimal parameterization for quenching of centrals. Furthermore, this suggests that stellar and supernova feedback (both of which correlate primarily with mass in stars, as the integral of the star formation rate over time) cannot be responsible for central galaxy quenching. The lack of impact of local density on the quenching of centrals at fixed central velocity dispersion further implies that these systems are not being quenched via environmental processes, which are thought to affect satellites more (see Section 7, where we discuss satellites). Bulge mass is also clearly not the most fundamental correlator to central galaxy quenching since it exhibits little variation in the dominant $f_{\rm Quench} - \sigma_{c}$ relation, although it does perform significantly better than any of the other parameters considered in this work (see, e.g., Bluck et al. 2014; Teimoorinia et al. 2016). However, there is at least one theoretically proposed quenching mechanism which is perfectly consistent with our data.

\begin{figure*}
\includegraphics[width=0.49\textwidth]{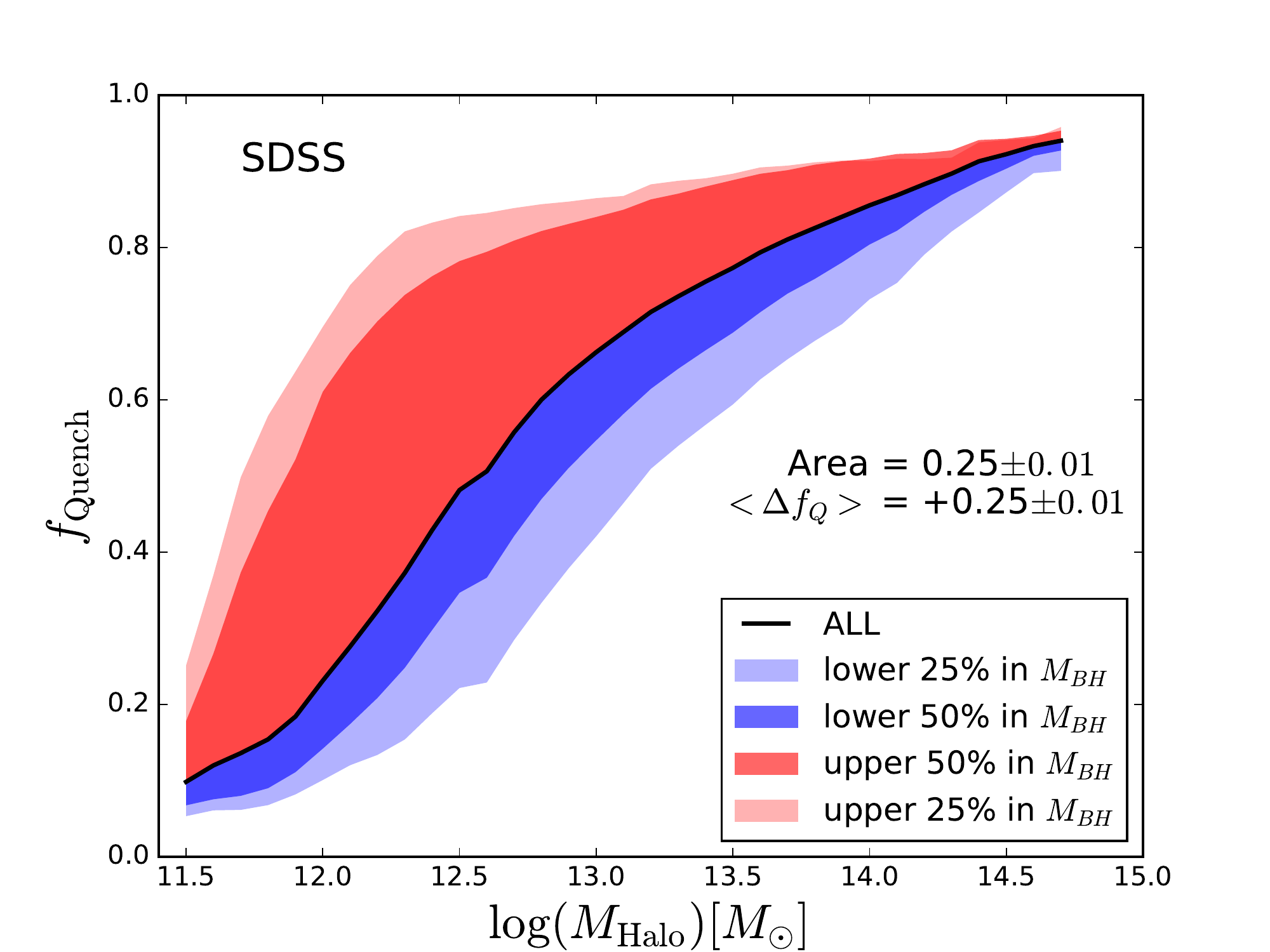}
\includegraphics[width=0.49\textwidth]{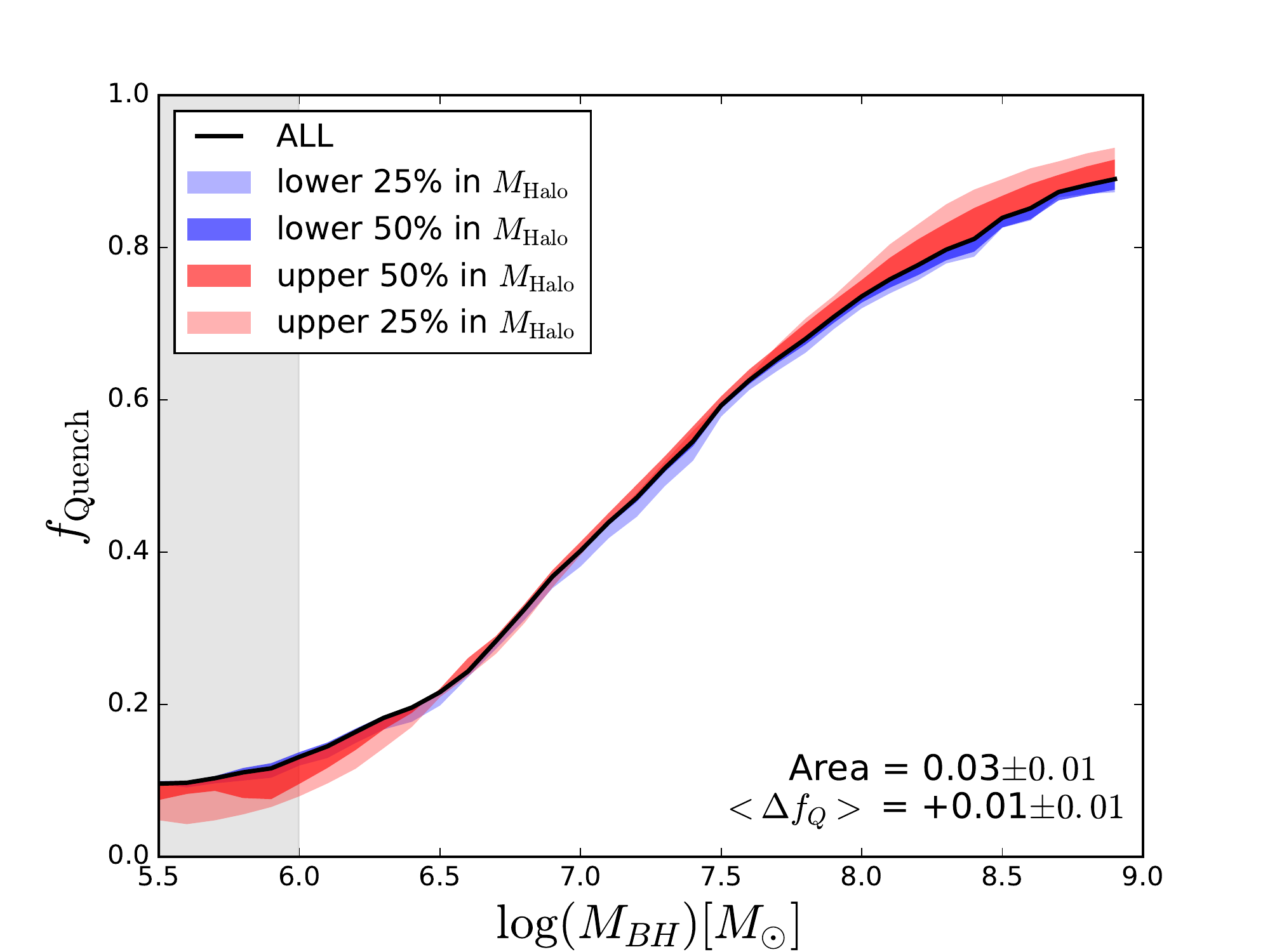}
\includegraphics[width=0.49\textwidth]{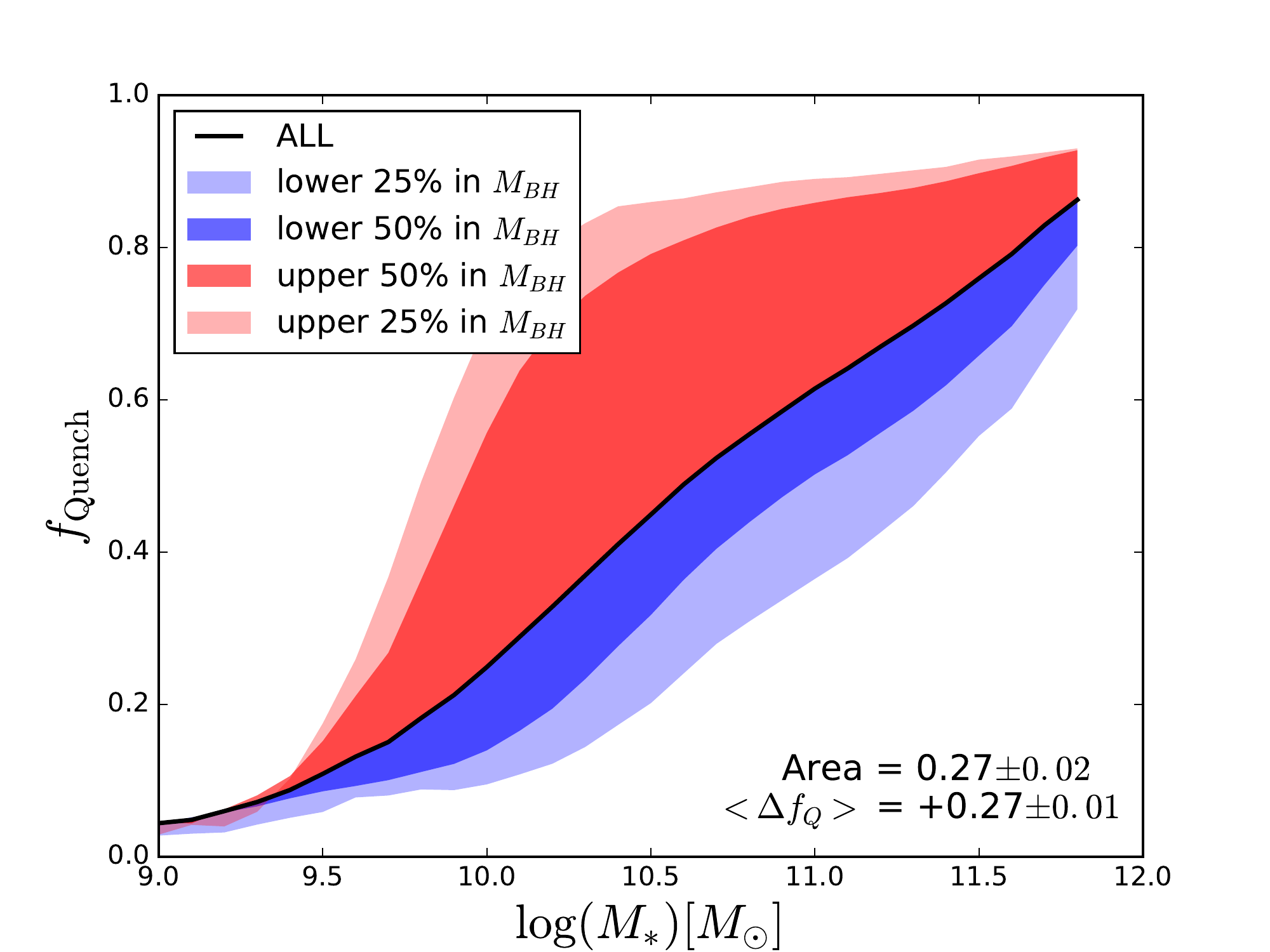}
\includegraphics[width=0.49\textwidth]{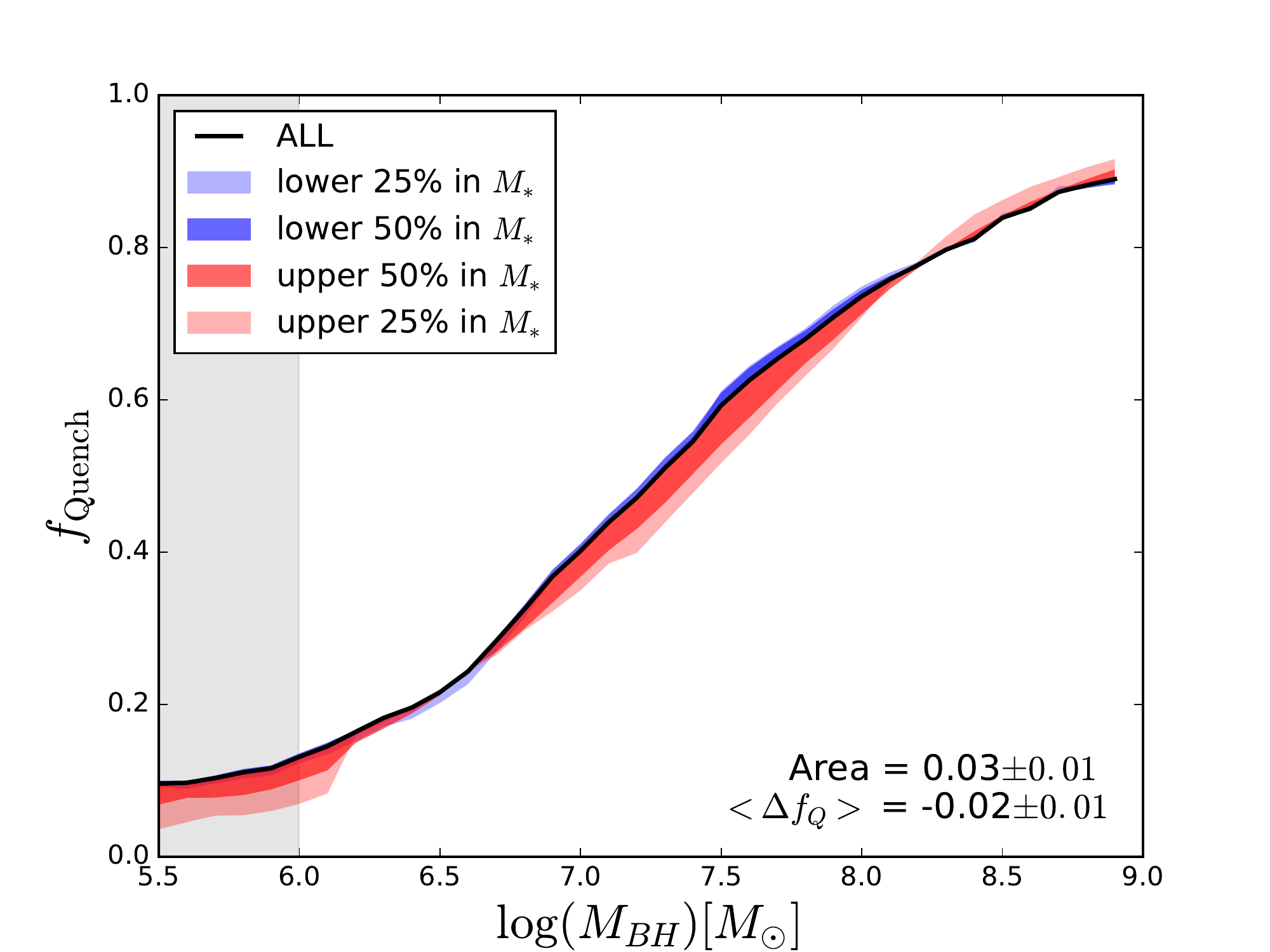}
\caption{The quenched fraction dependence on estimated black hole mass (right panels) and, for comparison, halo mass (top left) and stellar mass (bottom left). The black hole masses are estimated as a function of central velocity dispersion, using the scaling law from Saglia et al. (2016). Hence, these plots are very similar in nature to Fig. 4. However, this parameterization allows for a more direct comparison with the model predictions, shown in Figs. 9 \& 10.}
\end{figure*}

\begin{figure*}
\includegraphics[width=0.99\textwidth]{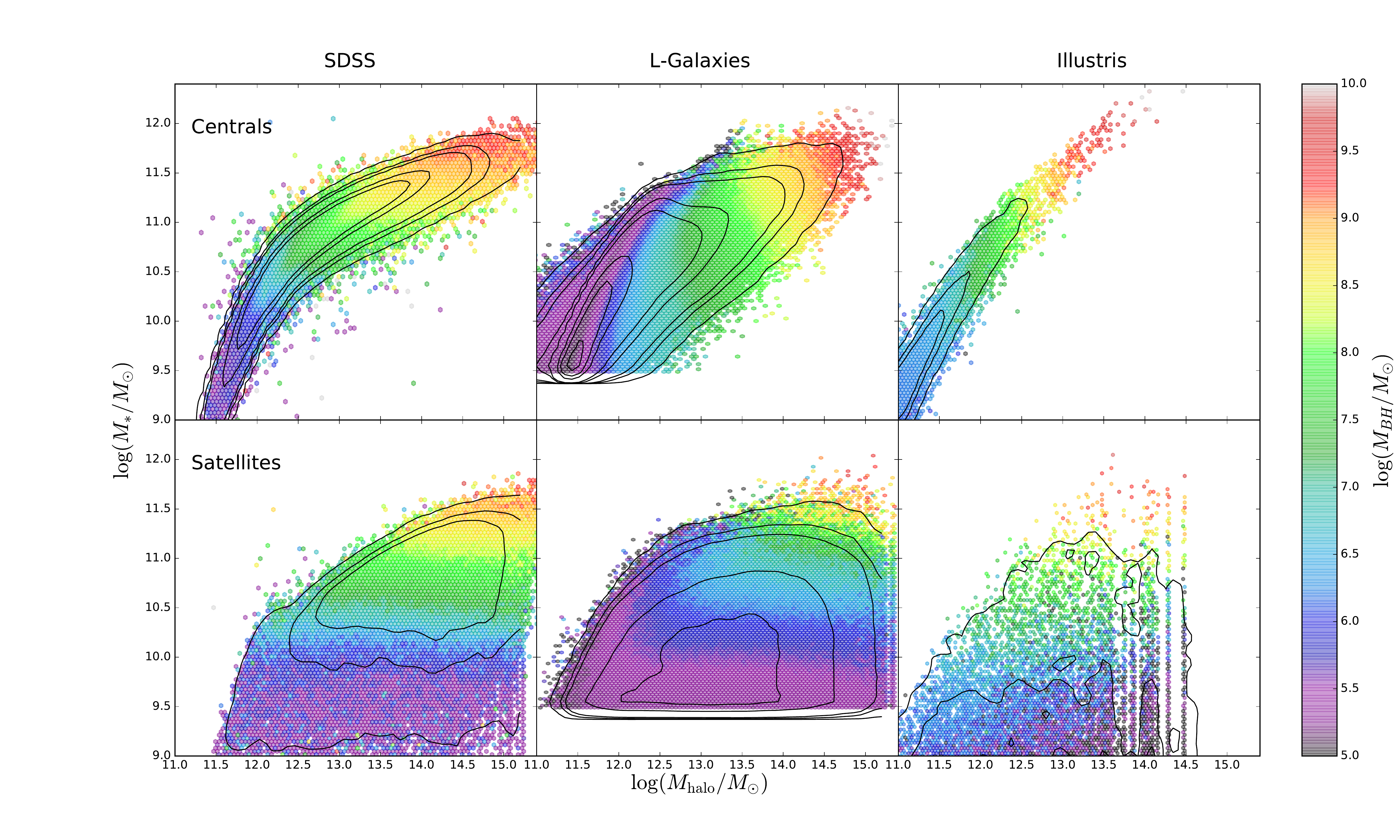}
\caption{The stellar mass - halo mass relation for SDSS (left), L-Galaxies (middle) and Illustris (right) galaxies. Each hexagonal bin is colour coded by the average supermassive black hole mass of galaxies contained within the binnings (as indicated by the colour bar). The top row shows our results for central galaxies, and the bottom row shows the results for satellite galaxies. Black solid lines show density contours for each case. Note the strong positive correlation between halo and stellar mass for centrals, which is mostly absent for satellites. For central galaxies black hole mass increases as a function of both stellar and halo mass (diagonal lines of constant mass), whereas for satellites black hole mass is primarily a function of stellar mass (horizontal lines of constant mass).}
\end{figure*}

\begin{figure*}
\includegraphics[width=0.49\textwidth]{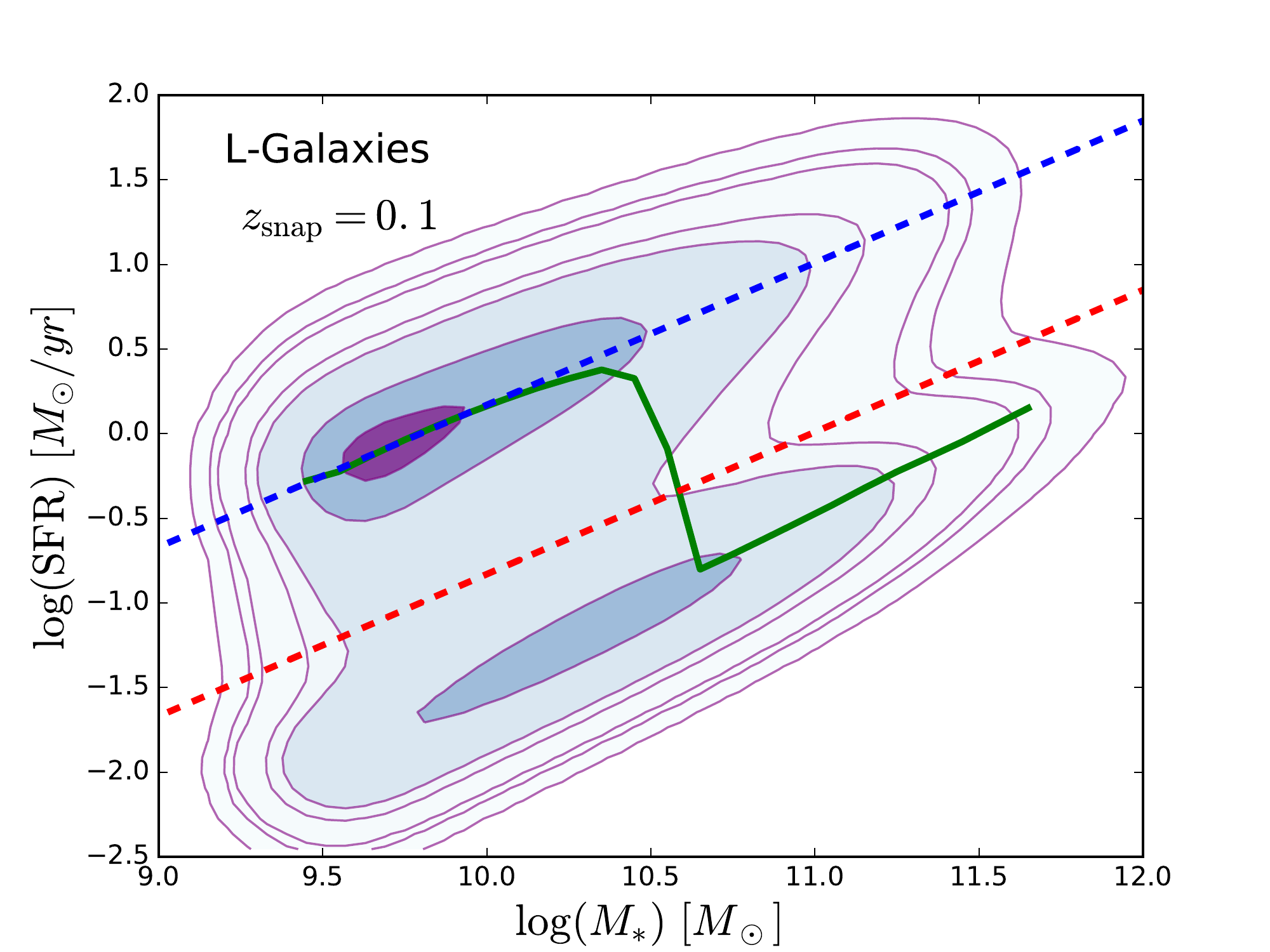}
\includegraphics[width=0.49\textwidth]{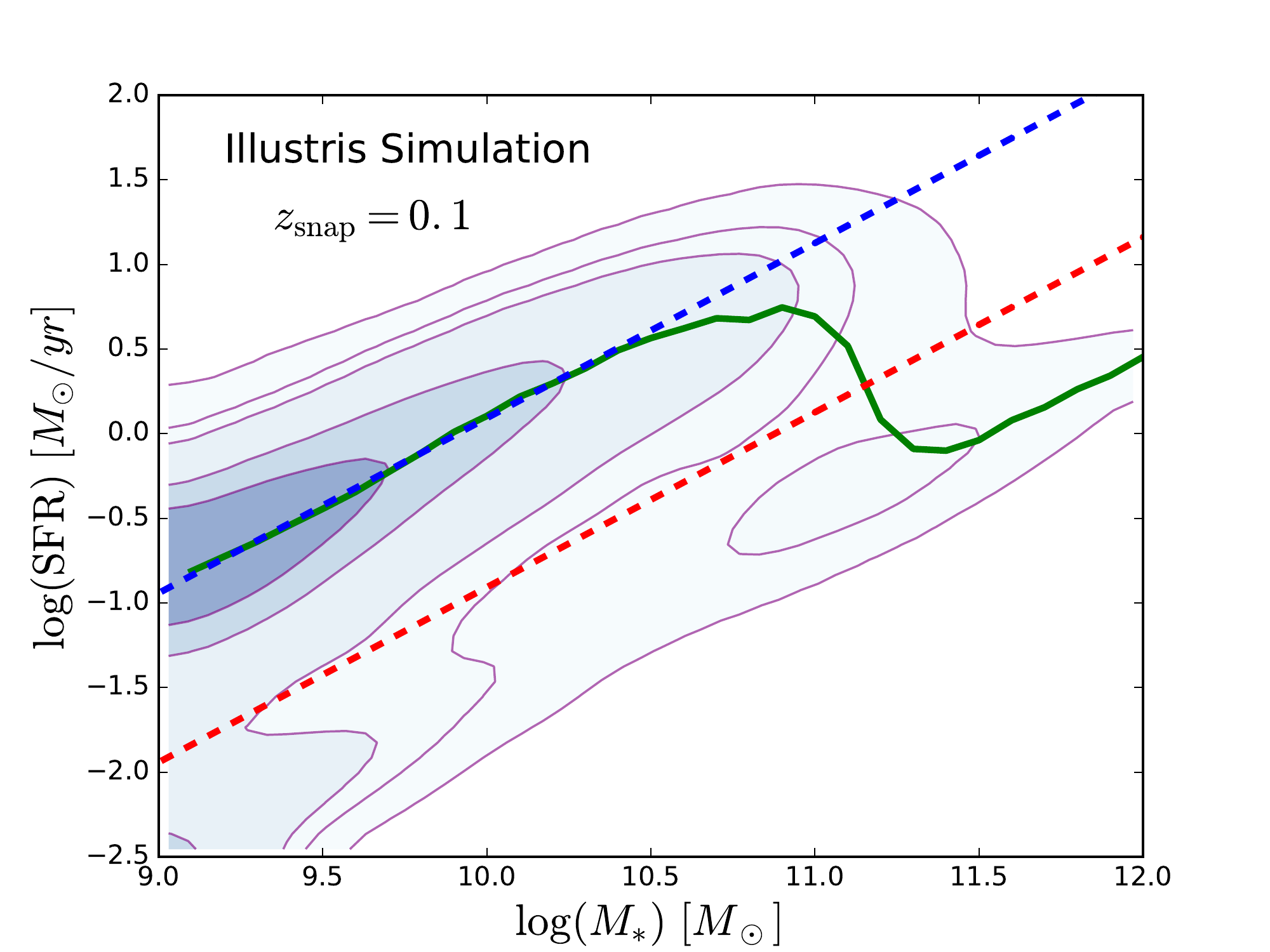}
\caption{The SFR - $M_{*}$ main sequence relationship in L-Galaxies (left panel) and the Illustris simulation (right panel), both taken at the $z$ = 0.1 snapshot. Solid green lines in each plot represent the median relation. The blue dashed lines show a linear fit to the median relation, at $M_{*} < 10^{10} M_{\odot}$. This fit is given for L-Galaxies by: $\log ({\rm SFR} [M_{\odot}/yr]) = (0.84\pm0.04) \times \log(M_{*}[M_{\odot}]) - (8.2 \pm 0.3)$; and for Illustris by: $\log ({\rm SFR} [M_{\odot}/yr]) = (1.03\pm0.06) \times \log(M_{*}[M_{\odot}]) - (10.2 \pm 0.7)$. Quenched galaxies are defined (as in the observational data, see Fig. 1) to lie one order of magnitude or greater below the star forming main sequence, indicated by a red dashed line. A minimum value of sSFR is applied as in the observational data, which is responsible to the passive contour peaks.}
\end{figure*}

The strong observed correlations between central velocity dispersion and dynamically measured supermassive black hole mass (e.g., Ferrarese \& Merritt 2000; McConnell et al. 2011; McConnell \& Ma 2013; Saglia et al. 2016) offer an intriguing possibility to explain our observational trends. In many (if not most) semi-analytic models and cosmological hydrodynamical simulations quenching of central galaxies is governed by AGN feedback, in either the radio (e.g., Croton et al. 2006; Bower et al. 2008) or quasar (e.g., Hopkins et al. 2008, 2010) mode. In this paradigm the mass of the black hole is the key predictor of whether a central galaxy will be quenched or star forming. In general, the probability that a galaxy will be quenched ($P_{Q}$) is proportional to the energy available to do the quenching, above some activation threshold, thus:

\begin{equation}
P_{Q} = W_{Q} - \phi_{\rm act} = \epsilon E_{BH} - \phi_{\rm act}
\end{equation}

\noindent where $W_{Q}$ indicates the work done to the galaxy and halo to quench star formation, and $\phi_{\rm act}$ is the activation energy required to have a measurable effect on the star forming state of the galaxy. In the model where AGN feedback provides this energy, the work can be set equal to some coupling efficiency ($\epsilon$) multiplied by the energy released in forming the black hole ($E_{BH}$). Effectively $\epsilon$ accounts for energy lost to the Universe via radiation, and hence does not impact the galaxy or halo. $\epsilon$ may vary in value from 0 to 1, and is poorly constrained at present. It may also ultimately turn out to be dependent on the environment in which the galaxy resides, particularly the temperature of the hot gas halo (e.g., Henriques et al. 2015). 

Following the Soltan argument (Soltan 1984; Silk \& Rees 1998; Fabian 1999), the total energy released in forming a black hole is proportional to its mass:

\begin{equation}
E_{BH} = \int_{z=0}^{z=zf} L(t) dt = \int_{z=0}^{z=zf} \mu c^{2} \frac{dM_{BH}(t)}{dt} dt
\end{equation}

\begin{equation}
\approx \langle \mu \rangle c^{2} M_{BH}
\end{equation}

\noindent where, $L(t)$ is the time dependent bolometric luminosity of the AGN, and $\mu$ is the fraction of accreted matter converted into energy (often estimated to be $\sim$ 0.1, Elvis et al. 2003; Shankar et al. 2009). Thus, the total energy available from AGN feedback to do work on a galaxy, quenching star formation, is given by

\begin{equation}
W_{Q} = \epsilon \langle \mu \rangle c^{2} M_{BH}
\end{equation}

\noindent where all terms apart from $M_{BH}$ may be taken to be approximately constant. Putting all of this together, and assuming that the probability of being quenched is approximately equal to the fraction of quenched galaxies in a given population of galaxies, we recover (as in Bluck et al. 2014):

\begin{equation}
f_{Quench} \propto M_{BH} = f(\sigma_{c}) \sim \sigma_{c}^{\alpha}
\end{equation}
 
\noindent with the last step inferred from observations. $\alpha$ is an observationally determined coefficient, often found to be $\sim$ 4 - 5 (e.g., McConnell \& Ma 2013; Saglia et al. 2016). Therefore, in the general prescription for AGN driven quenching, the quenched fraction is predicted to scale primarily with black hole mass and hence (observationally) central velocity dispersion, which is essentially exactly what we observe. To look at this in more detail, we explore two types of AGN quenching models in the next sub-section, and compare their predictions to our observations.

\subsection{Comparison to a Hydrodynamical Simulation and Semi-Analytic Model}

\subsubsection{Details on Illustris and L-Galaxies}

\begin{figure*}
\includegraphics[width=0.49\textwidth]{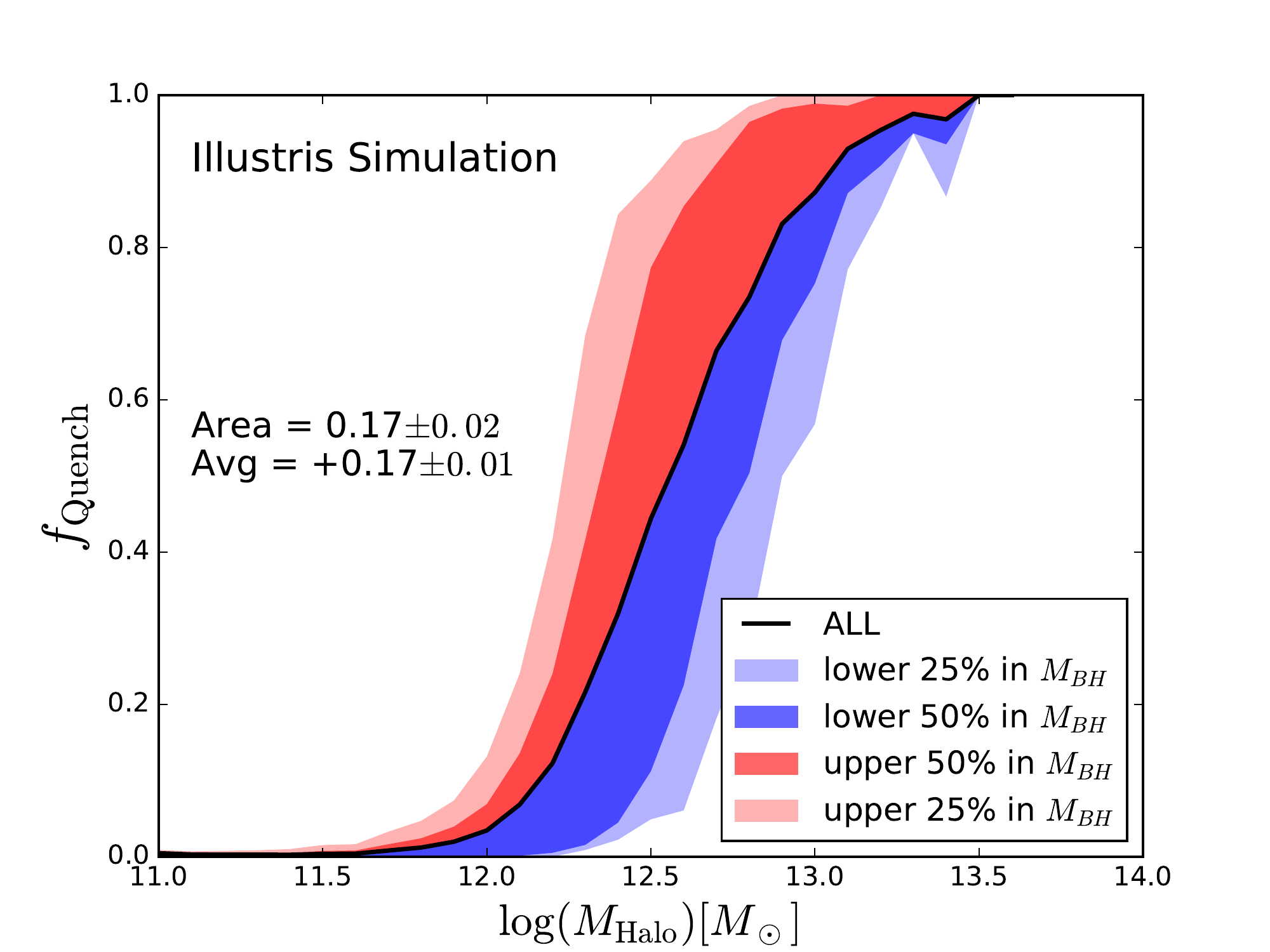}
\includegraphics[width=0.49\textwidth]{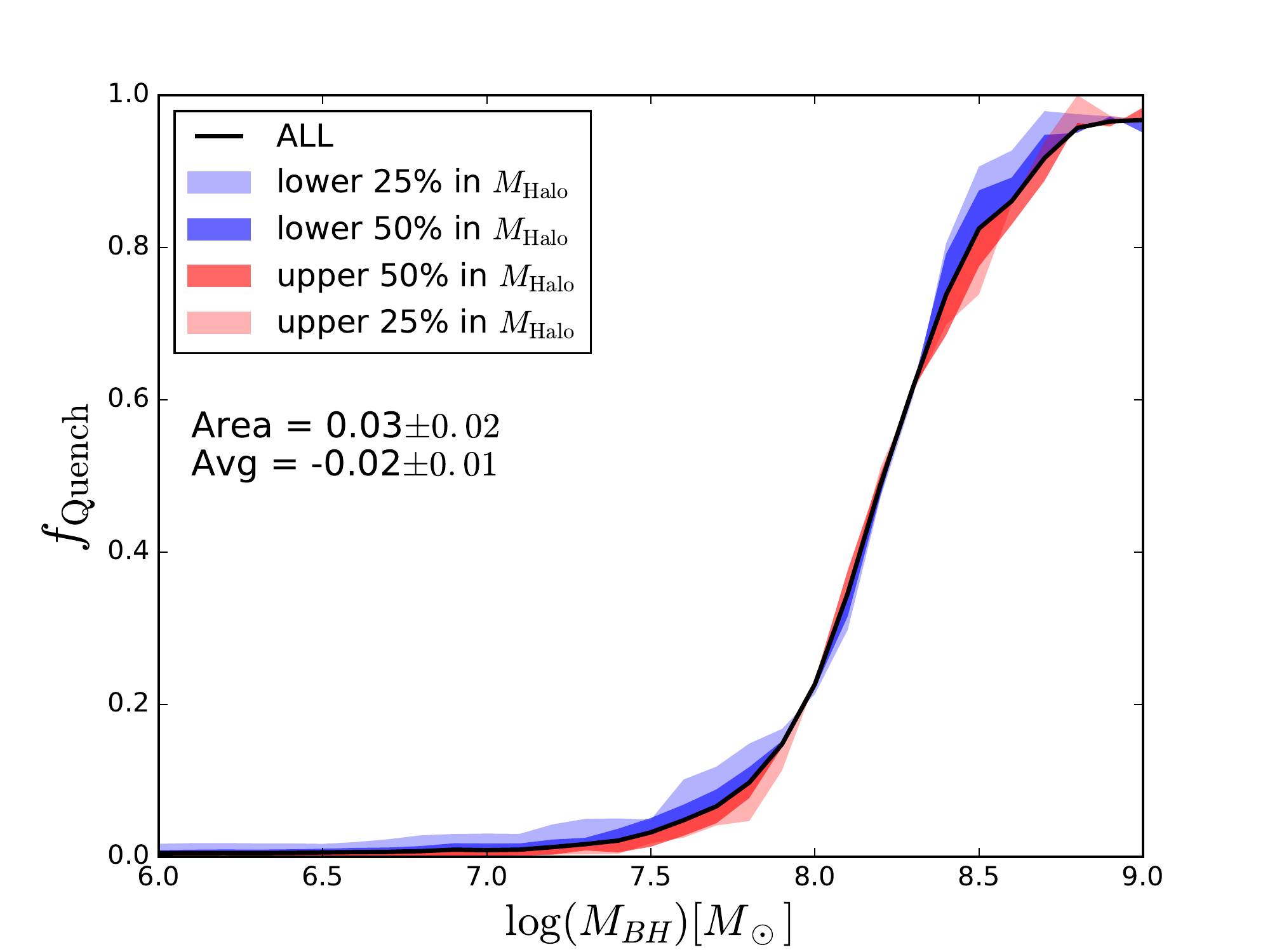}
\includegraphics[width=0.49\textwidth]{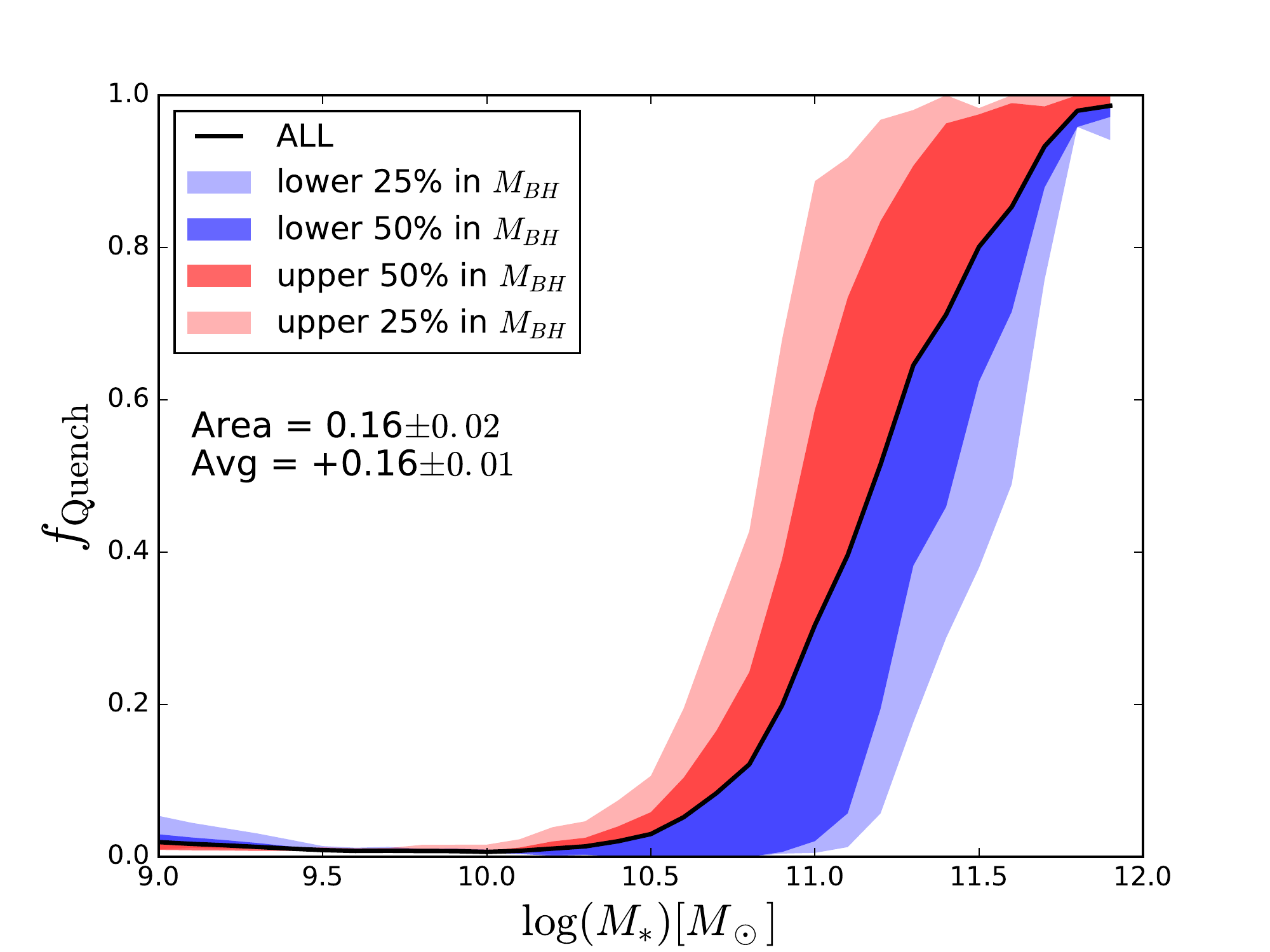}
\includegraphics[width=0.49\textwidth]{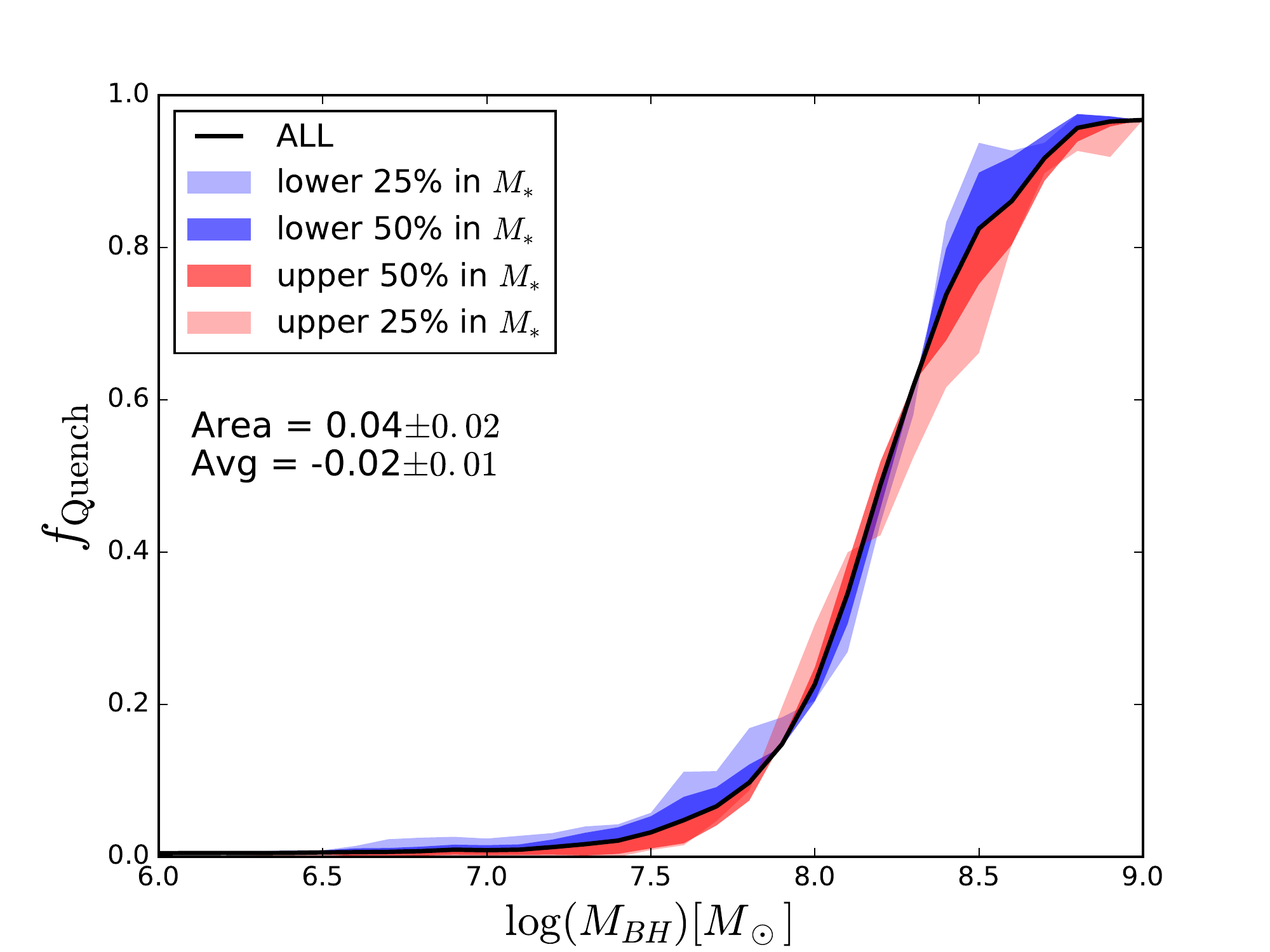}
\caption{The quenching of central galaxies in Illustris. The right panels show the quenched fraction - black hole mass relation, subdivided by halo mass (top) and stellar mass (bottom). The left panels show the quenched fraction - halo mass (top) and stellar mass (bottom) relations, each subdivided by black hole mass range. It is clear that the quenched fraction of central galaxies in Illustris is more accurately constrained by black hole mass than by halo or stellar mass, in agreement with the observations.}
\end{figure*}

In this sub-section we explore the predictions for central galaxy quenching from a semi-analytic model (SAM) and a cosmological hydrodynamical simulation. Specifically, we analyse the latest version of the Munich model (L-Galaxies: Henriques et al. 2015; earlier versions: Croton et al. 2006; De Lucia et al. 2009; Guo et al. 2011) and the Illustris simulation (Vogelsberger et al. 2014a,b). The details of the simulation and model are given in the above references. Both derive the properties of galaxies given theoretical prescriptions for galaxy formation and evolution, in a cosmological setting. The SAM constructs galaxies from an MCMC optimised set of free parameters applied to a coupled set of differential equations, modelling the physical processes that shape the evolution of different baryonic components on a fixed dark matter merger tree, from the Millennium Simulation (Springel et al. 2005). Thus, L-Galaxies inherits the resolution limits from the Millennium simulation and hence does not populate haloes with galaxies $< 10^{9.5} M_{\odot}$. Illustris probes the evolution of gas and dark matter together in a hydrodynamical simulation, and relies on semi-analytic prescriptions only for the sub-grid physics, typically for star formation and baryonic feedback. Both models quench massive central and satellite galaxies via AGN feedback, and lower mass satellite galaxies via environmental processes, such as ram pressure and tidal stripping, and the lack of primordial infall.

More specifically, `mass quenching' in L-Galaxies is driven by radio mode feedback (e.g., Croton et al. 2006), with the probability of a galaxy being quenched given by (Henriques et al. 2015):

\begin{equation}
P_{Q} \sim \dot{M}_{BH} = k_{\rm AGN} \bigg(\frac{M_{\rm hot}}{10^{11} M_{\odot}}  \bigg)  \bigg( \frac{M_{BH}} {10^{8} M_{\odot}} \bigg)
\end{equation}

\noindent where $M_{\rm hot}$ is the hot gas mass in the halo and $M_{BH}$ is the current mass of the central black hole. $k_{\rm AGN}$ is a free parameter to be fixed in the model. The mass of the black hole in the model grows primarily due to cold gas accretion triggered by merger events, and is proportional to both the mass ratio of the merger and the cold gas mass of the merger event. A fixed fraction of gas from the merger is channelled into the black hole, stars in the bulge and stars in the stellar halo. The specific fractions are determined from an MCMC minimisation comparing to a variety of observational inputs, including multi-epoch stellar mass functions. Mass growth from binary black hole mergers and hot gas accretion is also included, but is sub-dominant. Full details on the mass growth of black holes in L-galaxies is given in Henriques et al. (2013, 2015).

In Illustris there are three types of AGN feedback implemented: winds from the `quasar mode', mechanical heating of the halo from jets in the `radio mode', and radiative heating and ionisation of gas around the supermassive black hole. Of these three mechanisms, radio mode feedback is by far the most important mechanism for quenching galaxies in Illustris. Full details on the methods for implementing AGN feedback in Illustris are given in Sijacki et al. (2007),  Vogelsberger et al. (2013) and Torrey et al. (2014). Briefly, in the radio mode prescription, the energy contained within a jet induced bubble ($E_{\rm bub}$) in the hot gas halo is given by:

\begin{equation}
E_{\rm bub} = \mu \epsilon_{m} c^{2} \delta M_{BH}
\end{equation}

\noindent where $\mu$ is the radiative efficiency, i.e. the fraction of mass converted to energy via black hole accretion, and $\epsilon_{m}$ is the mechanical efficiency, i.e. the fraction of released energy which goes into the mechanical heating of the bubble, and hence halo. The bubble expands, shock heating the hot gas halo, and hence transferring its energy into increased temperature of the halo. The black hole mass growth, $\delta M_{BH}$, is modelled via Bondi accretion

\begin{equation}
\dot{M}_{BH} \propto M^{2}_{BH} \hspace{0.1cm}  \rho_{\rm gas} 
\end{equation} 

\noindent where $\rho_{\rm gas}$ is the density of gas around the black hole. The gas density and sound speed are determined based on the nearest gas particle neighbours which typically estimates gas properties on the spacial scale of the gravitational softening (i.e. $\sim$ 1 kpc for Illustris). The black hole growth is Eddington limited, and thus if the above prescription yields super Eddington accretion rates, the growth is taken as Eddington instead. Black hole mergers also contribute to the growth of black holes in Illustris. The formation of the stellar bulge and black hole are modelled quite independently in the Illustris simulation and hence relations between these two components may be taken as predictions rather than necessary consequences of the implementation, unlike in many semi-analytic models. Full details on the prescriptions for black hole growth in Illustris are given in Vogelsberger et al. (2013, 2014a,b).

As with L-Galaxies, the radio mode AGN feedback prescription in Illustris leads to a quenching probability which is primarily a function of black hole mass, i.e.

\begin{equation}
P_{Q} \sim f(M_{BH}) .
\end{equation}

\noindent However, both models have additional dependencies for AGN feedback: hot gas mass in L-Galaxies and gas density around the black hole in Illustris. In the following subsections we explore the quenching predictions for central galaxies from L-Galaxies and Illustris, and compare these to our observational results.

\begin{figure*}
\includegraphics[width=0.49\textwidth]{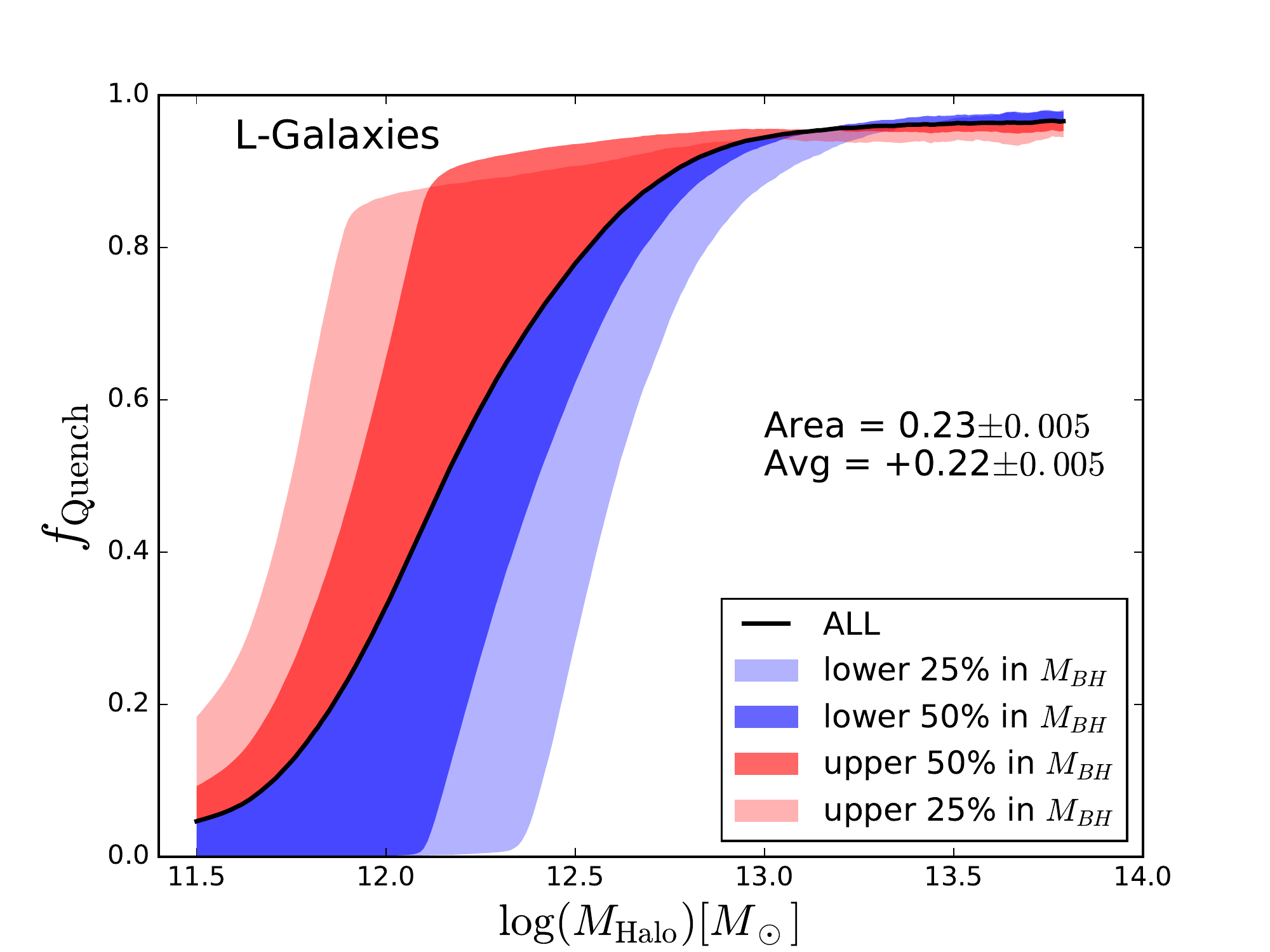}
\includegraphics[width=0.49\textwidth]{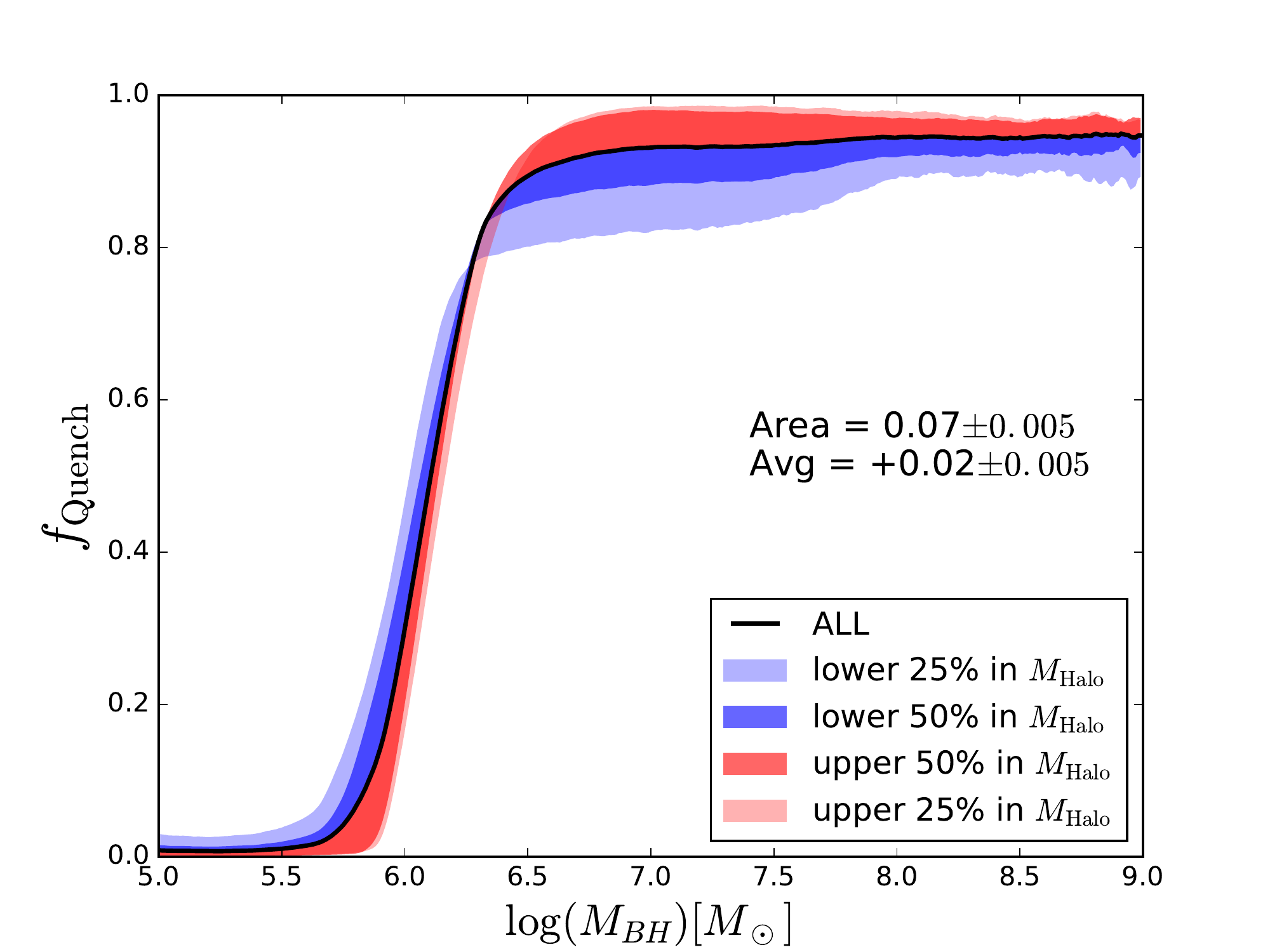}
\includegraphics[width=0.49\textwidth]{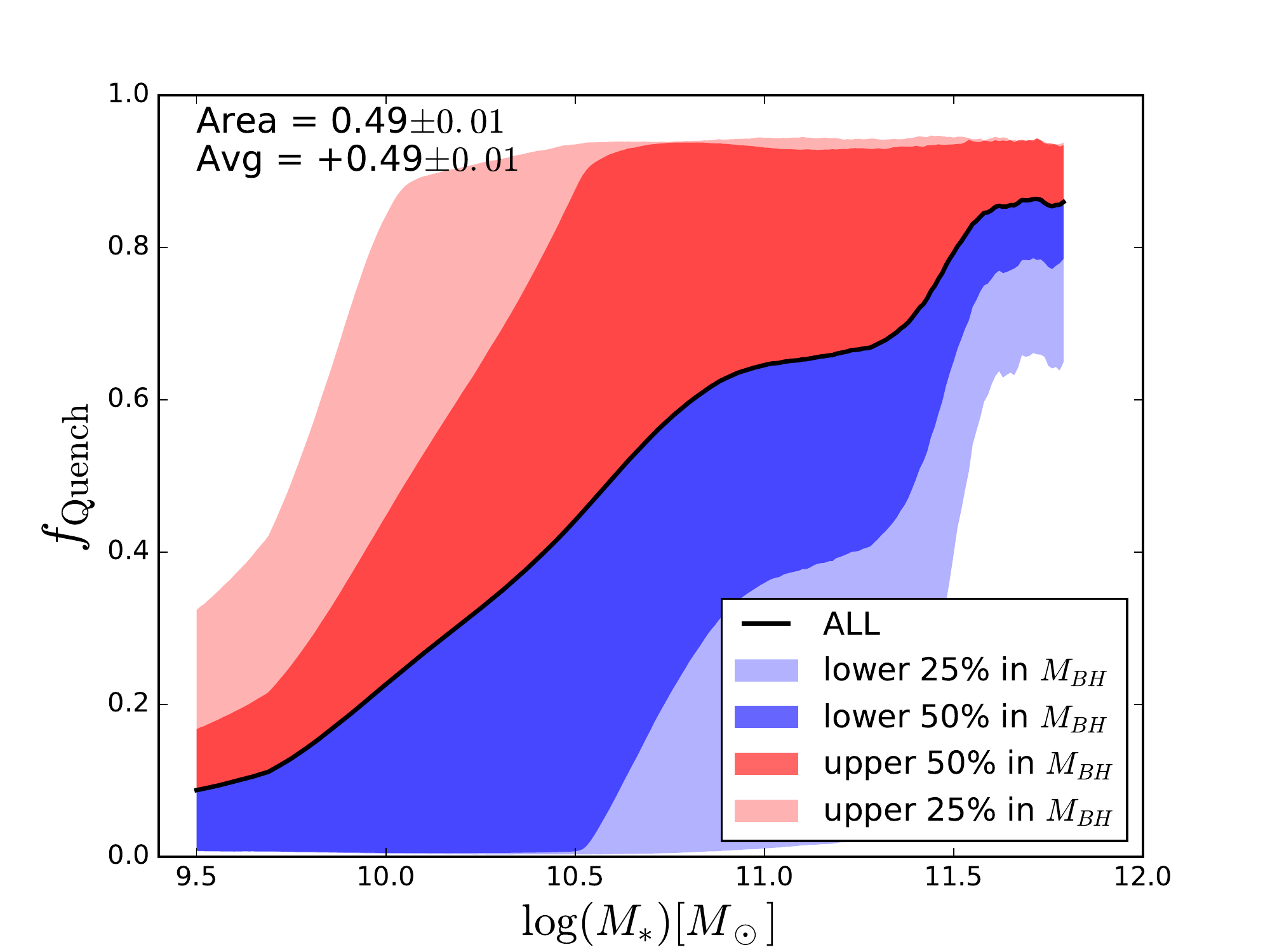}
\includegraphics[width=0.49\textwidth]{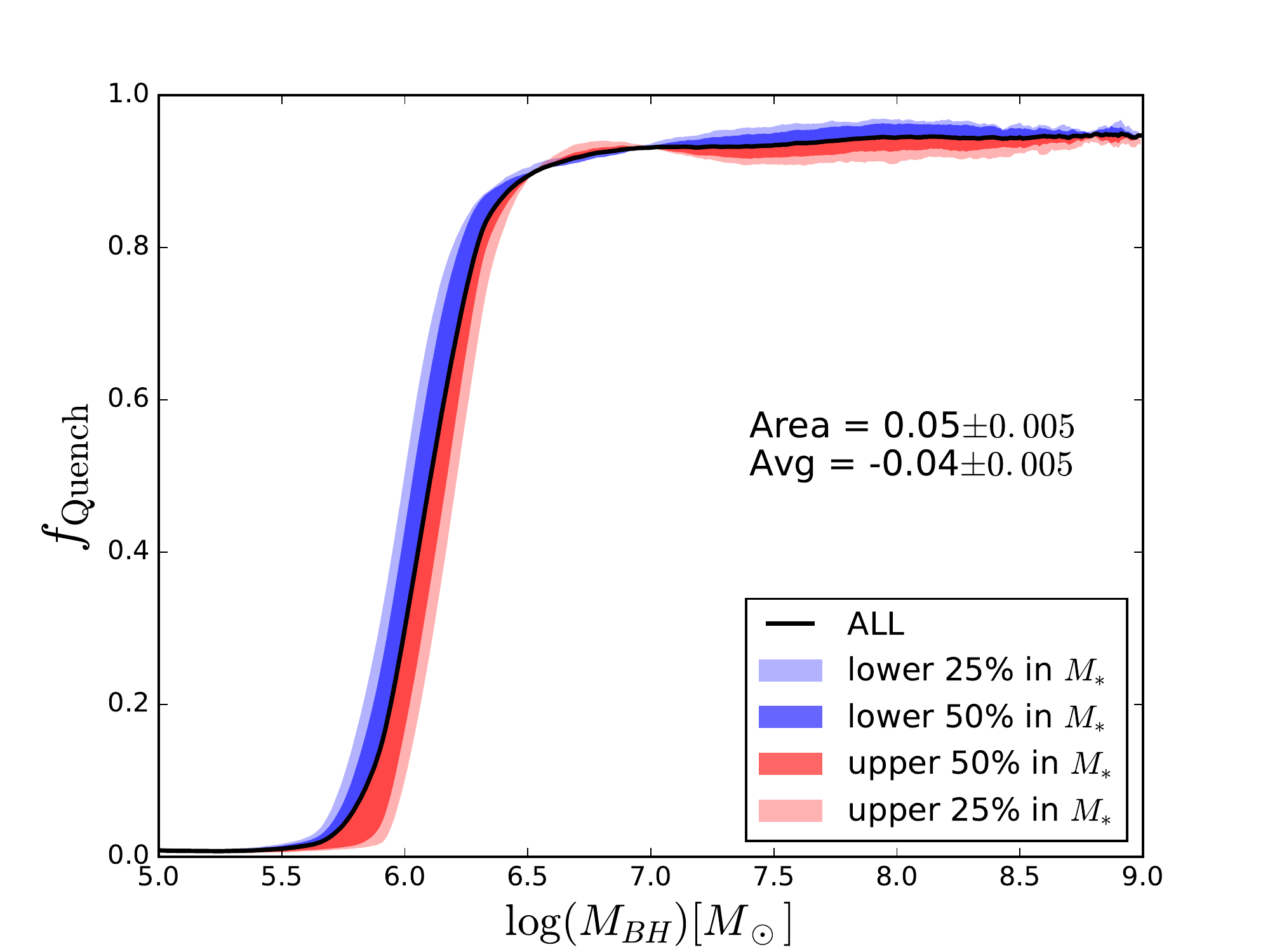}
\caption{The quenching of central galaxies in the Munich model, L-Galaxies. The right panels show the quenched fraction - black hole mass relation, subdivided by halo mass (top) and stellar mass (bottom). The left panels show the quenched fraction - halo mass (top) and stellar mass (bottom) relations, each subdivided by black hole mass range. It is clear that the quenched fraction of central galaxies in L-Galaxies is more accurately constrained by black hole mass than by halo or stellar mass. This trend is even more evident here than in the observations or in Illustris.}
\end{figure*}

\subsubsection{Estimating Black Hole Masses for SDSS Galaxies}

In order to compare our observational results for the SDSS to the predictions for central galaxy quenching from L-Galaxies and Illustris, we first must estimate central supermassive black hole masses for our observed galaxies. This is because black hole masses in the models are a fundamental output of the catalogues, whereas central velocity dispersions are not. The reason for this is that central velocity dispersions span the intermediate range between what can be modelled only via sub-grid physical prescriptions and the processes for which there is sufficient resolution to simulate the evolution directly. We estimate black hole masses using the well known $M_{BH} - \sigma$ relation (e.g., Ferrarese \& Merritt 2000). Specifically, we use a new parameterization from Saglia et al. (2016) for their full morphological sample, calculating:

\begin{equation}
\log(M_{BH}[M_{\odot}]) = 5.25 \times \log(\sigma_{c}{\rm [km/s]}) - 3.77
\end{equation}

\noindent This gives a formal scatter with 96 dynamically measured black hole masses of 0.46 dex. This relation also leads to a significantly tighter fit than the best parameterizations with bulge mass, bulge effective radius or central stellar mass density. Moreover, dynamically measured black hole mass is much more tightly correlated with central velocity dispersion than global galaxies properties, such as total stellar mass and morphology, as well as for environmental properties, such as halo mass or local density (e.g., Hopkins et al. 2007). Thus, in order to make comparisons with black hole masses in models and simulations, we use the (same) $M_{BH} - \sigma_{c}$ scaling relation in all cases. 

Given our large sample size, and hence that we typically have several tens of thousands of galaxies per bin in our analyses, the statistical error on the black hole estimate will be negligible. However, issues from systematics in the $M_{BH} - \sigma_{c}$ relation will likely dominate the total error. For example, Kormendy \& Ho (2013) and Saglia et al. (2016) find that classical bulges and pseudo bulges exhibit very different $M_{BH} - \sigma_{c}$ scaling laws, with different average scatter. Additionally, McConnell \& Ma (2013) find differences in the scaling laws for early- and late-type galaxy morphologies (ETGs and LTGs, respectively).

For our purposes here, the intent is merely to provide an estimate of what given values of central velocity dispersion correspond to in terms of central black hole mass, under the assumption that the scaling law can be applied across the diverse set of galaxy types in the SDSS. Thus, in general, the details of our comparison to models will be scaling law dependent. However, we find for a variety of reasonable choices of scaling law (including those which fit ETGs and LTGs, and classical and pseudo bulges separately) our final results are almost identical, and hence our conclusions are robust to uncertainties in the choice of scaling law parameterization. See Appendix B for examples of different scaling laws and their impact on our analysis.

In Fig. 6 we show the quenched fraction relationship with black hole mass instead of central velocity dispersion, where the former is estimated as a function of the latter. Additionally, we compare in this figure he quenched fraction black hole mass relation to the halo mass and stellar mass relations, using the area statistics technique (see Section 5). Although this exercise is done primarily to facilitate comparison to models, it does also reveal some interesting general features. As with central velocity dispersion, the quenched fraction -- black hole mass relation exhibits very little (if any) variation from changing stellar or halo mass. However, increasing black hole mass at a fixed stellar or halo mass dramatically increases the quenched fraction (compare rows in Fig. 6).  The fraction of quenched galaxies reaches 50\% at an estimated black hole mass of $M_{BH} \sim 2 \times 10^{7} M_{\odot}$.

In Fig. 7 we show the stellar mass - halo mass relation for observed SDSS and simulated L-Galaxies and Illustris galaxies. The halo masses for each dataset are given as the $M_{200}$ virial mass of the group or cluster. Hexagonal bins are coloured by the mean value of supermassive black hole mass. For the simulated data, black hole masses are taken as the sub-grid output mass in their respective catalogues. For SDSS galaxies we estimate black hole masses from central velocity dispersions (outlined above, see also Appendix B). In general, there is a strong positive relationship between stellar mass and halo mass for centrals but no discernible relationship for satellites. This is as expected, since the mass of a satellite is not constrained tightly by its parent halo, but the mass of a central is. Black hole masses increase with both stellar and halo mass for centrals, but are primarily a function of stellar mass alone for satellites. These features are qualitatively similar between the observations and the models, however a detailed look at Fig. 7 reveals many subtle differences. For example, the `knee' in the $M_{*} - M_{\rm halo}$ relation for centrals is much more pronounced in the observational data than in either L-Galaxies or Illustris. Additionally the $M_{*} - M_{\rm halo}$ relation for centrals is noticeably tighter in Illustris than in the observations, and much less tight in L-Galaxies than in the observations or Illustris. For the remainder of the analyses in this section we focus on the quenching of central galaxies in the models and how this compares to observations.

\subsubsection{Central Galaxy Quenching Predictions from L-Galaxies \& Illustris}

In Fig. 8 we show the z = 0.1 snapshot main sequence relation for L-Galaxies and Illustris, as contour plots. Given that the normalisation of the main sequence is offset between these two models and, indeed, the observational data (see Fig. 1), a simple prescription for defining quenched galaxies based on a fixed sSFR cut would be ill advised. The solid green line indicates the median value of SFR at each stellar mass, and it is very close to a linear relationship at low stellar masses ($M_{*} < 10^{10} M_{\odot}$). By fitting the median relationship, we find that the straight line fit (shown as a blue dashed line on each panel) fairly well describes the main sequence relation, i.e. it goes through the middle of the contours in each case. We then define a quenched galaxy to lie at least one order of magnitude in SFR below the main sequence line, as with the observational data (this is indicated by a red dashed line). In the models, passive galaxies have arbitrarily low SFRs and hence, unaltered, would form a dilute `tail' not visible as a concentrated peak in a contour plot. The reason the passive population in the SDSS appears as a concentrated peak in Fig. 1 is due to there being a minimum sSFR (= $10^{-3} {\rm Gyr}^{-1}$) used to estimate the passive galaxies' SFR in the Brinchmann et al. (2004) star formation rate determination. We apply this same minimum sSFR threshold in Fig. 8 to better compare with the observational data.

In Fig. 9 the dependence of central galaxy quenching on stellar mass, halo mass and black hole mass is shown for the Illustris simulation. An area statistics approach is used, as in Section 5.2. There are strong correlations between the fraction of quenched galaxies and each of these parameters. However, at fixed black hole mass there is very little impact on the quenched fraction from varying either stellar or halo mass (right panels). Whereas, at fixed stellar or halo mass there is a large positive affect on the quenched fraction from varying black hole mass (left panels). This is as expected in the simulation, since central galaxies are quenched primarily due to radio mode AGN feedback, which is correlated directly with $M_{BH}$, and only indirectly with $M_{*}$ and $M_{\rm halo}$. These results may be compared to the equivalent plots for SDSS galaxies (under the assumption of the Saglia et al. 2016 scaling law) in Fig. 6.

The smaller areas and average mean differences from black hole mass compared to stellar or halo mass agrees well with our observational findings (e.g., Fig. 6), indicating that quenching via AGN feedback is at least consistent with our results. However, some of the details of the Illustris quenching prediction are different to the SDSS data. In Illustris, 50\% of central galaxies are quenched at a black hole mass of $M_{BH} \sim 2 \times 10^{8} M_{\odot}$, whereas in the observations this occurs at a significantly lower mass of $M_{BH} \sim 2 \times 10^{7} M_{\odot}$. Furthermore, the quenching of centrals in Illustris occurs more abruptly than in the SDSS, with a noticeably steeper gradient on the $f_{\rm Quench} - M_{BH}(\sigma_{c})$ relationship (compare Fig. 9 \& 6, right panels). The comparison of observational data with the Illustris simulation shows that, although its implementation in the models may not be fully accurate, the qualitative impact of AGN feedback is consistent with the observations.

In Fig. 10 the dependence of central galaxy quenching on stellar, halo and black hole mass is shown for the Munich semi-analytic model (L-Galaxies), again using the area statistics prescription (see Section 5.2). Although there are strong correlations between quenched fraction and each of these galaxy properties, it is clear that by far the tightest relationship is with black hole mass. This agrees qualitatively with the observational findings of this work (see Section 5 and Fig. 6). This trend, however, is even larger than witnessed in the Illustris simulation, or in the observational data (especially for stellar mass). In L-Galaxies, 50\% of centrals are quenched at $M_{BH} \sim 10^{6} M_{\odot}$, a significantly lower mass than in the SDSS or the Illustris simulation. Furthermore, quenching is even more abrupt in this model than in Illustris, which is itself more abrupt in its quenching than observed in the SDSS. Part of the reason for this may be that we are only estimating black hole masses (from central velocity dispersion) and hence the dependence between black hole mass and quenching may get stronger, and perhaps more steep, if we had more direct means to measure black hole masses. However, given that quenching is less abrupt in the hydrodynamical simulation than in the SAM, it may also be that a more realistic description of how the energy released from the black hole couples to the hot gas halo may reduce the steepness of the dependence of black hole mass on quenched fraction.

\begin{figure*}
\includegraphics[width=0.49\textwidth]{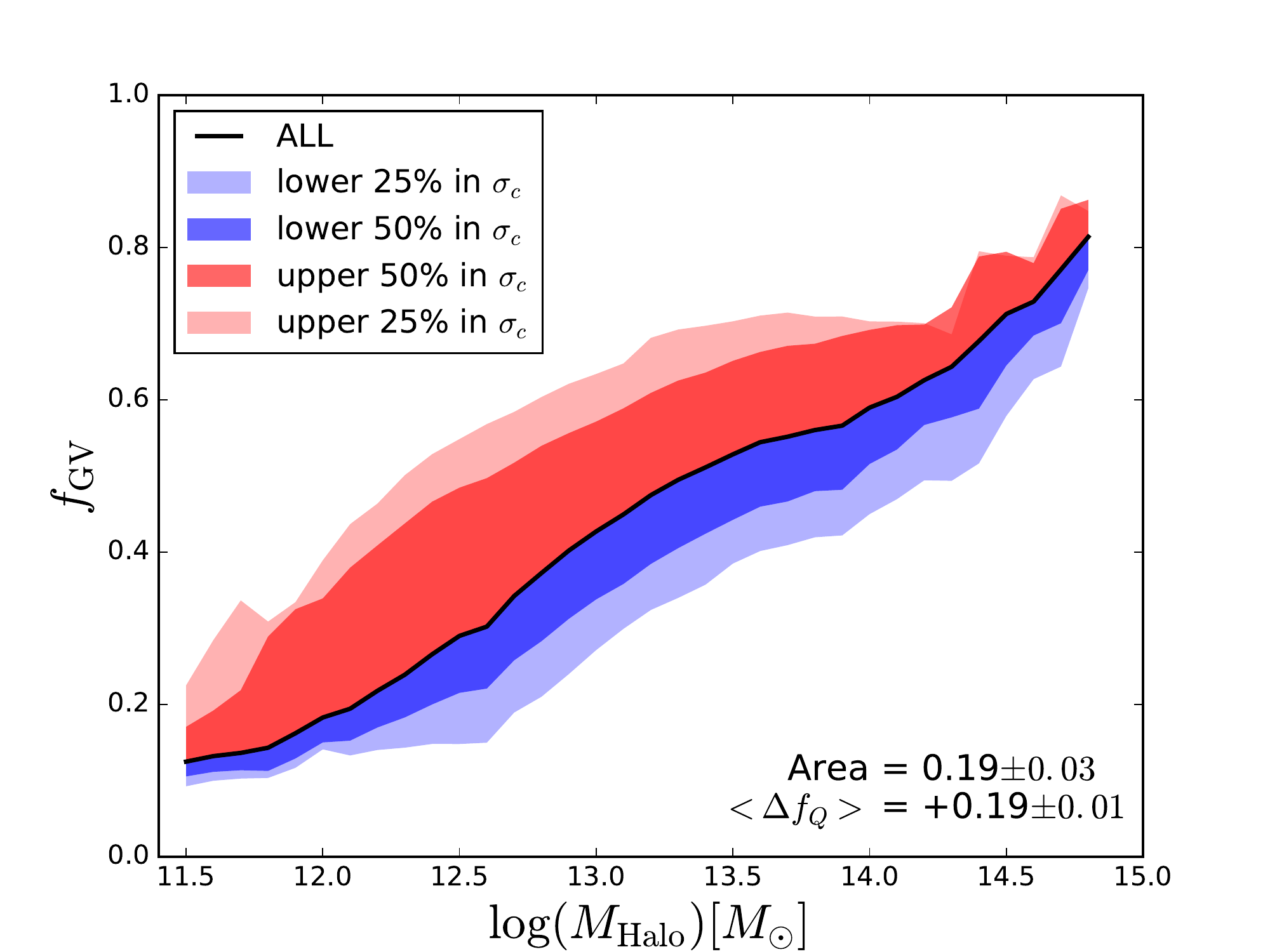}
\includegraphics[width=0.49\textwidth]{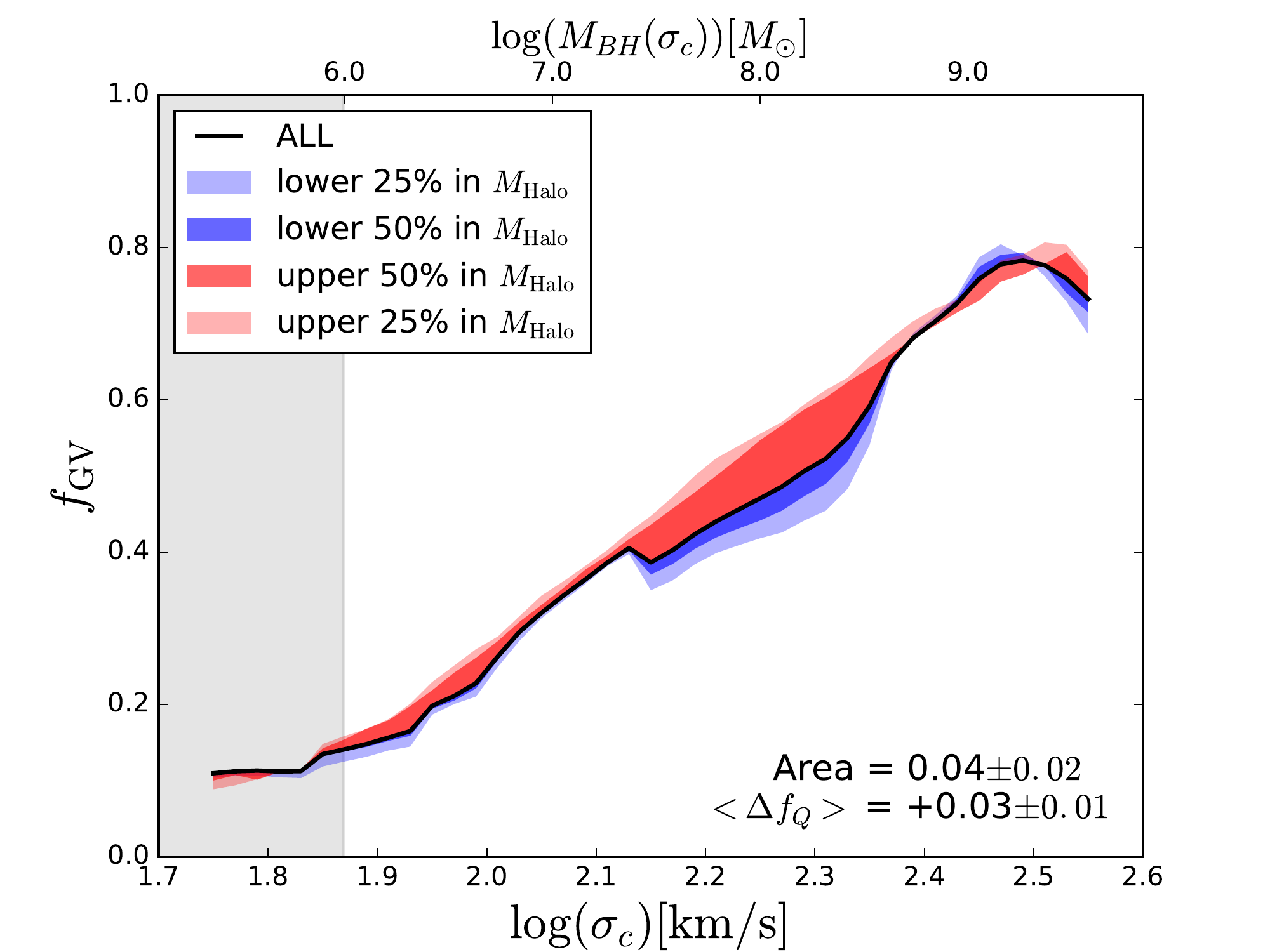}
\includegraphics[width=0.49\textwidth]{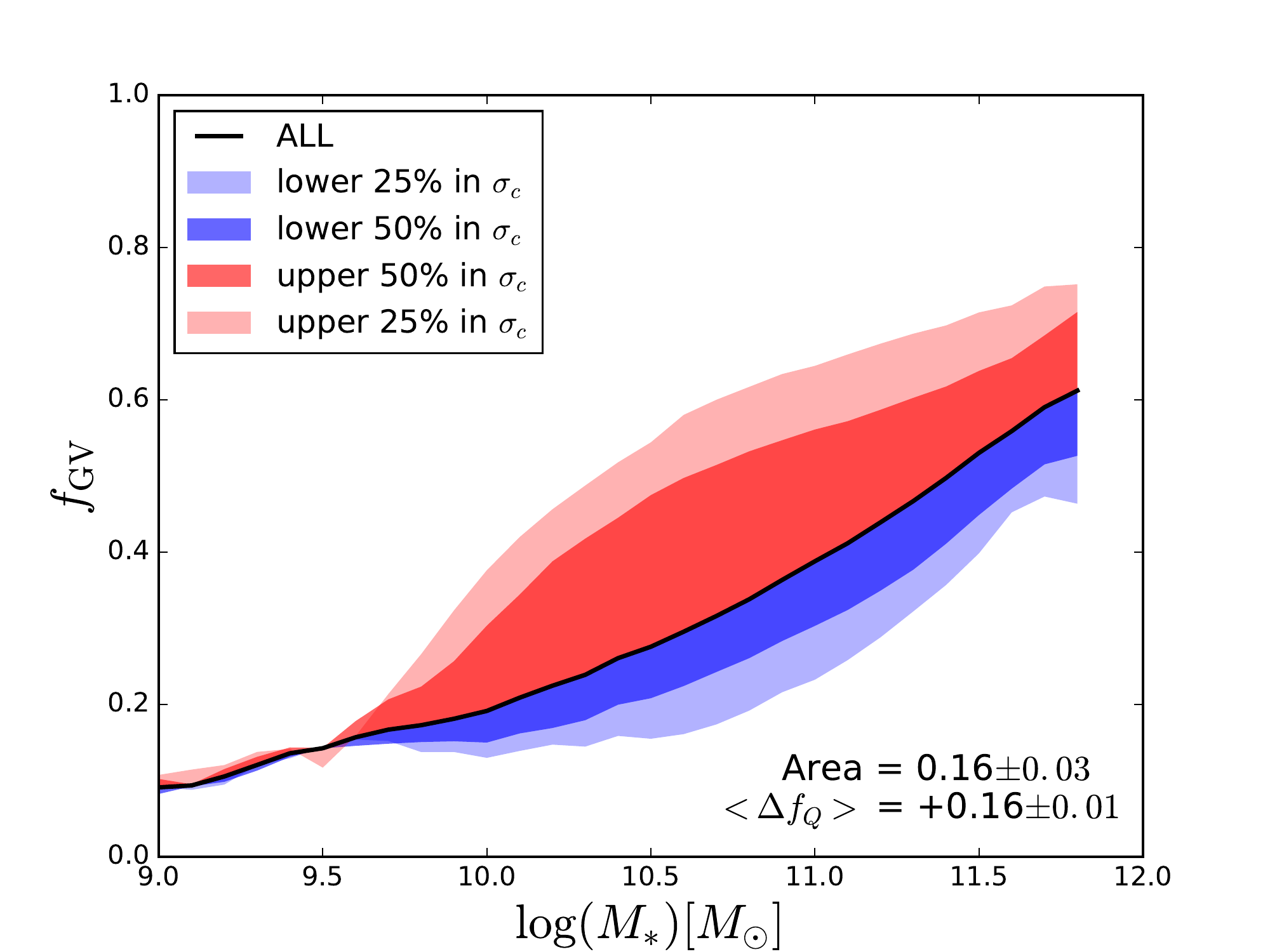}
\includegraphics[width=0.49\textwidth]{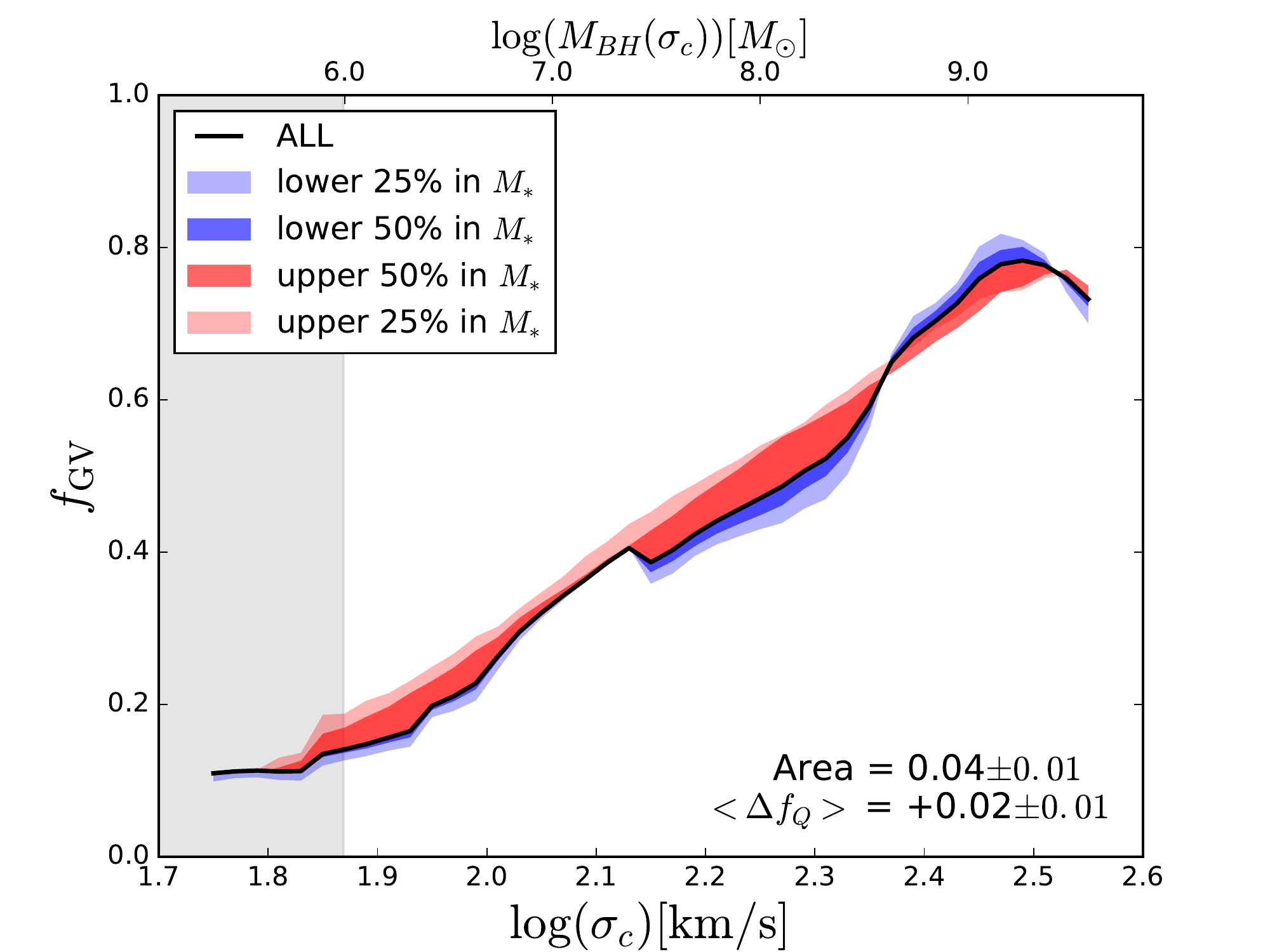}
\caption{Area statistics plot for green valley fractions. The left panels show the green valley fraction as a function of central velocity dispersion, split in the top panel by halo mass and in the bottom panel by stellar mass. The right top panel shows the green valley fraction as a function of halo mass and the right bottom panel shows the green valley fraction as a function of stellar mass, each split by central velocity dispersion range. It is clear that varying stellar or halo mass at fixed central velocity dispersion leads to very little difference in the green valley fraction, whereas varying central velocity dispersion at fixed stellar or halo mass leads to a significant positive impact on the green valley fraction. Thus, central velocity dispersion is the most important variable for {\it quenching} as well as quenched galaxies.}
\end{figure*}

L-Galaxies also predicts a turn over in the way halo mass affects central galaxy quenching at fixed black hole mass (Fig. 10, top right panel). At lower black hole masses increasing the halo mass leads to a decrease in the quenched fraction; whereas, at higher black hole masses, increasing the halo mass results in increasing the quenched fraction. We do see a hint of this feature in the observational data as well, for some scaling laws (see Appendix B, Fig. B3), but the effect is significantly smaller than in the model. In the model, the turn over is explained by the probability of a galaxy being quenched depending on hot gas mass, which increases with increasing halo mass. The comparison with observational data suggests that this dependency may be too strong in the current implementation.

To summarise the model comparisons, the quenching of central galaxies in Illustris and L-Galaxies are both significantly more tightly constrained by black hole mass than stellar or halo mass. Assuming the scaling relation of Saglia et al. (2016), we find the exact same result in the SDSS observations. Thus, given that the quenching of centrals in Illustris and in the Munich model is governed by radio mode AGN feedback, our observational results are consistent with models that quench central galaxies via AGN feedback. However, the details of central galaxy quenching in both the SAM and the hydrodynamical simulation disagree with several detailed features of the observational data.  Specifically, L-Galaxies has too efficient quenching as a function of black hole mass (lower average black hole mass, at a 50\% quenched fraction, than observed), and Illustris has too inefficient quenching as a function of black hole mass (higher average black hole mass, at a 50\% quenched fraction, than observed).

\subsection{Alternative Explanations to AGN Feedback \& Green Valley Fraction Test}

The term `quenching' is somewhat confusing, as it may refer to either (or both) of 1) the process(es) which initially cause a galaxy to depart the main sequence and progress towards being passive; or 2) the process(es) which keep a galaxy passive, after its initial cessation of star formation. If our purpose is to understand quenching in the first definition, it is essential to measure the galaxy properties at the time of first quenching, not some arbitrary time later. This is important because the evolution of galaxy properties with redshift can cause systematic differences in their values, which may align with quenching and have no causal connection. For instance, the size of a galaxy of a given mass evolves with redshift as $(1+z)^{-a}$, where $a \sim 1$ (e.g., Newman et al. 2012). This results in galaxies which are quenched earlier in the history of the Universe having smaller sizes for their masses, and hence higher central mass concentrations and presumably central velocity dispersions. Therefore, it is possible that the observed strong correlation between quenched fraction and central velocity dispersion is in some sense tautologous, when the size evolution is taken into account (as first argued for in Carollo et al. 2013; Lilly \& Carollo 2016).

It remains to be seen in detail if size evolution alone can lead to the observed trends witnessed in this work. However, in this section we consider a few cases from the literature which suggest that this may not be the ultimate origin of the link between inner galactic structure and quenching, and furthermore propose a test of this idea using the green valley of the SDSS data. The problem arises because in our sample of passive galaxies, most of the galaxies quenched for the first time several billion years ago, when their properties would have (potentially significantly) different values. The peak of quenching in the Universe appears to be at $z \sim 1 - 2$, given the sharp decline in the star formation rate density over this epoch (e.g., Lilly et al. 1996; Madau et al. 1998; Cucciati et al. 2012). Lang et al. (2014) probed this redshift range using CANDELS\footnote{Cosmic Assembly Near-Infrared Extragalactic Legacy Survey} data, finding that the quenched fraction correlates more strongly with bulge mass than total stellar mass or morphology (an equivalent result at high redshifts to the low redshift result of Bluck et al. 2014). This suggests that the requirement of a dense central region for a galaxy to be quenched was in place at early cosmic times, and hence likely contemporaneous to quenching (in the first definition). Furthermore, the fact that galaxies are less bulge dominated for their stellar masses, and hence more likely late-types morphologically, at earlier cosmic times (e.g., Buitrago et al. 2013; Mortlock et al. 2013) suggests that the importance of a large bulge structure for central galaxy quenching cannot be the result of evolutionary systematics.

However, Barro et al. (2013, 2014) find evidence at high redshifts ($z \sim 2 - 3$) for a substantial population of star forming galaxies with highly compact cores, and high central velocity dispersions. Thus, the relationship between quenching and central mass concentration (and hence central velocity dispersion and black hole mass) may not come into place until more moderate redshift ranges. This notwithstanding, they also find that a high fraction of these galaxies ($>$ 30 \%) are currently undergoing a luminous AGN phase, which could be precisely the mechanism by which these galaxies become passive at $z \sim$ 1.  At low redshifts, in the SDSS, Mendel et al. (2013) performed a spectral decomposition of passive galaxies, into recently and not-recently quenched systems. They find that the requirement for galaxies to have a high S\'{e}rsic index in order to be quenched is equally true in the recently and not-recently quenched sample, suggesting that a high central light/ mass concentration is achieved contemporaneously (or prior) to quenching and hence is not an artefact of evolutionary systematics (e.g., Carollo et al. 2013; Lilly et al. 2016).

In this work, we focus on SDSS data where we are limited to $z < 0.2$, which is a requirement additionally of our desire to measure accurate bulge - disk decompositions and central velocity dispersions. Thus, we cannot probe the peak of quenching in the Universe. However, with a few assumptions, we can investigate the quenching of galaxies today. Assuming that galaxies in the green valley are on a trajectory towards the quenched red sequence, and that they spend in general only a small amount of time in transit ($\sim 1 - 2$ Gyr), the fraction of galaxies entering the green valley at a given time can be used as a proxy for the {\it quenching} fraction, i.e. the recently quenched or to-be quenched fraction of galaxies in a given population. These assumptions are consistent with recent observations (e.g., Schawinski et al. 2014). Moreover we emphasise that they only have to be true {\it on average} in that most galaxies are moving in one direction (from blue to red) and most do so quickly relative to the evolutionary changes of galaxies under consideration here, e.g. size and structural evolution (with scaling times typically $>$ 3 Gyr at late cosmic times). 

\begin{figure*}
\includegraphics[width=0.49\textwidth]{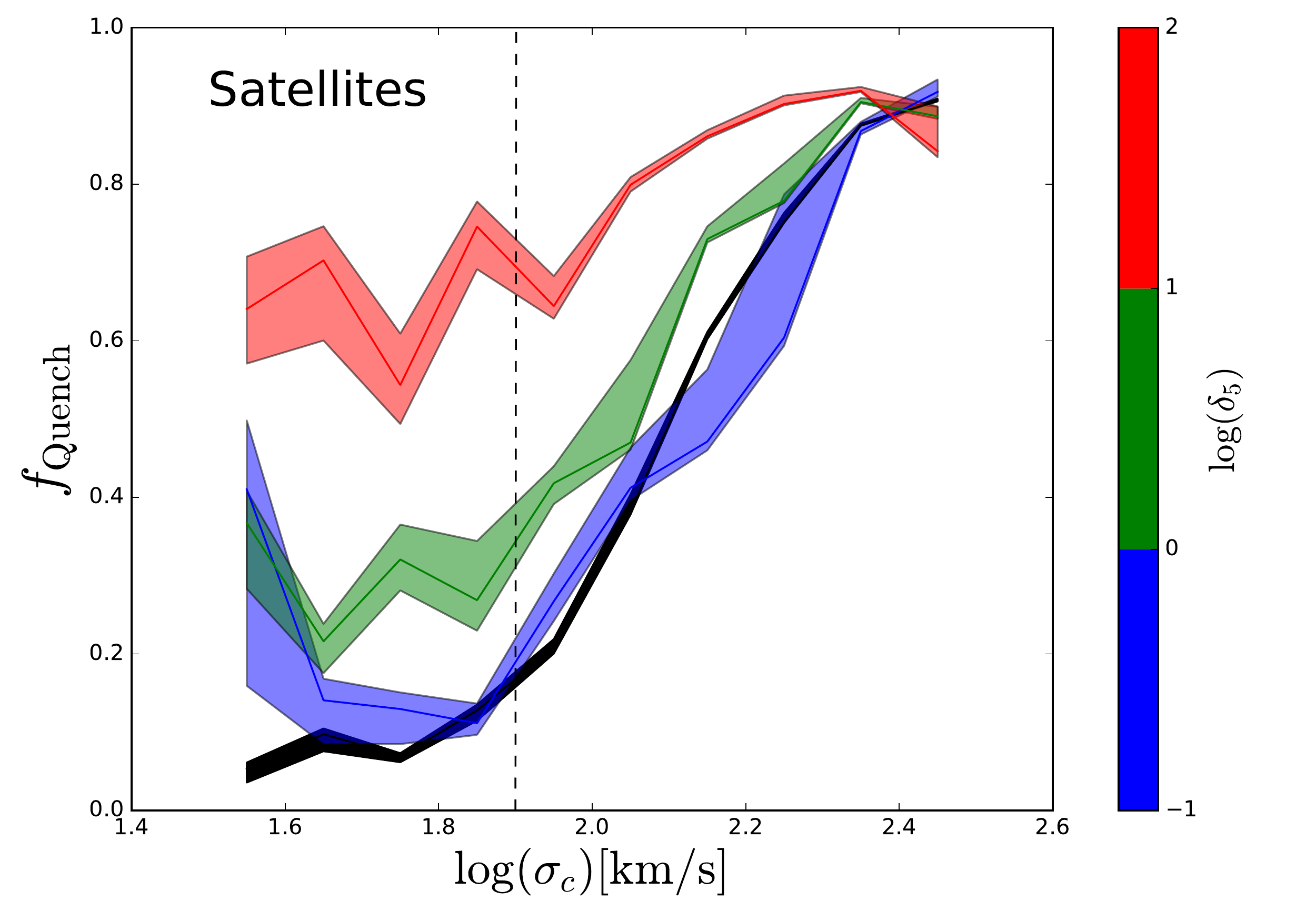}
\includegraphics[width=0.49\textwidth]{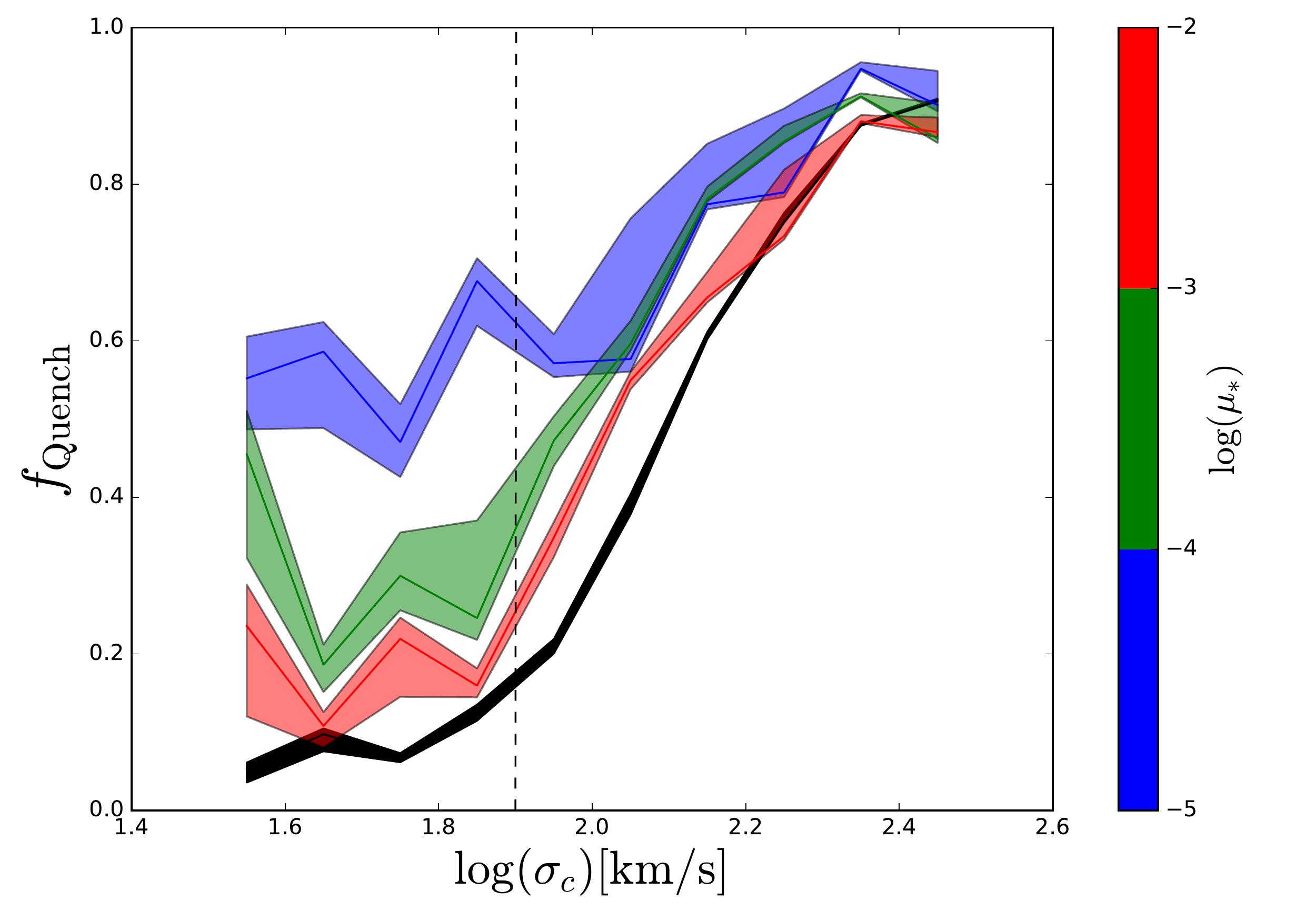}
\includegraphics[width=0.49\textwidth]{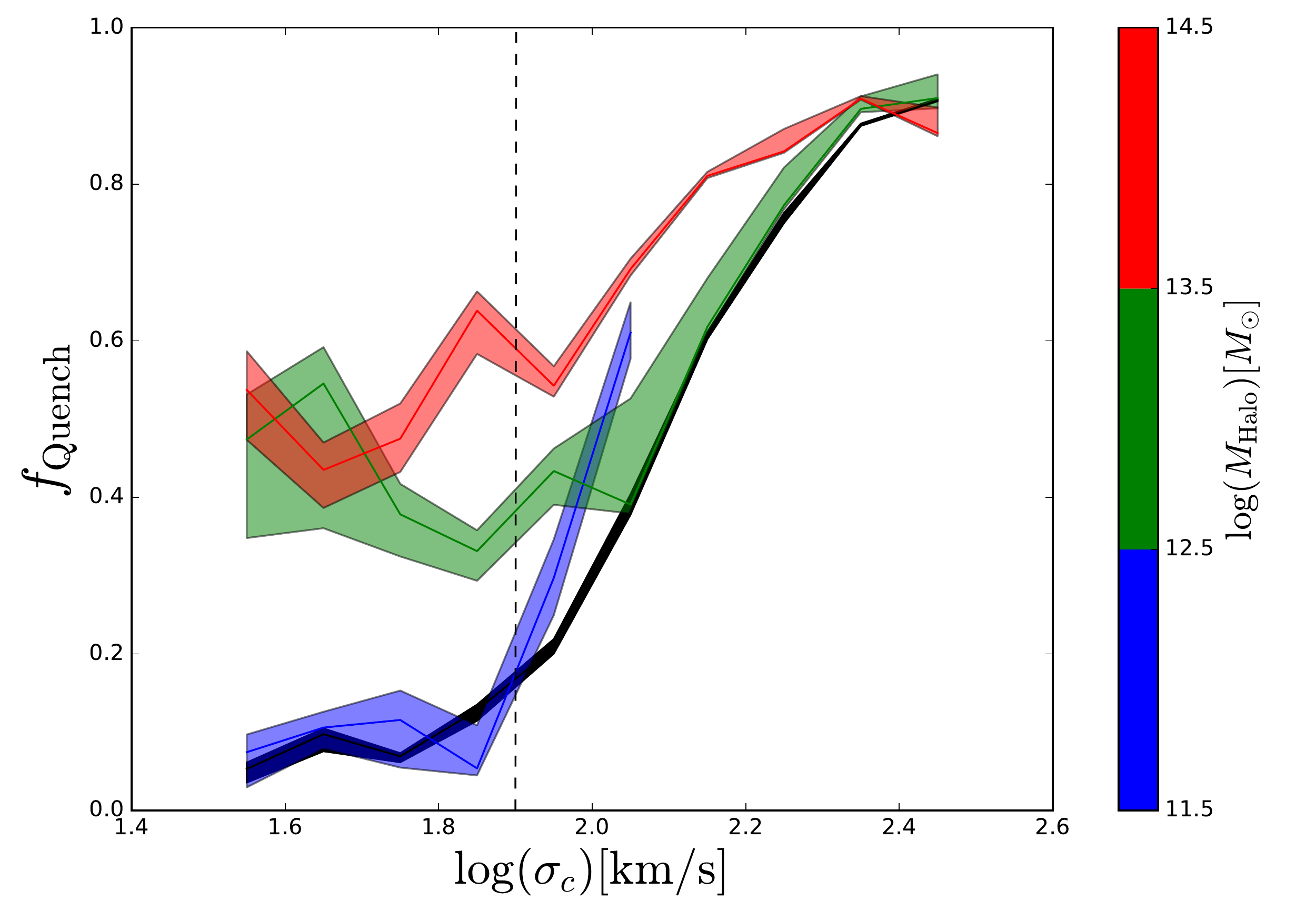}
\includegraphics[width=0.49\textwidth]{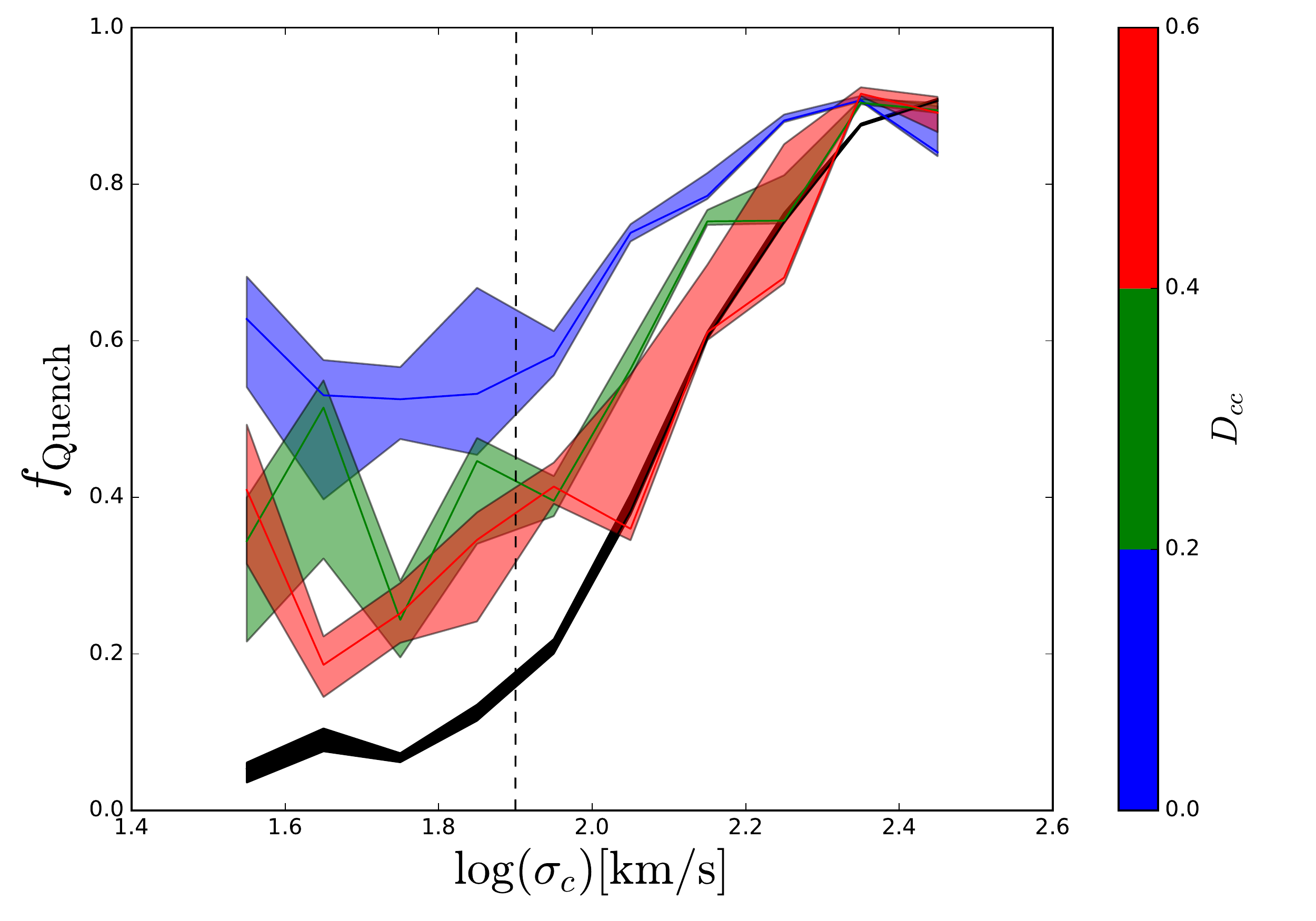}
\caption{The quenched fraction - central velocity dispersion relationship for satellite galaxies, divided by local density ($\delta_{5}$), mass ratio ($\mu_{*}$), halo mass ($M_{\rm halo}$) and group-centric distance ($D_{cc}$), shown as colour bars. Each of the environmental parameters engender a significant perturbation on the $f_{\rm Quench} - \sigma_{c}$ relation for satellites, unlike for centrals (see Figs. 3 \& 5). The impact of environment on quenching is clearly more pronounced at low central velocity dispersions and is absent entirely at high values of $\sigma_{c}$. The black line indicates the unbinned $f_{\rm Quench} - \sigma_{c}$ relation for central galaxies, shown for comparison. The 1 $\sigma$ error from the jack-knife technique is shown as the shaded region for each binning in each plot.}
\end{figure*}

Specifically, we define the green valley fraction, in analogy to the quenched fraction (eq. 7), as:

\begin{equation}
f_{GV} = \frac{w_{i}/V_{\rm max} (-1.2 < \Delta SFR < -0.6)}{w_{i}/V_{\rm max} (\Delta SFR > -1.2)} \approx \frac{N_{GV}}{N_{MS} + N_{GV}}
\end{equation}

\noindent where the range of the green valley is indicated in Fig. 1, and represents roughly speaking the trough or valley between the dominant blue and red peaks. The normalisation of the green valley fraction excludes passive systems and in this sense represents the fraction of galaxies which have recently departed the main sequence at each binning in galaxy properties. This statistic allows us to get a better understanding of the first definition of quenching (outlined above), i.e. which properties matter most when quenching first takes effect. 

In Fig. 11 we compare halo mass, stellar mass and central velocity dispersion as drivers of initial galaxy quenching, via an area statistics technique applied to the green valley fraction. Here again, we find that central velocity dispersion, and hence estimated black hole mass (shown as a upper x-axis using the scaling law of Saglia et al. 2016), correlate significantly tighter with the green valley fraction than stellar or halo mass. This suggests that the formation of a dense inner structure is important to quenching contemporaneously (or prior) to the initial onset of star formation cessation. Thus, evolutionary systematics cannot fully explain the trends witnessed in this work, in agreement with the conclusions of Mendel et al. (2013). 

For the second definition of quenching, which refers to the process(es) which keep galaxies quenched, it is relevant to study the properties of quenched galaxies at all epochs. Certainly at low redshifts, effectively all quenched central galaxies have high central velocity dispersions and hence black hole masses, but not all quenched centrals have high stellar or halo masses (see Figs. 4 \& 6). This is also true of contemporaneously {\it quenching} galaxies in the green valley (Fig. 11). Taken together, it is highly likely that the observed tight correlation between central velocity dispersion and galaxy quenching is causal in nature (as considered in Section 6.1), and not an artefact of evolutionary systematics on the galaxy population. However, a key test to this paradigm is still to be performed, which involves measuring the redshift evolution in the quenched fraction dependence on the various galaxy properties considered in this work, which has to date been only partially realised (e.g., Barro et al. 2013, 2014; Lang et al. 2014).

\section{Results for Satellites}

\begin{figure*}
\includegraphics[width=0.49\textwidth]{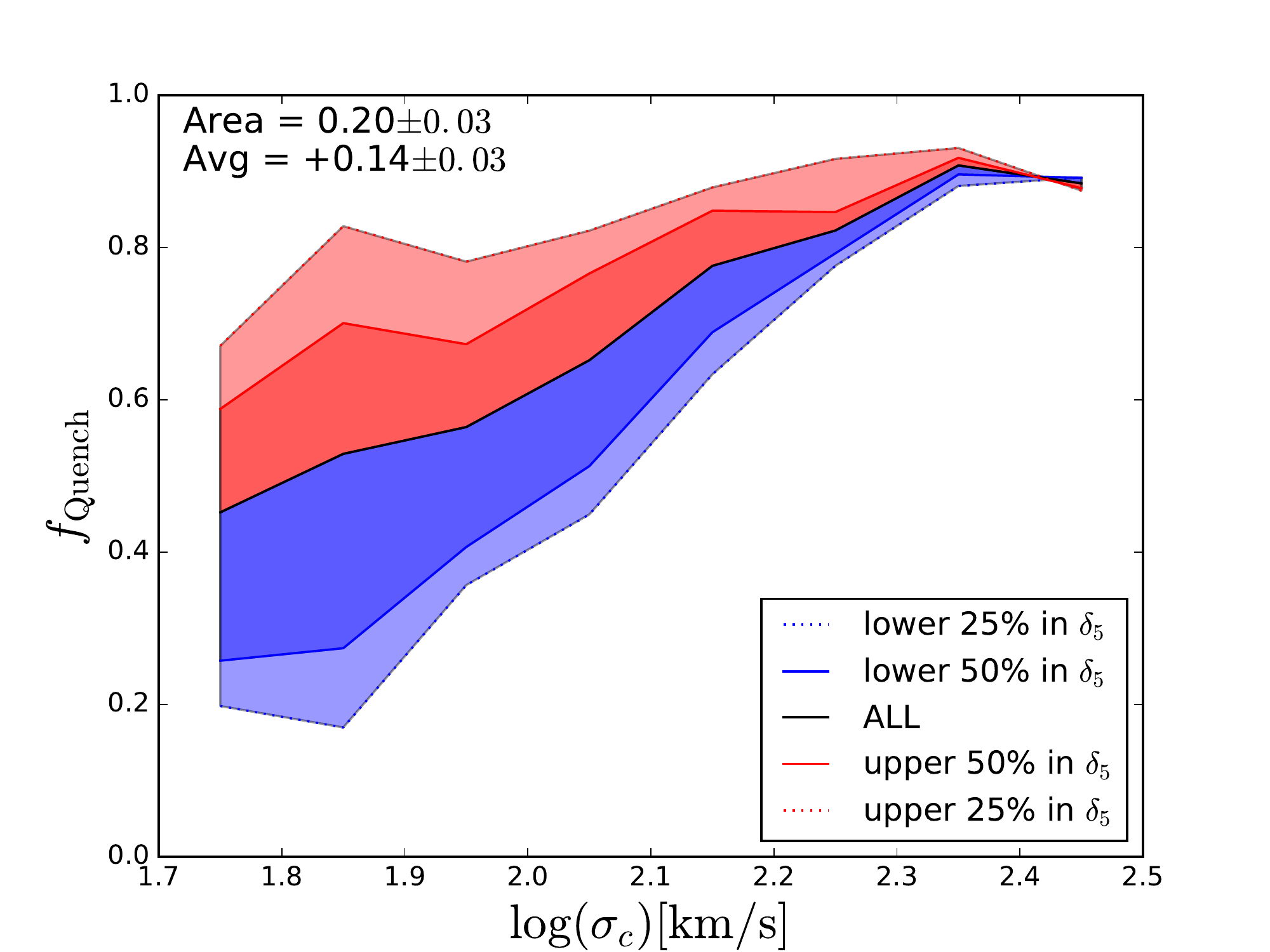}
\includegraphics[width=0.49\textwidth]{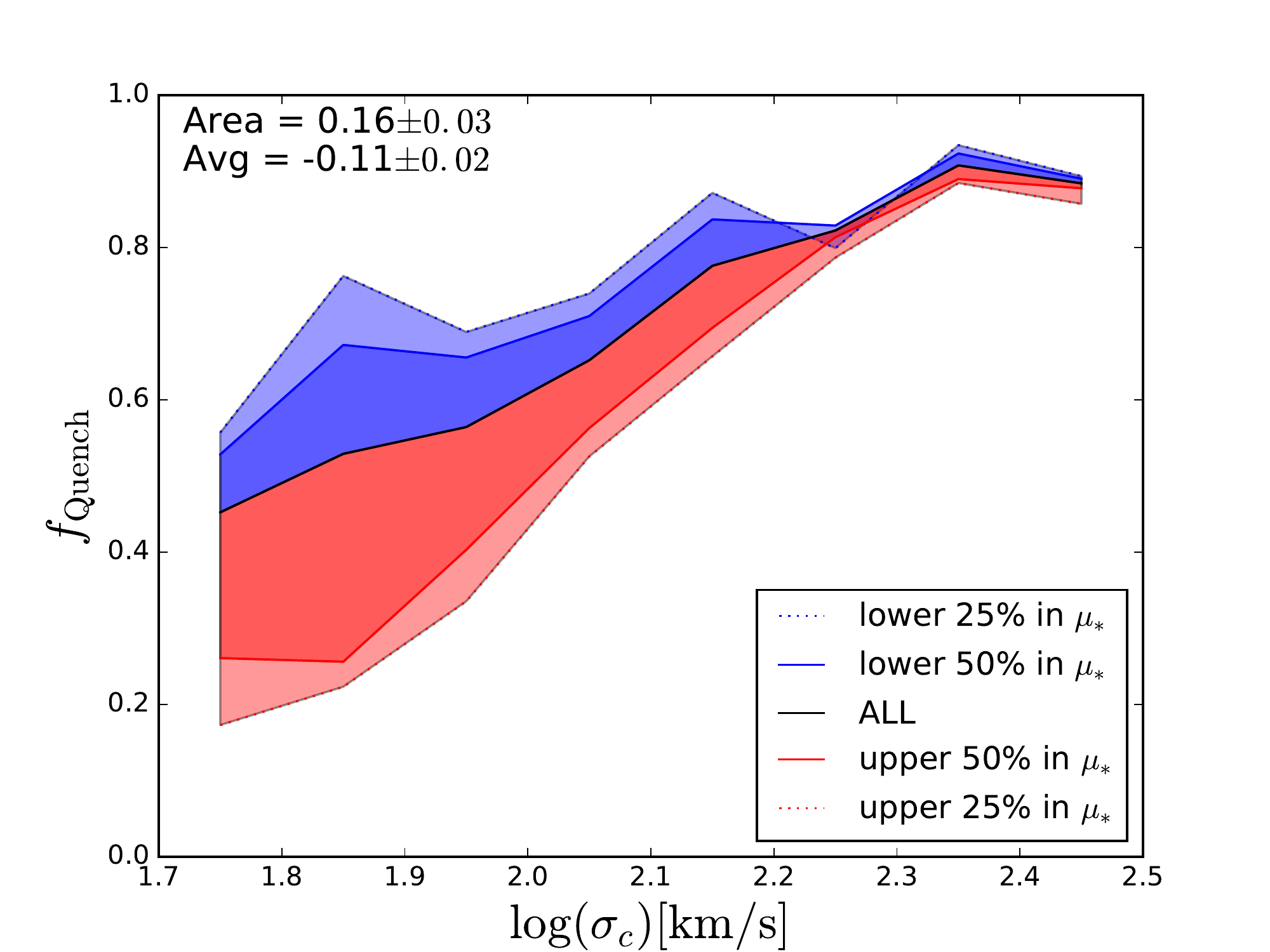}
\includegraphics[width=0.49\textwidth]{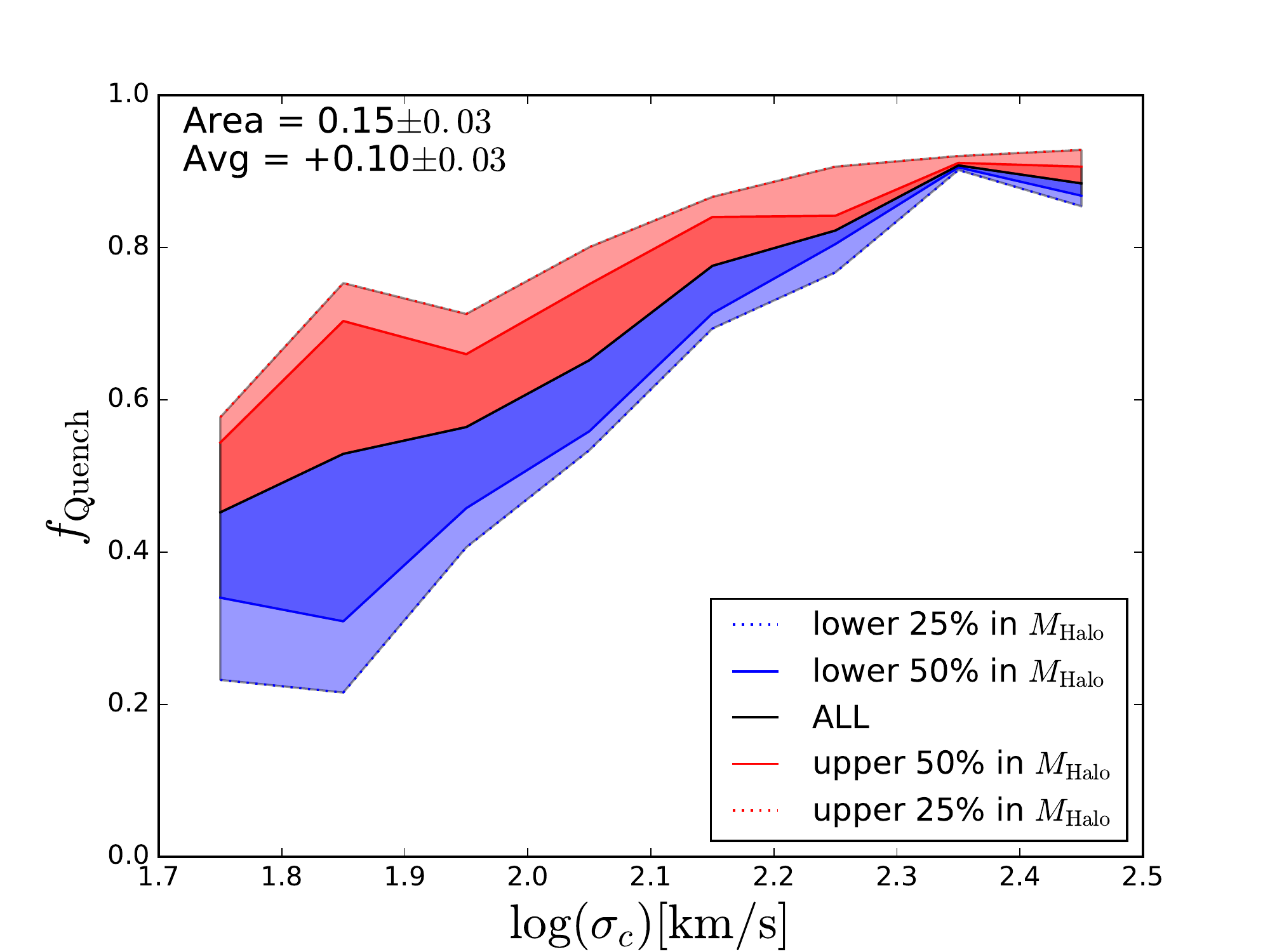}
\includegraphics[width=0.49\textwidth]{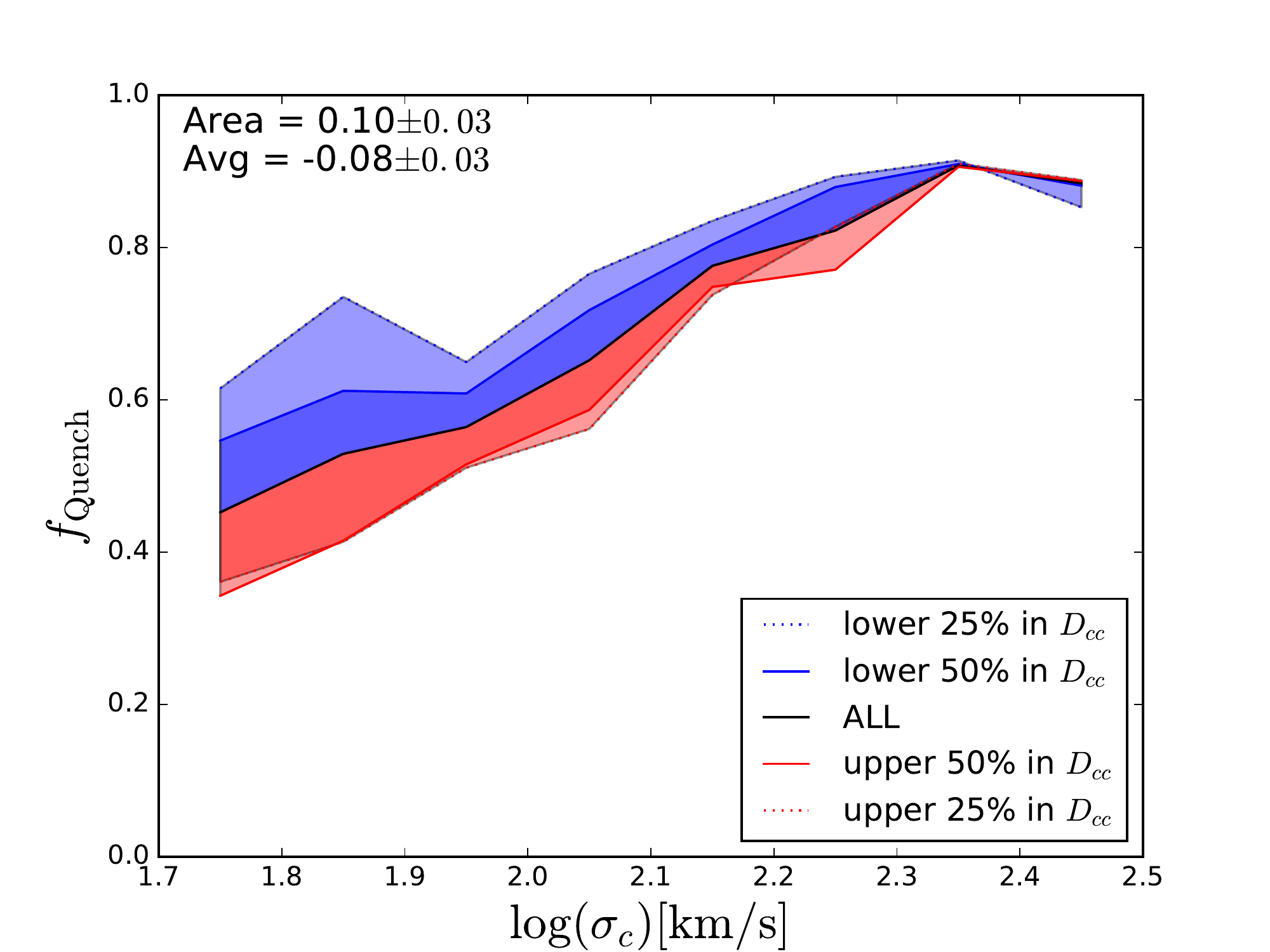}
\caption{Area statistics plots for satellite galaxies. Each panel shows the $f_{\rm Quench} - \sigma_{c}$ relation for satellites, split by percentile ranges in each of the following environmental parameters: local density ($\delta_{5}$), mass ratio ($\mu_{*}$), halo mass ($M_{\rm halo}$) and group-centric distance ($D_{cc}$). The area contained within the quenched fraction evaluated at upper and lower 50\% of the environmental variable, and the average difference in quenched fraction is shown on each plot. The largest impact on satellite quenching at fixed central velocity dispersion is seen for local density, with halo mass and mass ratio performing jointly second best, and group centric distance yielding the smallest perturbation on the $f_{\rm Quench} - \sigma_{c}$ relation.}
\end{figure*}

Many prior studies have found evidence for the quenched fraction of satellites exceeding that of centrals at a fixed stellar mass, particularly at lower stellar masses (e.g., Balogh et al. 2004; Cortese et al. 2006; Moran et al. 2007; van den Bosch et al. 2007, 2008; Tasca et al. 2009; Peng et al. 2012; Hirschmann et al. 2013; Wetzel et al. 2013; Bluck et al. 2014). Taken together, these studies additionally find that the quenching of satellites depends on environmental parameters, such as local (over-)density, halo mass, location within the halo, galaxy clustering, and combinations of these parameters. This is true both intrinsically and at a fixed stellar mass. However, in this work we have found that stellar mass is not the most fundamental parameter which governs the quenching of central galaxies (consistent with several other studies, e.g., Wuyts et al. 2011; Wake et al. 2012; Cheung et al. 2012; Fang et al. 2013; Bluck et al. 2014; Woo et al. 2015; Teimoorinia et al. 2016). This is important because the aim behind binning at a fixed stellar mass is to effectively control for the intrinsic drivers to quenching (the sole drivers for central galaxies) before investigating what additional (most probably environmental) processes quench satellite galaxies. 

Although perhaps unlikely, it is possible that satellite galaxies are quenched in the same way as centrals, but since the key parameter for driving central galaxy quenching is central velocity dispersion not stellar mass, this has not been witnessed in the literature. Some credence to this idea comes from the morphology - density relation (e.g., Dressler et al. 1980; Bamford et al. 2009; Tasca et al. 2009; Capellari et al. 2011), whereby galaxies are more bulge dominated (and hence have higher central velocity dispersions) for their stellar masses in denser regions of space. The only way to determine to what extent this explanation can account for satellite galaxy quenching is to measure the dependence of satellite quenching on environment at fixed central velocity dispersion.

\subsection{Fixed Binning Approach}

\begin{figure*}
\includegraphics[width=0.49\textwidth]{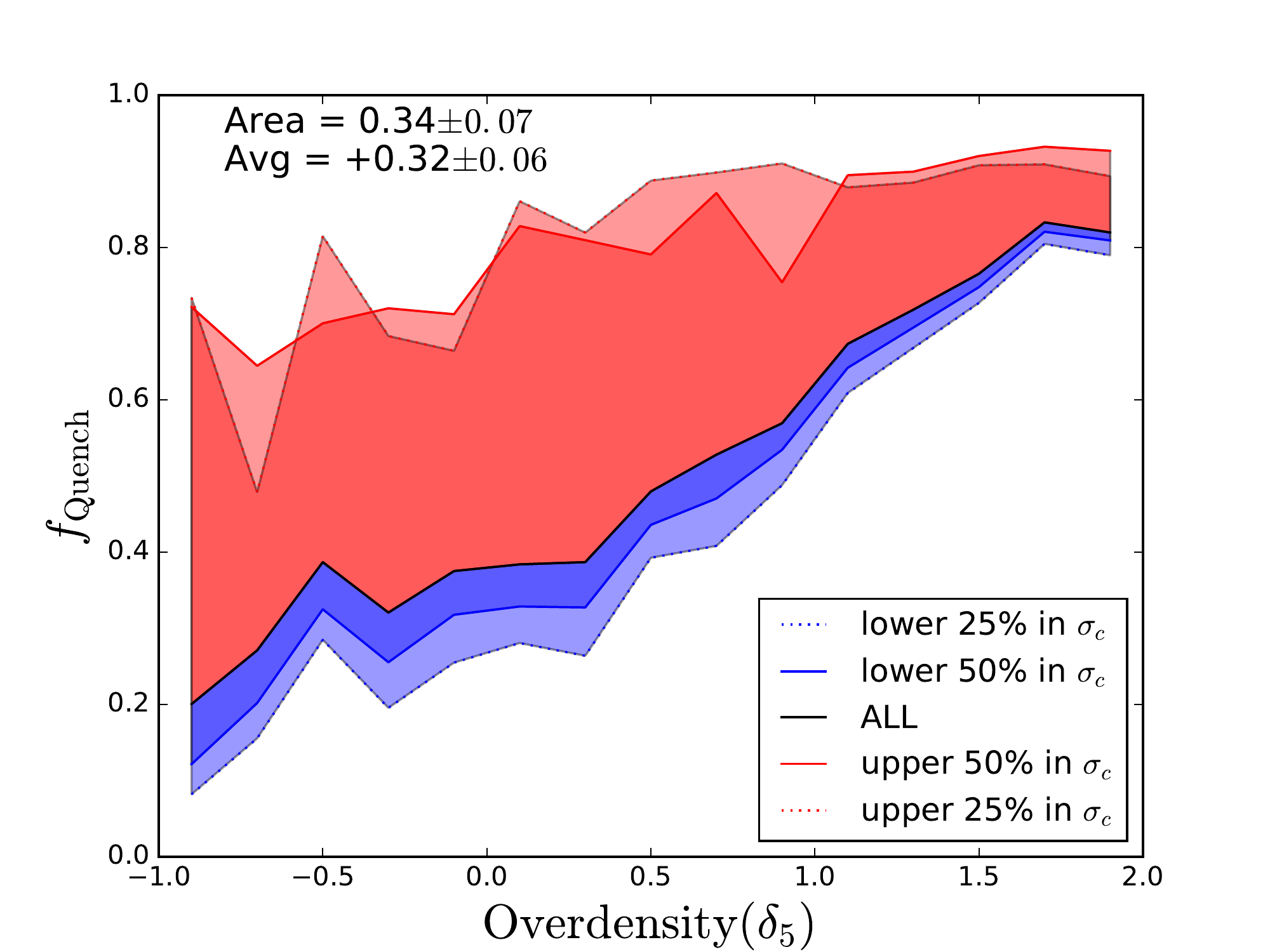}
\includegraphics[width=0.49\textwidth]{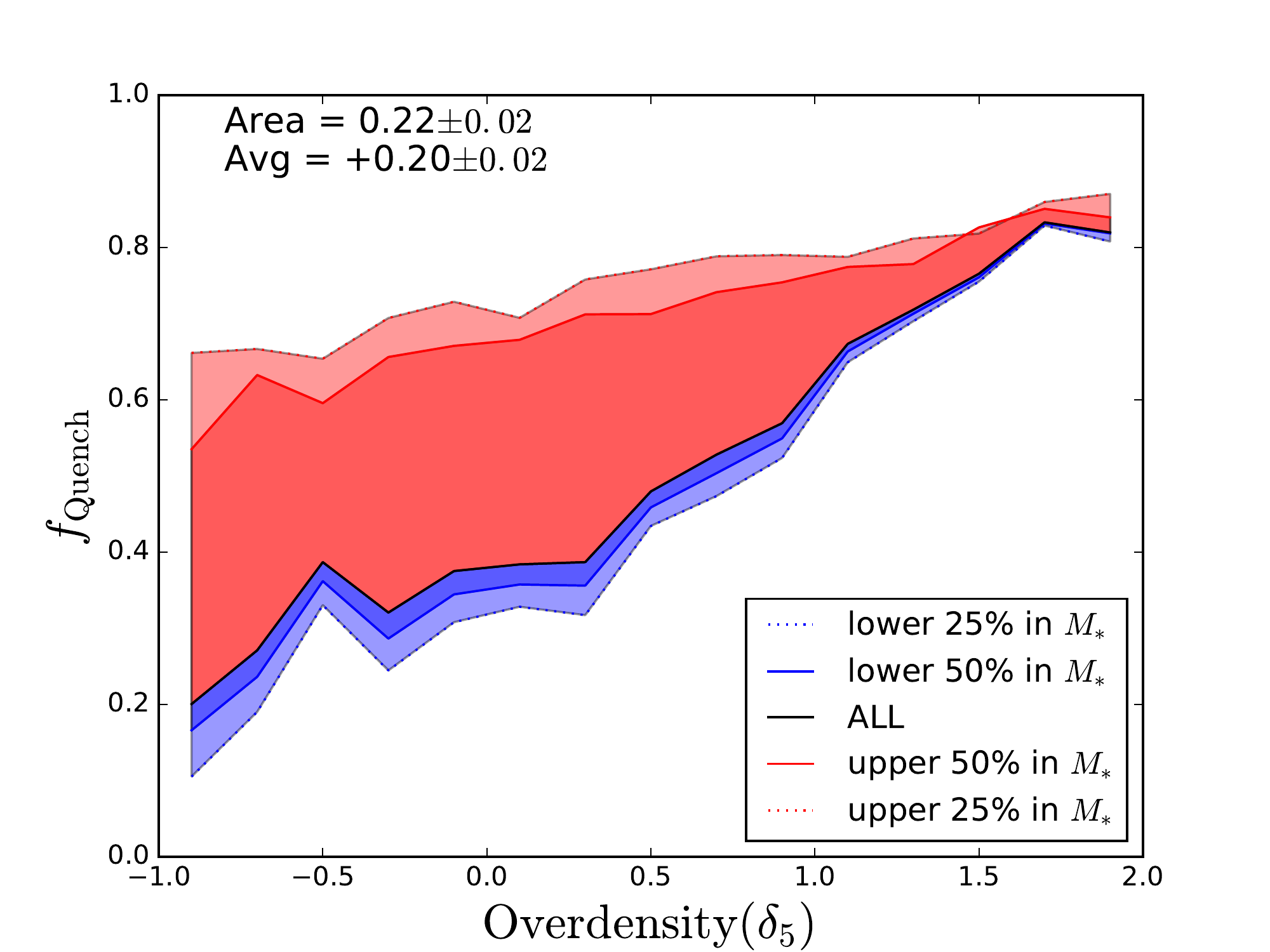}
\caption{The relationship between quenched fraction and over-density ($\delta_{5}$), divided by percentile ranges in central velocity dispersion (left) and stellar mass (right). The area and average difference statistics are shown on each plot. Central velocity dispersion affects the quenching of satellites significantly more than stellar mass, at fixed environment (here probed by local density). This indicates that $\sigma_{c}$ is more important to control for when assessing the role of environment in satellite quenching than $M_{*}$.}
\end{figure*}

In Fig. 12 we show the quenched fraction - central velocity dispersion relationship for satellite galaxies, split by over-density to the 5th nearest neighbour ($\delta_{5}$), mass ratio ($\mu_{*} = M_{*,{\rm sat}}/ M_{\rm halo}$), halo mass ($M_{\rm halo}$), and projected distance from the centre of the group or cluster ($D_{cc} = R/R_{\rm vir}$, where $R_{vir}$ is the virial radius of the group). It is evident that even at a fixed central velocity dispersion the quenching of satellites depends significantly on these environmental parameters. The impact of environment on the quenching of satellites is much more pronounced at lower central velocity dispersions and disappears entirely at the highest central velocity dispersions probed here. This suggests that satellites and centrals are quenched in the same manner at high masses and central velocity dispersions (see also Fig. 2), but satellites have additional routes to quenching at low masses and low central velocity dispersions, which are strongly correlated with environment.

For halo mass and local density, the plots for satellites in Fig. 12 can be compared to the equivalent plots for centrals in Figs. 3 and 5. Central galaxies exhibit very little variation in their quenched fraction at a fixed central velocity dispersion as a function of environment, whereas satellites do experience a significant boost in their quenched fraction from increasing the mass or density of their environments. These results are similar to what has been found previously for fixed stellar mass, and hence confirms that environmental effects are important in the quenching of satellites even at a fixed central velocity dispersion, the parameter which is most tightly correlated with the quenching of centrals. In low density environments, low mass haloes, high mass ratios and large distances from the centre of the group, the quenched fraction for satellites approaches that of centrals. This provides further evidence that what differentiates centrals and satellites in their quenching is primarily environment.

\subsection{Area Statistics Approach}

In this section we follow the area statistics approach (see Section 5.2), and apply this to satellites instead of centrals. In Fig. 13 we show the quenched fraction - central velocity dispersion relationship for satellites, splitting in percentile ranges of $\delta_{5}$, $\mu_{*}$, $M_{\rm halo}$ and $D_{cc}$. We find that there is a significant variation in the quenched fraction at fixed central velocity dispersion from each of these environmental parameters for satellites, unlike for halo mass and local density in centrals (compare to Figs. 4 \& 5). The areas contained within the upper  and lower 50th percentiles in quenched fraction, and the average difference in quenched fraction are shown on each plot in Fig. 13. 

Local density affects the quenched fraction of satellites more at a fixed central velocity dispersion than any other environmental parameter considered in this work. Mass ratio and halo mass perform similarly well to each other, ranking in the middle of the group in terms of their impact on satellite galaxy quenching. Distance from the centre of the group is the least constraining of the environmental parameters. These rankings are summarised, along with the area and average difference results, in Table 2. These measurements provide new constraints to simulations and models of satellite galaxy quenching through environmental processes, which will be considered in detail in an upcoming publication from this series (Bluck et al. in prep.). Whatever quenches satellites must correlate stronger with local density than mass ratio, halo mass or distance from the group centre. One possibility is that local density is simply a good {\it average} property, sensitive to both mass of, and location in, the halo. However, it is also feasible that this result informs us more directly about the quenching process, e.g. quenching by galaxy - galaxy tidal harassment and stripping of the satellite's hot gas halo may be a more significant route to quenching than, e.g., ram pressure stripping from the hot gas halo of the group (which correlates primarily with halo mass and/ or location within the halo). 

It is interesting to note that the directionality of the trends are clearly seen in Fig. 13. Increasing local density and halo mass both lead to increasing the quenched fraction at fixed central velocity dispersion. This is seen as red shaded regions lying above blue shaded regions, and values of $\Delta f_{Q} > 0$ (left panels). Increasing the group centric distance and mass ratio both lead to a decrease in the quenched fraction of satellites at fixed central velocity dispersion. This is seen as blue shaded regions lying above red shaded regions, and values of $\Delta f_{Q} < 0$ (right panels). These results are as expected in the paradigm where environment quenches satellites. 

Finally, in Fig. 14, we show the effect of varying central velocity dispersion (left panel) and stellar mass (right panel) at fixed local density, the environmental parameter which is found to affect satellite galaxy quenching the most. Varying central velocity dispersion at fixed local density affects the quenched fraction of satellites significantly more than varying stellar mass. For brevity we only show the case with local density, but this trend is true for all of the environmental parameters considered here. Thus, central velocity dispersion is a more significant intrinsic parameter for the quenching of satellites, than stellar mass. This confirms our prior assumption that the parameter which matters most for central galaxy quenching should be the {\it intrinsic} parameter which is most important for satellite quenching. Therefore, in order not to overestimate the impact of environment on satellite quenching, future studies must fix the central velocity dispersion wherever possible, before making an environmental comparison, or a comparison with central galaxies in terms of quenching.

\begin{table}
\caption{Summary of area and mean difference statistics for satellites (taken from Fig. 13)}
 \label{tab2}
 \begin{center}
 \begin{tabular}{@{}cccc}
  \hline
\hline
Rank & $[\alpha, \beta]$ & Area  & $\langle\Delta f_{Q}\rangle$ \\ 
\hline
1 & $[\sigma_{c}, \delta_{5}]$ & 0.20 $\pm$ 0.03 & +0.14 $\pm$ 0.03 \\ 
\hline
2= & $[\sigma_{c}, \mu_{*}]$ & 0.16 $\pm$ 0.03 & -0.11 $\pm$ 0.02 \\ 
\hline
2= & $[\sigma_{c}, M_{\rm halo}]$ & 0.15 $\pm$ 0.03 & +0.10 $\pm$ 0.02 \\ 
\hline
4 & $[\sigma_{c}, D_{cc}]$ & 0.10 $\pm$ 0.03 & -0.8 $\pm$ 0.03 \\ 
\hline
\end{tabular}
\end{center}
Note: $\alpha$ and $\beta$ are defined as in eqs. 8 \& 9, they correspond to the x-axis variable and the percentile (colour) variable in Fig. 13, respectively.
\end{table}

\section{Conclusions}

In this paper, we explore the dependence of central and satellite galaxy quenching on a variety of physical galaxy and environmental properties. We start with a sample of $\sim$ half a million SDSS galaxies (80\% centrals and 20\% satellites) at z $<$ 0.2. We quantify the quenched fraction dependence on galaxy properties at fixed central velocity dispersion, which has previously been found to be the tightest correlator to quenching for central galaxies (Teimoorinia, Bluck \& Ellison 2016). 

At a fixed central velocity dispersion, we find that satellite galaxies are more frequently quenched than central galaxies, with inner satellites (within 0.1 virial radii of their centrals) being more frequently quenched than the general satellite population (see Fig. 2). This effect is more pronounced at lower central velocity dispersions and disappears entirely by $\sigma_{c} >$ 250 km/s.  Furthermore, the $f_{\rm Quench} - \sigma_{c}$ relationship is steep for centrals, varying from $\sim$ 0.05 -- 0.95 across the range we probe, and progressively less steep for satellites and inner satellites. Qualitatively, this result is consistent with central galaxies being quenched by AGN feedback, given the tight observed relationships between central velocity dispersion and black hole mass, and the dependence of AGN-driven quenching on black hole mass in simulations and models. However, the quenching of satellites and inner satellites cannot be driven by AGN feedback at low central velocity dispersion, suggesting that other (most probably environmental) processes must be important in their quenching when they depart from centrals at low masses.

For central galaxies, we confirm the prior result that central velocity dispersion is more predictive of quenching than any of the following properties: stellar mass, halo mass, bulge mass, disk mass, $B/T$ structure and local density ($\delta_{5}$), e.g. Wake et al. (2012); Teimoorinia et al. (2016). Moreover, we find that varying stellar, halo or bulge mass or local density (by even three orders of magnitude) has little if any effect on the quenched fraction at fixed central velocity dispersion for centrals. This indicates that these parameters cannot be causally connected to central galaxy quenching, which provides powerful new constraints on the mechanism(s) which may be responsible for causing quenching in these galaxies. 

In Section 5.2 we develop a new technique for ascertaining and quantifying the impact on quenching of varying one parameter at a fixed other parameter. In particular, we define two statistics, the area contained within the upper and lower 50\% range in the quenched fraction from varying a secondary parameter at fixed first parameter, and the average difference between quenched fraction at upper and lower 50\% range. The former indicates the tightness of the quenched fraction dependence on the primary variable, and the latter additionally indicates the directionality of the trend. For centrals, we find a strong positive effect on the quenched fraction from varying central velocity dispersion at fixed values of all of the other variables considered in this work. Most of the other variables have little to no effect on quenching at fixed central velocity dispersion. However, $B/T$ and disk mass do have a statistically significant effect, although smaller in magnitude to central velocity dispersion. This is most probably due to these parameters correlating with gas mass and hence being related to the amount of work which needs to be done to quench the galaxy.

Given the lack of impact on the quenched fraction of halo mass and stellar mass, it is highly improbable that either halo mass quenching, stellar or supernova feedback can be responsible for central galaxy quenching. However, the strong observed correlations between central velocity dispersion and supermassive black hole mass do present an interesting opportunity for explanation of our results via AGN feedback. In Section 6, we compare the quenching of centrals in Illustris and L-Galaxies to our observational (SDSS) results. In both models, the quenched fraction - black hole mass relationship is significantly tighter than the stellar mass or halo mass relation, qualitatively in agreement with observations. However, we find a quenching threshold (defined as the black hole mass at which 50\% of galaxies are quenched) of $10^{6} M_{\odot}$ in the Munich model and $2 \times 10^{8} M_{\odot}$ in Illustris, compared to $2 \times 10^{7} M_{\odot}$ in the SDSS (assuming the scaling law of Saglia et al. 2016). This suggests that quenching via AGN feedback may be too efficient (as a function of black hole mass growth) in L-Galaxies and too inefficient in Illustris, compared to local galaxies. 

We also consider if evolutionary systematics (e.g., via size evolution) can give rise to the observed tightness of the $f_{\rm Quench} - \sigma_{c}$ relationship, without any causal connection. We perform a test using the green valley fraction (of {\it quenching}) galaxies. We find that central velocity dispersion remains a significantly tighter correlator to the quenched fraction than stellar or halo mass, even for galaxies currently undergoing transformation in their star forming state. This implies that evolutionary systematics, which can affect the quenched fraction, are not ultimately responsible for the dependence of central galaxy quenching on central velocity dispersion since this exists already in galaxies which are contemporaneously quenching. 

For satellites, we find that the environmental metrics we consider (i.e., local density, halo mass, satellite-halo mass ratio, and group centric distance) all have a significant effect on the quenched fraction at fixed central velocity dispersion (see Section 7), unlike for centrals which experience very little dependence on halo mass or local density at fixed central velocity dispersion. Using the area statistics approach we developed for centrals, we find that local density engenders the most significant perturbation on the $f_{\rm Quench} - \sigma_{c}$ relationship, followed jointly by halo mass and mass ratio, with group centric distance leading to the smallest impact on quenching. One possibility is that local density simply represents a good average quantity, sensitive to the mass of the group or cluster and the location of the satellite within it. However, it is possible that this ranking gives more direct information on what mechanisms are likely responsible for satellite galaxy quenching. For example, if galaxy - galaxy interactions dominate over galaxy - halo interactions, this would naturally lead to similar results to what we observe.

In summary, we find the tightest correlation between quenched fraction and central velocity dispersion for central galaxies, tighter than for any other parameter considered in this work. Moreover, the $f_{\rm Quench} - \sigma_{c}$ relationship is largely unaffected by varying other galaxy parameters for centrals, whereas the quenched fraction dependence on each of the other galaxy parameters is heavily affected by varying central velocity dispersion. The invariance of the dependence of central galaxy quenching on central velocity dispersion with other galaxy parameters suggests that this may be a causal relationship. If so, it is most likely explained by AGN feedback given the observed $M_{BH} - \sigma_{c}$ relation. Furthermore, our observational results are qualitatively in agreement with the predictions from a hydrodynamical simulation and semi-analytic model, both of which quench galaxies via AGN feedback in the radio mode. However, the details of our comparison do motivate further work in the implementation of quenching in the model and simulation, since the former is too efficient in its quenching and the latter is not efficient enough. Finally, we find that central velocity dispersion is the most significant {\it intrinsic} parameter for satellite quenching; although environment has a much larger impact on satellites than centrals. Thus, additional quenching mechanisms are clearly needed for satellite galaxies over centrals, which must be strongly related to environment, particularly local galaxy density which performs best out of the environmental parameters we consider.

\section*{Acknowledgments}

We thank Steven Bamford, Richard Bower, Marcella Carollo, Chris Conselice, Rob Crain, Will Hartley, Lars Hernquist, Simon Lilly, Avi Loeb, Laura Sales and Peter Thomas for helpful discussions on this work. We thank Kevin Schawinski for giving us the green valley test idea. We also thank the anonymous referee for many helpful and insightful comments on this work, which have contributed greatly to the final version. We gratefully acknowledge funding from the Swiss National Foundation for Sciences and the National Science and Engineering Research Council (NSERC) of Canada, particularly for Discovery Grants awarded to SLE and DRP. JM is supported by NSF grant AST-1516364. ES gratefully acknowledges funding by the Emmy Noether programme from the Deutsche Forschungsgemeinschaft (DFG) as well as funding through a Canadian Institute for Advanced Research (CIFAR) global scholarship and AIP Schwarzschild fellowship.

Funding for the SDSS and SDSS-II has been provided by
the Alfred P. Sloan Foundation, the Participating Institutions, the National Science Foundation, the U.S. Department of Energy, the National Aeronautics and Space Administration, the
Japanese Monbukagakusho, the Max Planck Society, and the
Higher Education Funding Council for England. The SDSS
Web Site is http://www.sdss.org/

The SDSS is managed by the Astrophysical Research Consortium for the Participating Institutions. The Participating
Institutions are the American Museum of Natural History, Astrophysical Institute Potsdam, University of Basel, University of Cambridge, Case Western Reserve University, University of Chicago, Drexel University, Fermilab, the Institute for Advanced Study, the Japan Participation Group, Johns
Hopkins University, the Joint Institute for Nuclear Astrophysics, the Kavli Institute for Particle Astrophysics and
Cosmology, the Korean Scientist Group, the Chinese Academy
of Sciences (LAMOST), Los Alamos National Laboratory,
the Max-Planck-Institute for Astronomy (MPIA), the Max-Planck-Institute for Astrophysics (MPA), New Mexico State
University, Ohio State University, University of Pittsburgh, University of Portsmouth, Princeton University, the United States Naval Observatory, and the University of Washington.

\appendix

\section{Area Statistics Approach Example}

\begin{figure*}
\includegraphics[width=0.33\textwidth]{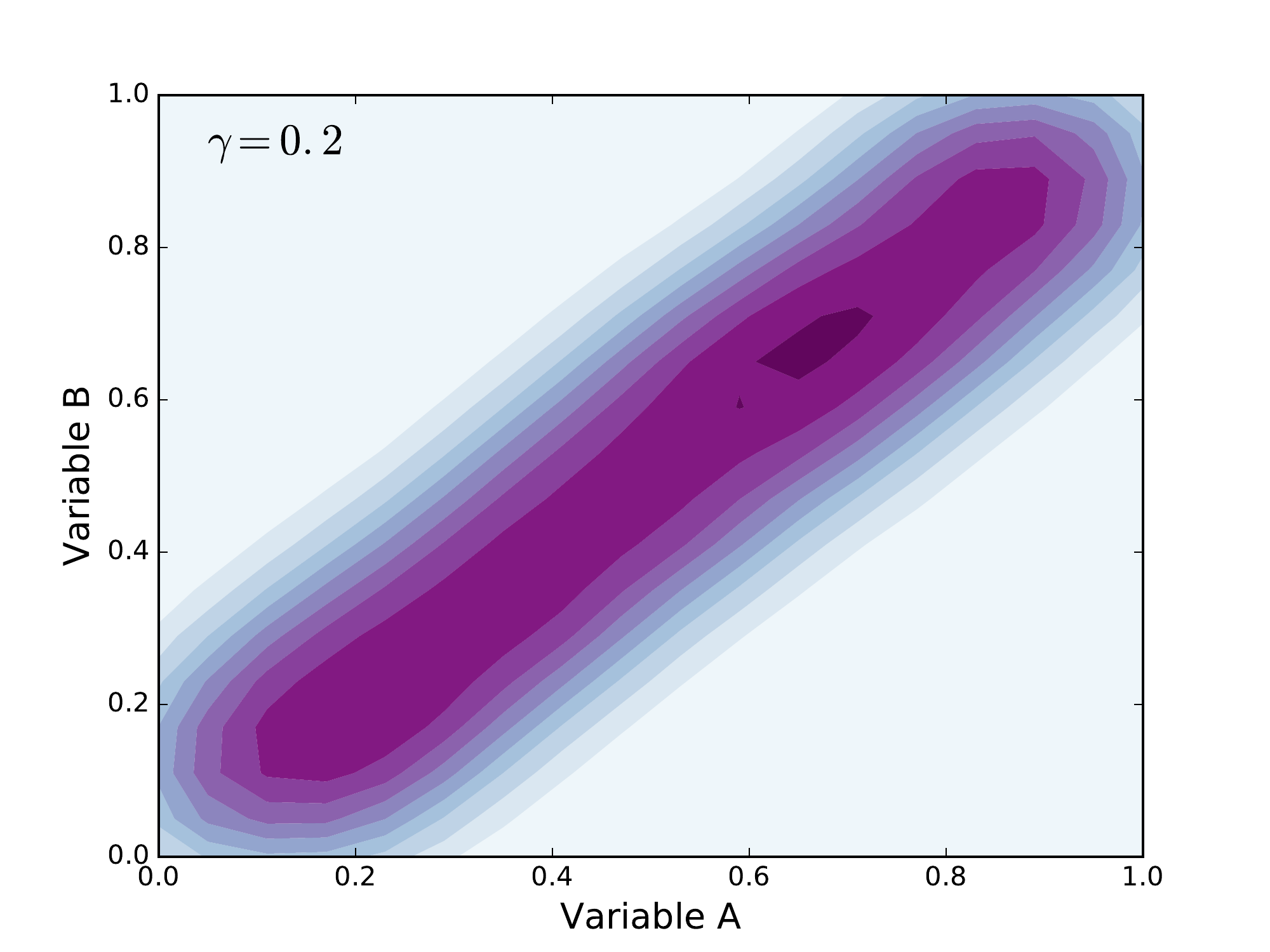}
\includegraphics[width=0.33\textwidth]{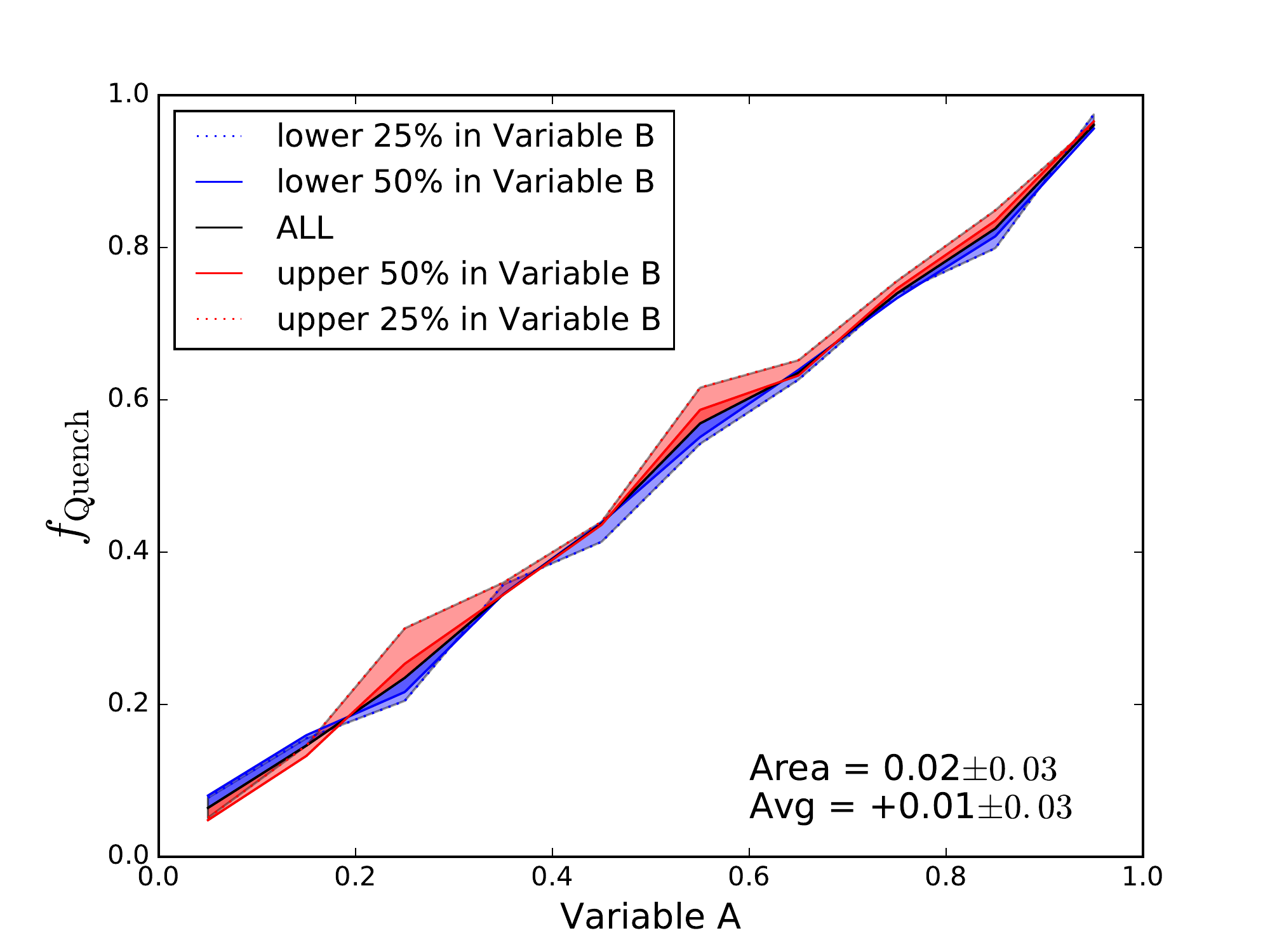}
\includegraphics[width=0.33\textwidth]{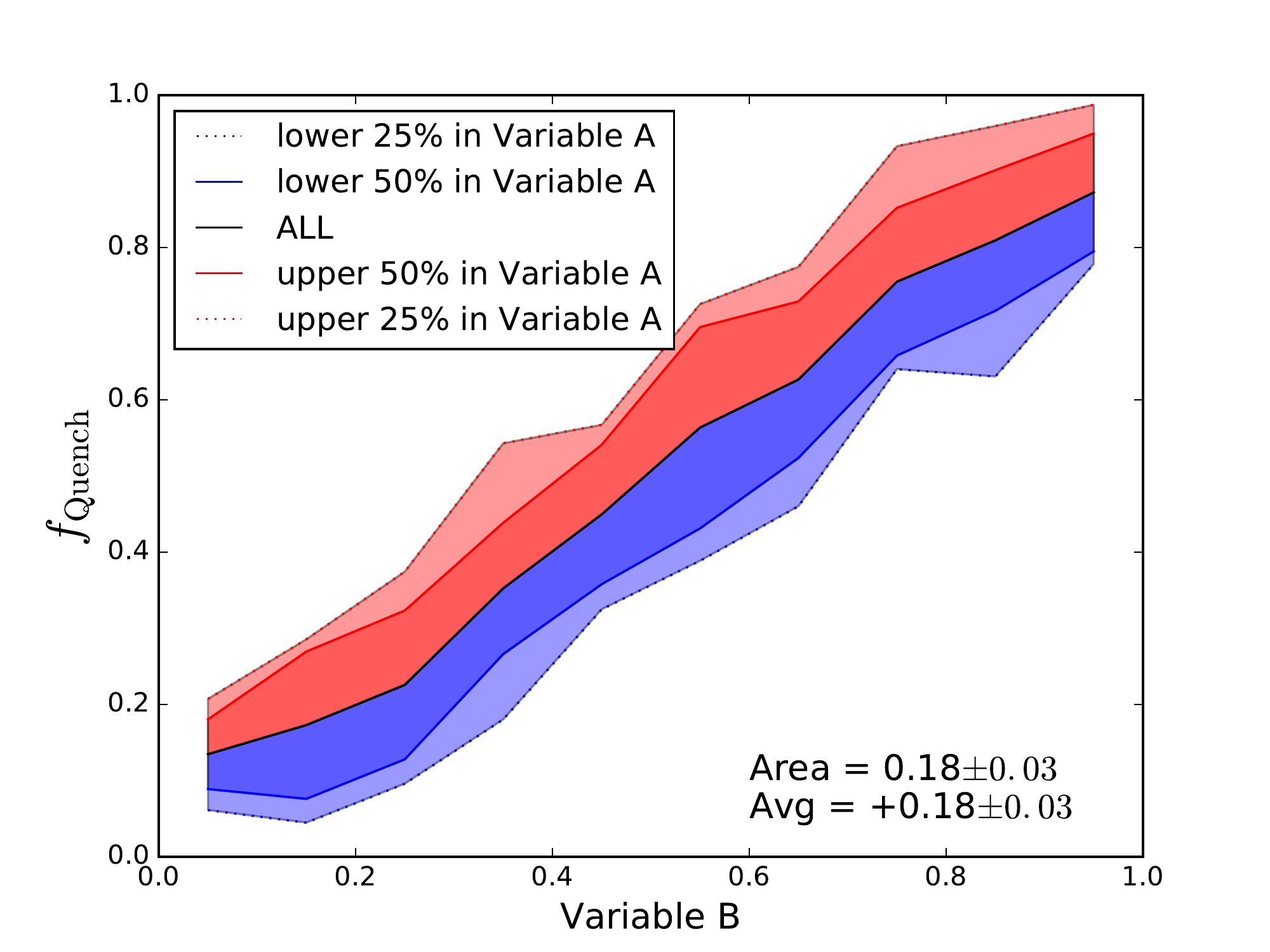}
\includegraphics[width=0.33\textwidth]{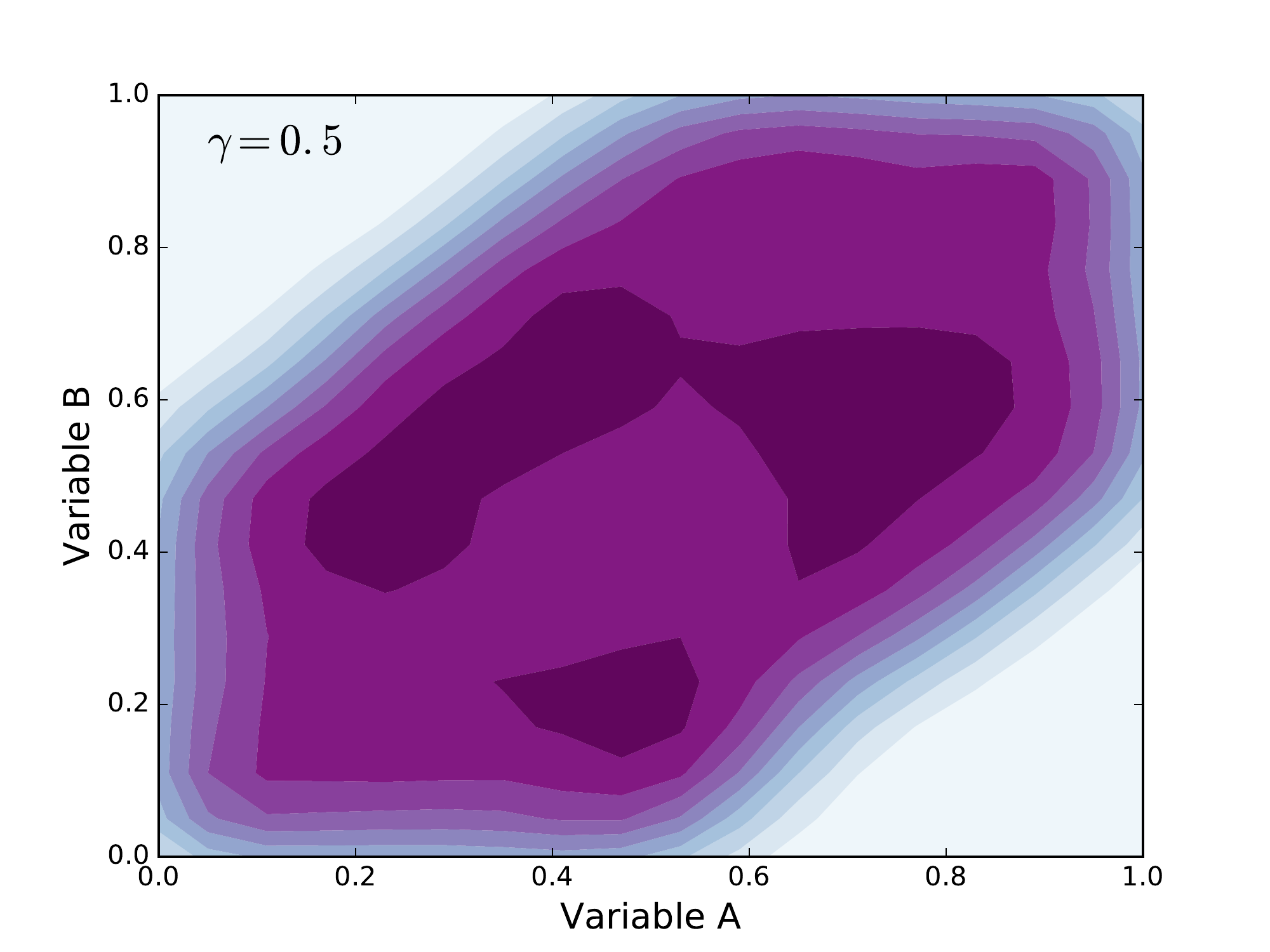}
\includegraphics[width=0.33\textwidth]{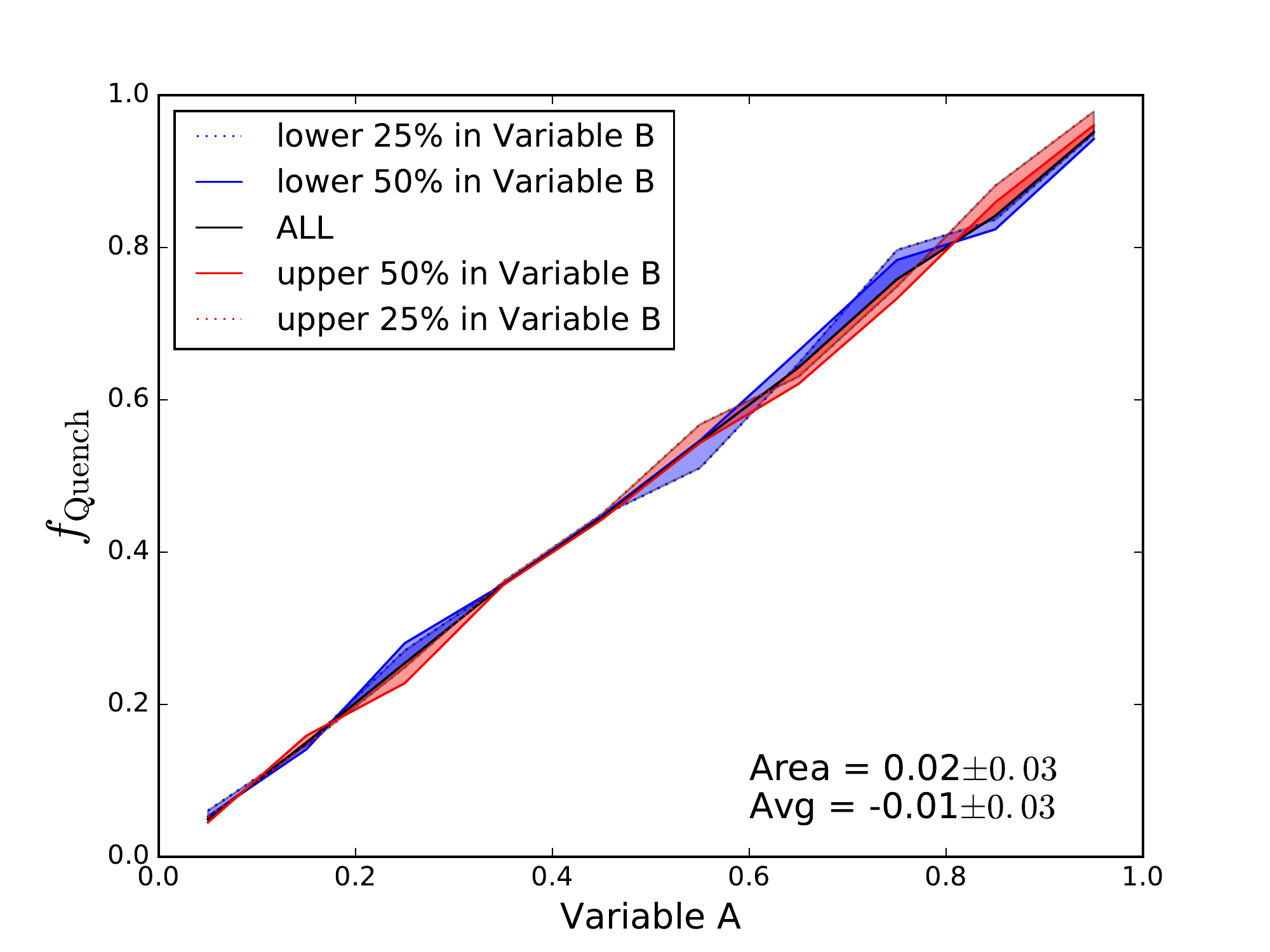}
\includegraphics[width=0.33\textwidth]{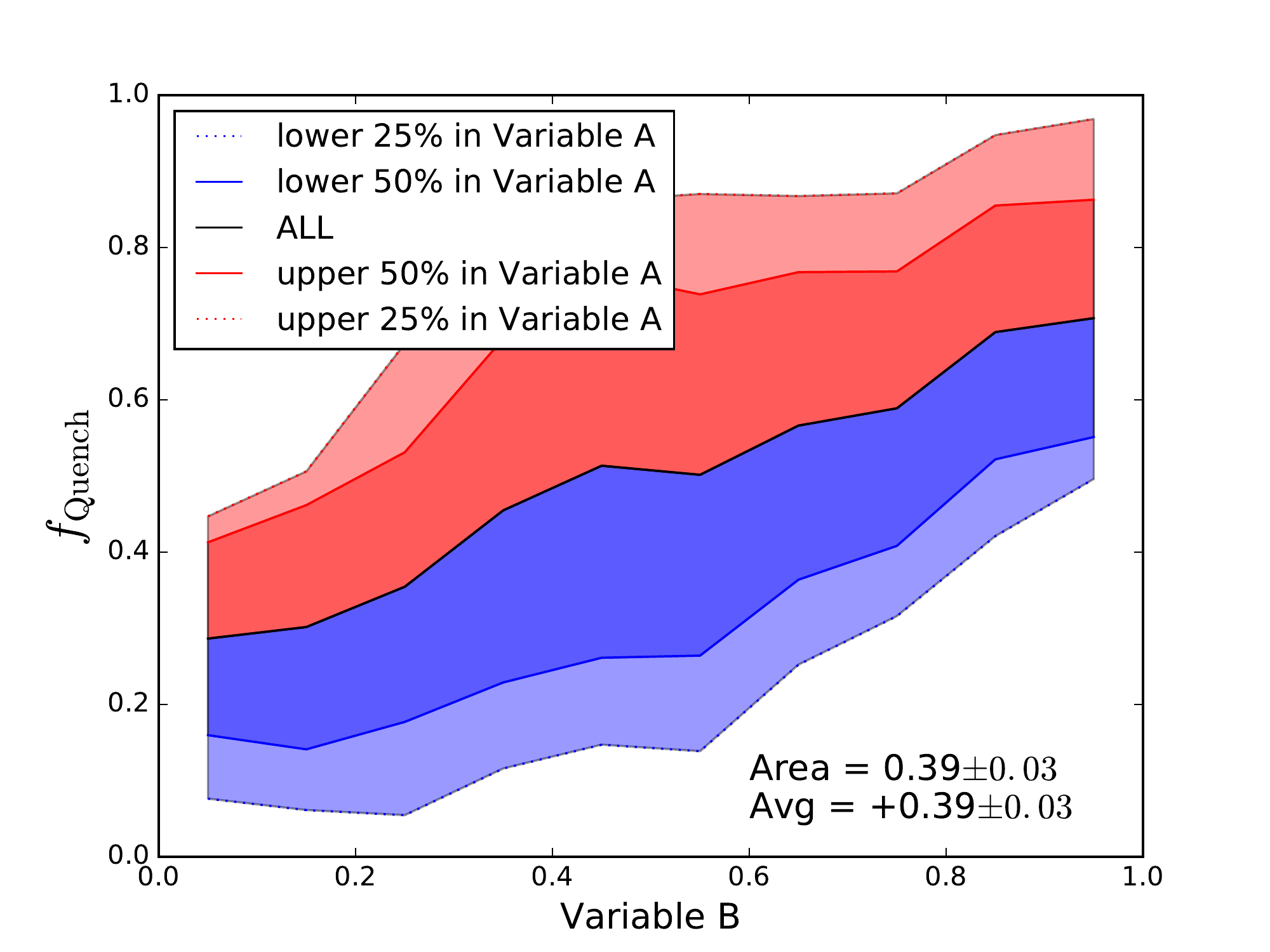}
\caption{Left panels show contour plots of the relationship between simulated variables $A$ and $B$ for correlation parameter, $\gamma$ = 0.2 (top) and 0.5 (bottom). The middle panels show the quenched fraction relationship with $A$, at fixed increasing percentile binnings of variable $B$. The right panels show the quenched fraction relationship with variable $B$, at fixed increasing percentile binnings of variable $A$. Both the area and the average difference between quenched fractions (indicated on the plots) are larger for variable $B$ acting as the primary variable than for variable $A$. This demonstrates that we can recover the input dependence of quenching on variable $A$ using our area statistics approach, for both highly correlated and weakly correlated cases (top and bottom, respectively).}
\end{figure*}

To demonstrate how the area statistics approach works in practice, we consider in this appendix a few simple cases to build intuition with the technique. First, we consider a case where the probability of a galaxy being quenched is proportional to some observable variable, $A$. Hence:

\begin{equation}
P_{Q} \propto A .
\end{equation}

\noindent Additionally, variable $A$ is correlated with another variable $B$, such that:

\begin{equation}
B = f(A) + \gamma \times \mathcal{R} \{ -1:1 \} 
\end{equation}

\noindent where $\mathcal{R} \{ -1:1 \}$ indicates a random number between -1 and 1, and $\gamma$ is a coefficient related to the tightness of the relationship between $A$ and $B$, where higher values of $\gamma$ lead to weaker correlations. In a very simple case where $A$ and $B$ are both set to have values between 0 and 1, $P_{Q}$ and $f(A)$ can be set directly equal to $A$. To illustrate how our area statistics approach works, we generate a sample of a million random values for parameter $A$, and determine whether each case is `quenched' or `star forming' by a Monte Carlo method, using the probability of being quenched given by $A$. We additionally construct a million values of $B$, from $A$, using different thresholds for the correlation parameter, $\gamma$.

In Fig. A1, left panels, we show a contour plot of the constructed correlations between $A$ and $B$, for $\gamma$ = 0.2 (top) and 0.5 (bottom). In the middle panels we show the quenched fraction relationship with $A$, and on the right panels we show the quenched fraction relationship with $B$, as solid black lines. When $A$ and $B$ are highly correlated (top panels) the slope of the quenched fraction relationship is comparable with each variable, so it is hard to tell a priori whether it is ultimately $A$ or $B$ which affects quenching, or some combination of both or neither. We then proceed with our area statistics approach (in \S 5.2.1), by measuring the quenched fraction at various percentile ranges of $B$ at fixed $A$ in the middle panels, and at various percentile ranges of $A$ at fixed $B$ in the right panels. We then measure the area and average $\Delta f_{Q}$, as in eqs.  8 \& 9.

We find that the area is lower in both cases for the quenched fraction as a function of $A$ (Fig. A1, middle panels) than as a function of $B$ (Fig. A1, right panels), which recovers our input result, that the probability of quenching is determined by $A$ not $B$. We also note that the magnitude of the difference in area is also dependent on the correlation parameter ($\gamma$) and hence how tight the correlation between $A$ and $B$ is. Tighter correlations lead to smaller areas in general because the extent to which variable $A$ can vary at fixed variable $B$ is limited. In this example the areas and average differences are identical, because in this simple setup quenching depends only on one variable, in a positive manner only. If, for instance, quenching in our example depended on $(1-A)$ (instead of $A$), we would find that Area $\sim$ $-<\Delta f_{Q}>$. If quenching depends on both $A$ and $B$ equally, we would find equivalent areas for the middle and right panels, in each case of $\gamma$. For brevity we do not show all these examples here, but we mention them to add some further intuition to our method.

\section{The Impact of the Choice of Scaling Law on the Model Comparisons}

\begin{figure*}
\includegraphics[width=0.49\textwidth]{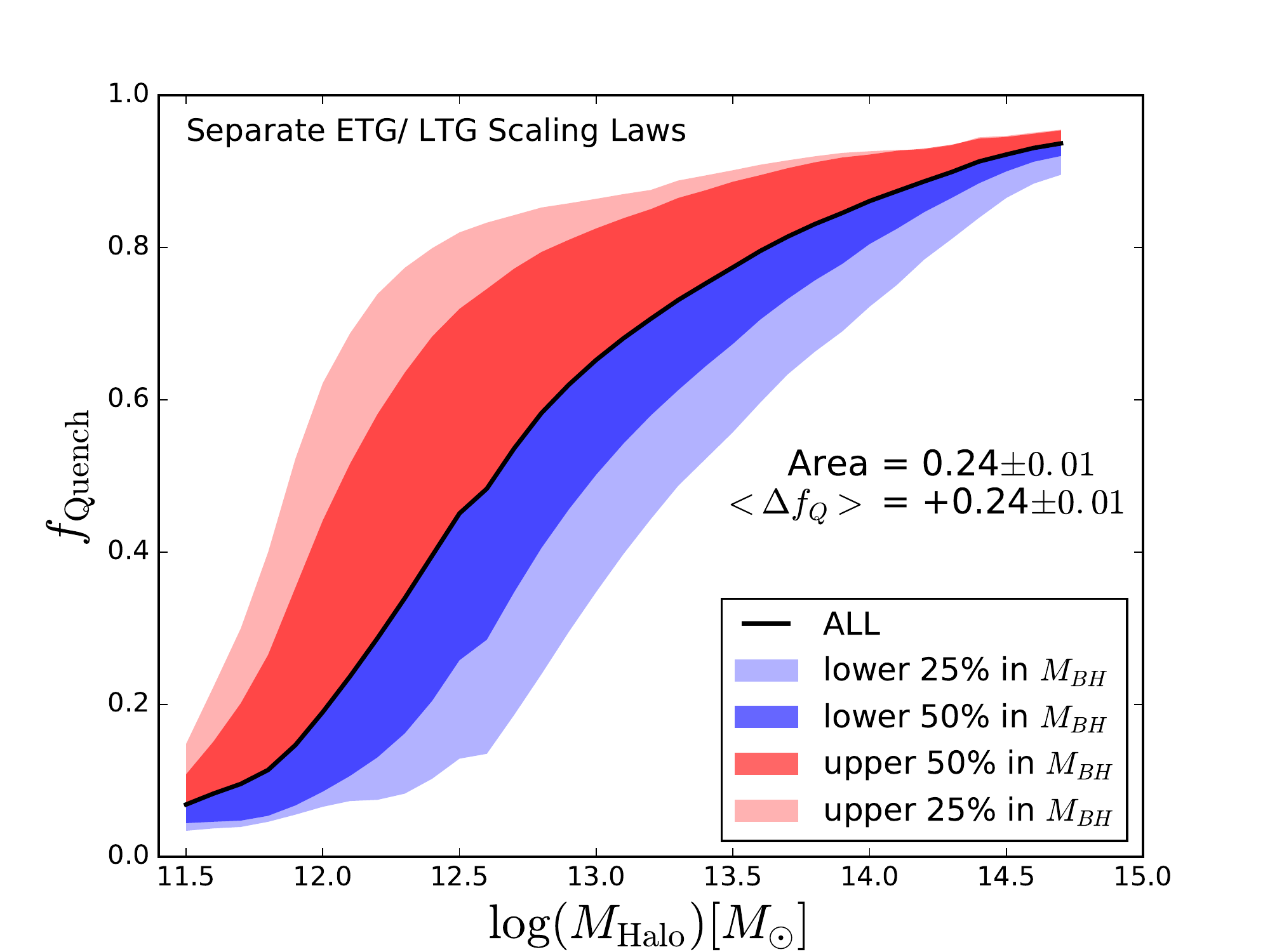}
\includegraphics[width=0.49\textwidth]{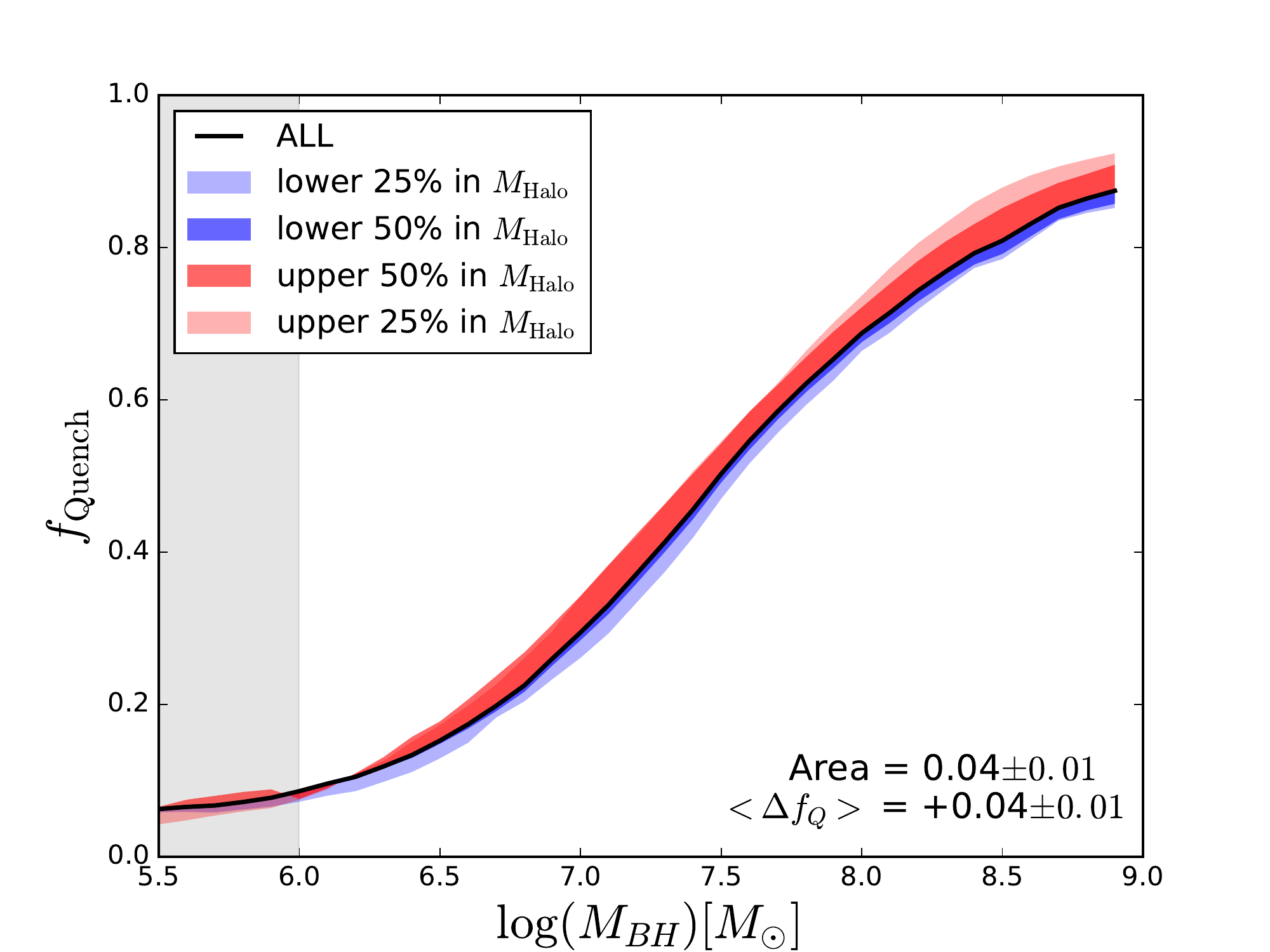}
\includegraphics[width=0.49\textwidth]{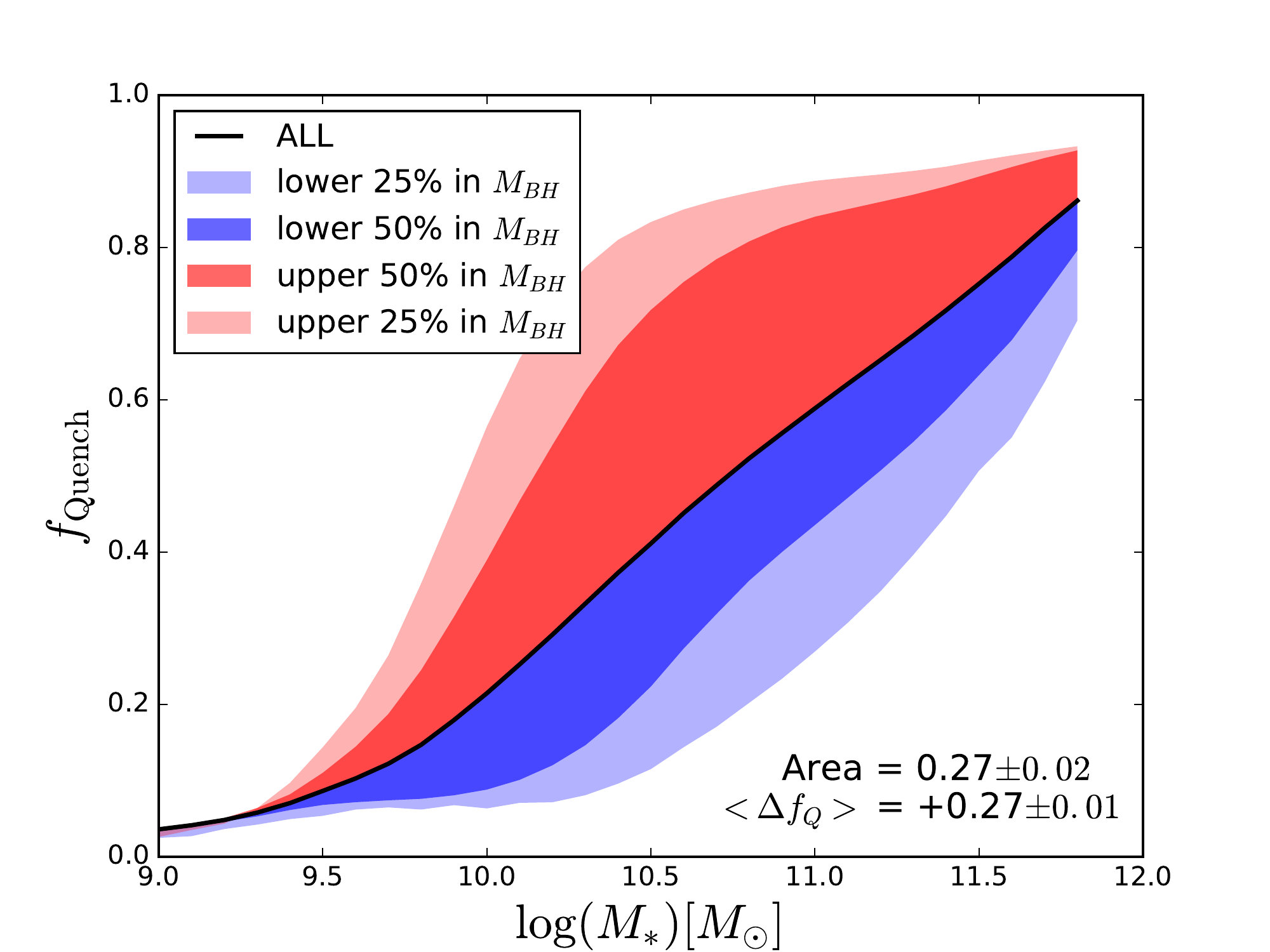}
\includegraphics[width=0.49\textwidth]{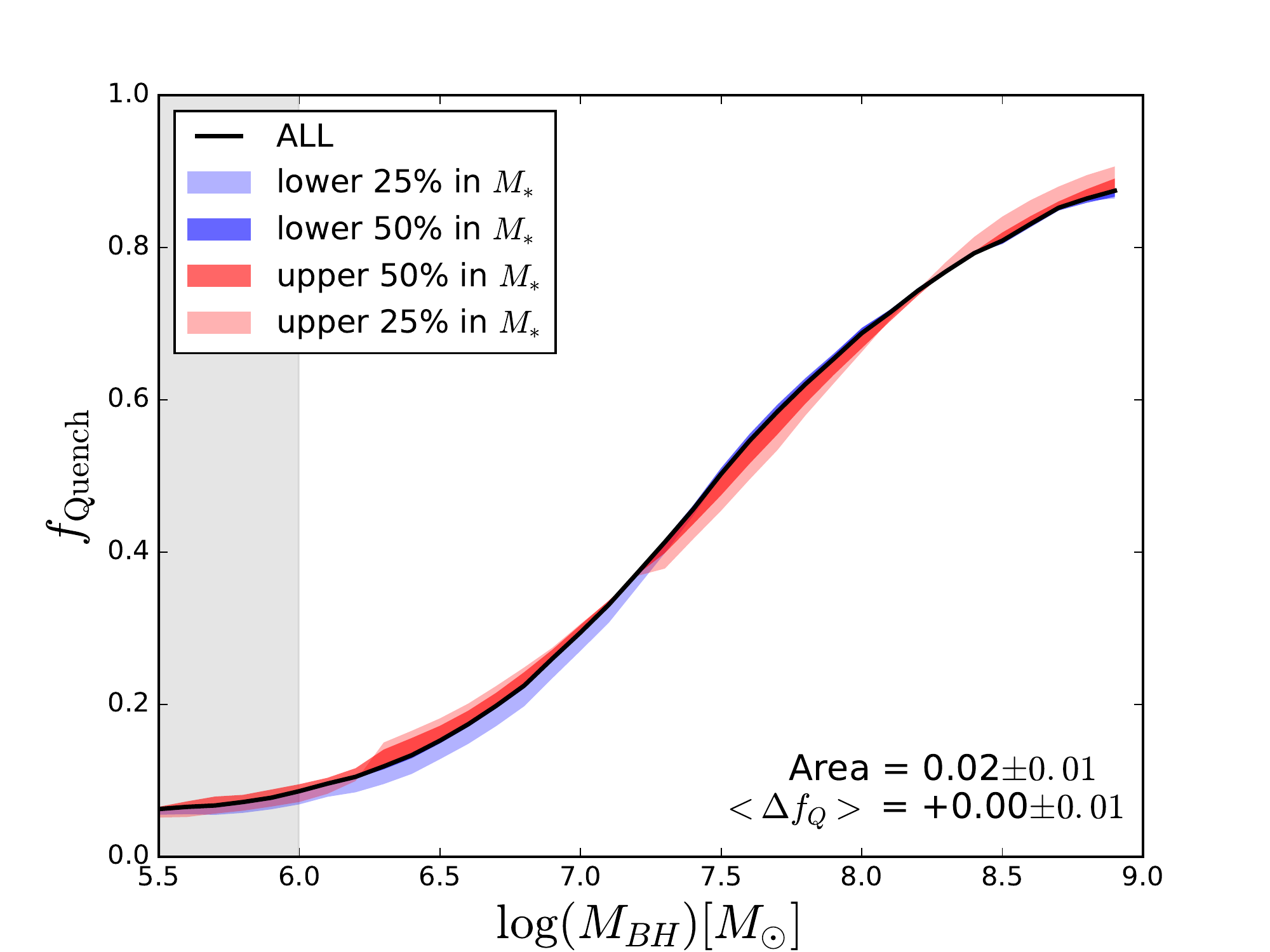}
\caption{Reproduction of Fig. 6 comparison of the relationship between quenched fraction and stellar, halo and black hole mass. For this figure black hole masses are estimated using separate scaling laws for ETGs and LTGs, for all inclination angles (as in McConnell \& Ma 2013).}
\end{figure*}

\begin{figure*}
\includegraphics[width=0.95\textwidth]{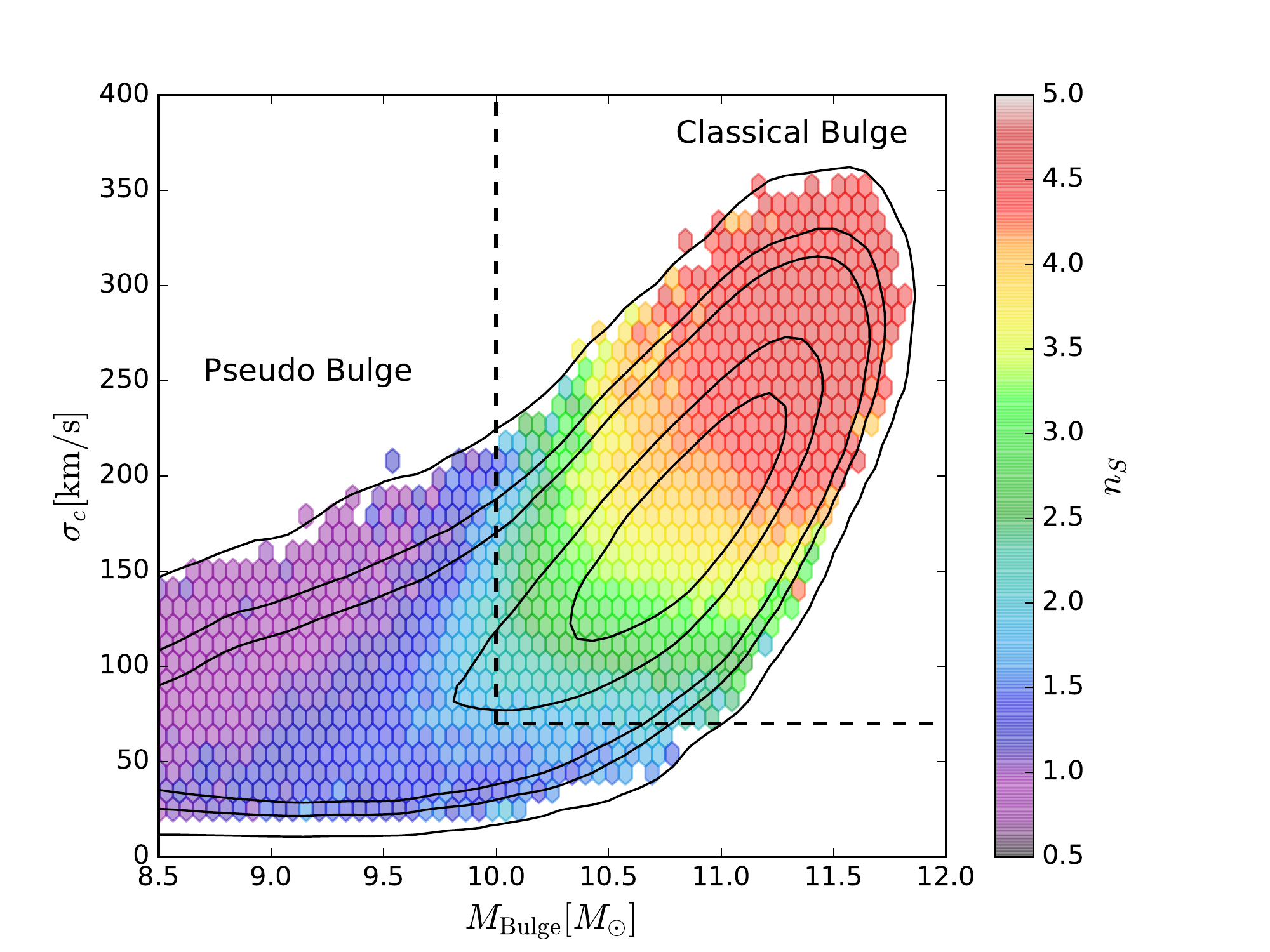}
\caption{Bulge mass - central velocity dispersion relation for SDSS galaxies. The hexagonal binned regions of this plot are colour coded by mean S\'{e}rsic index for the galaxy. Classical bulges are selected to have a steep correlation between bulge mass and velocity dispersion, and tend to have high S\'{e}rsic indices. Pseudo bulges are selected to have a weak (or null) correlation between bulge mass and velocity dispersion, and tend to have low S\'{e}rsic indices.}
\end{figure*}

\begin{figure*}
\includegraphics[width=0.49\textwidth]{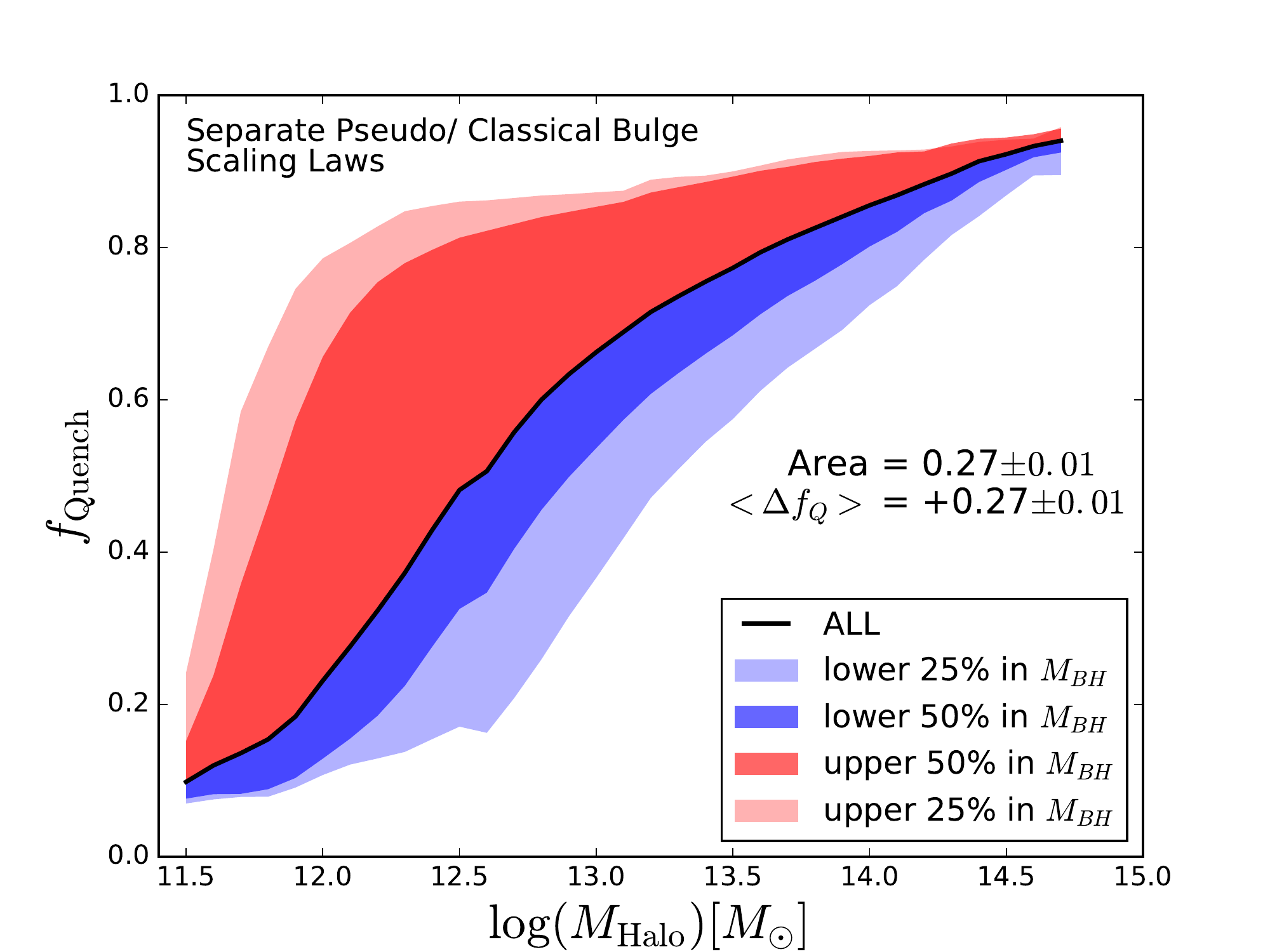}
\includegraphics[width=0.49\textwidth]{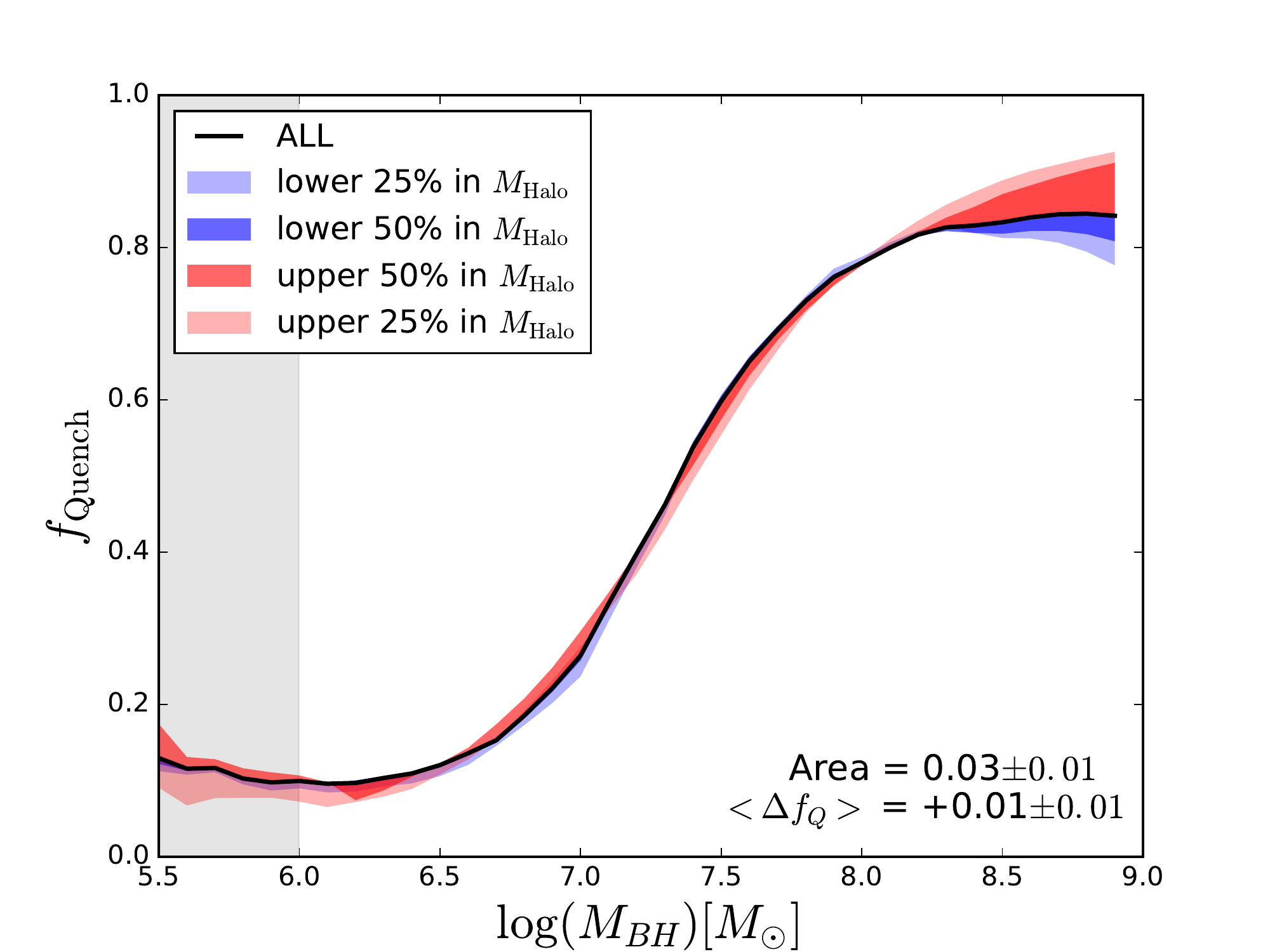}
\includegraphics[width=0.49\textwidth]{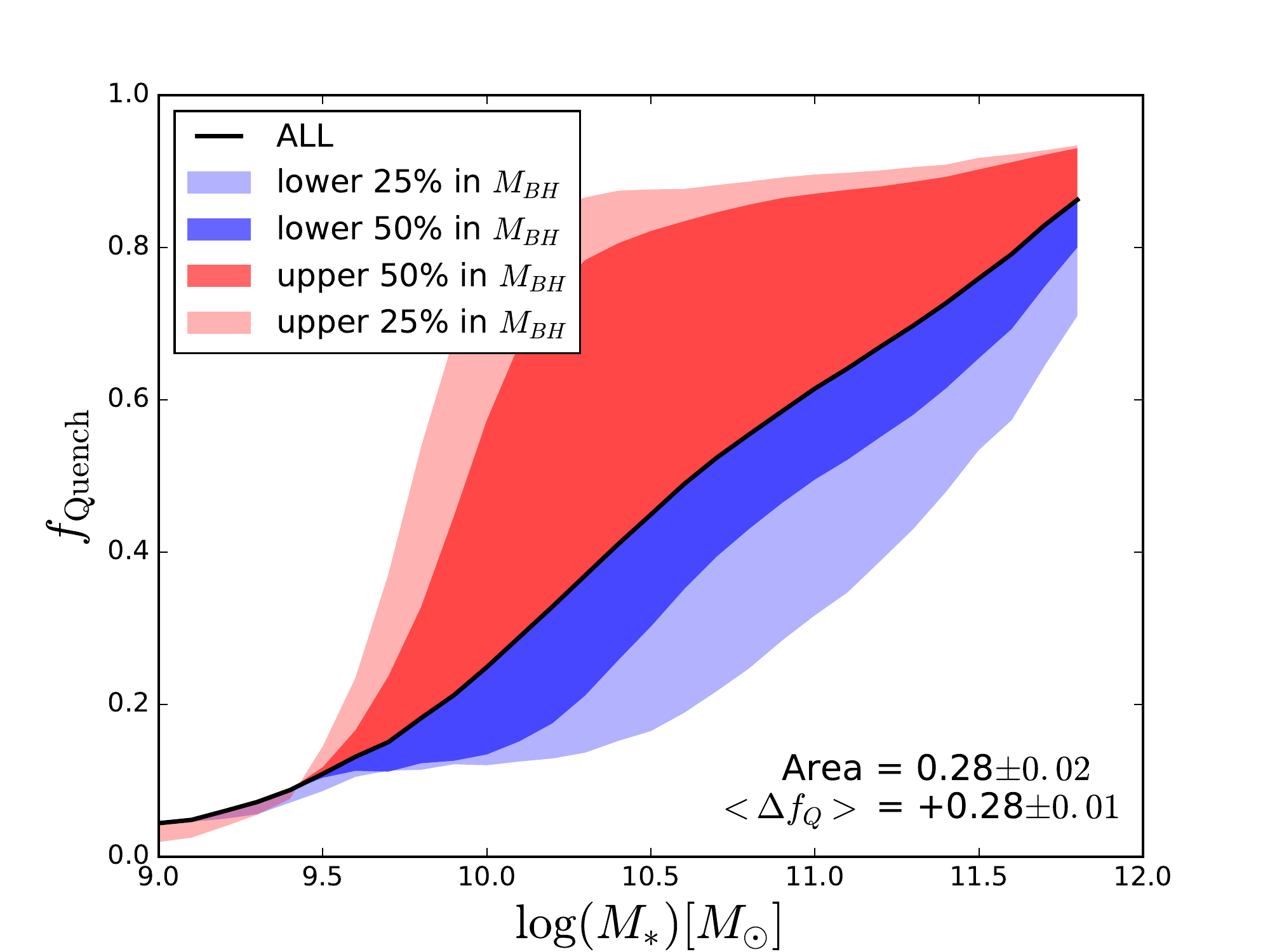}
\includegraphics[width=0.49\textwidth]{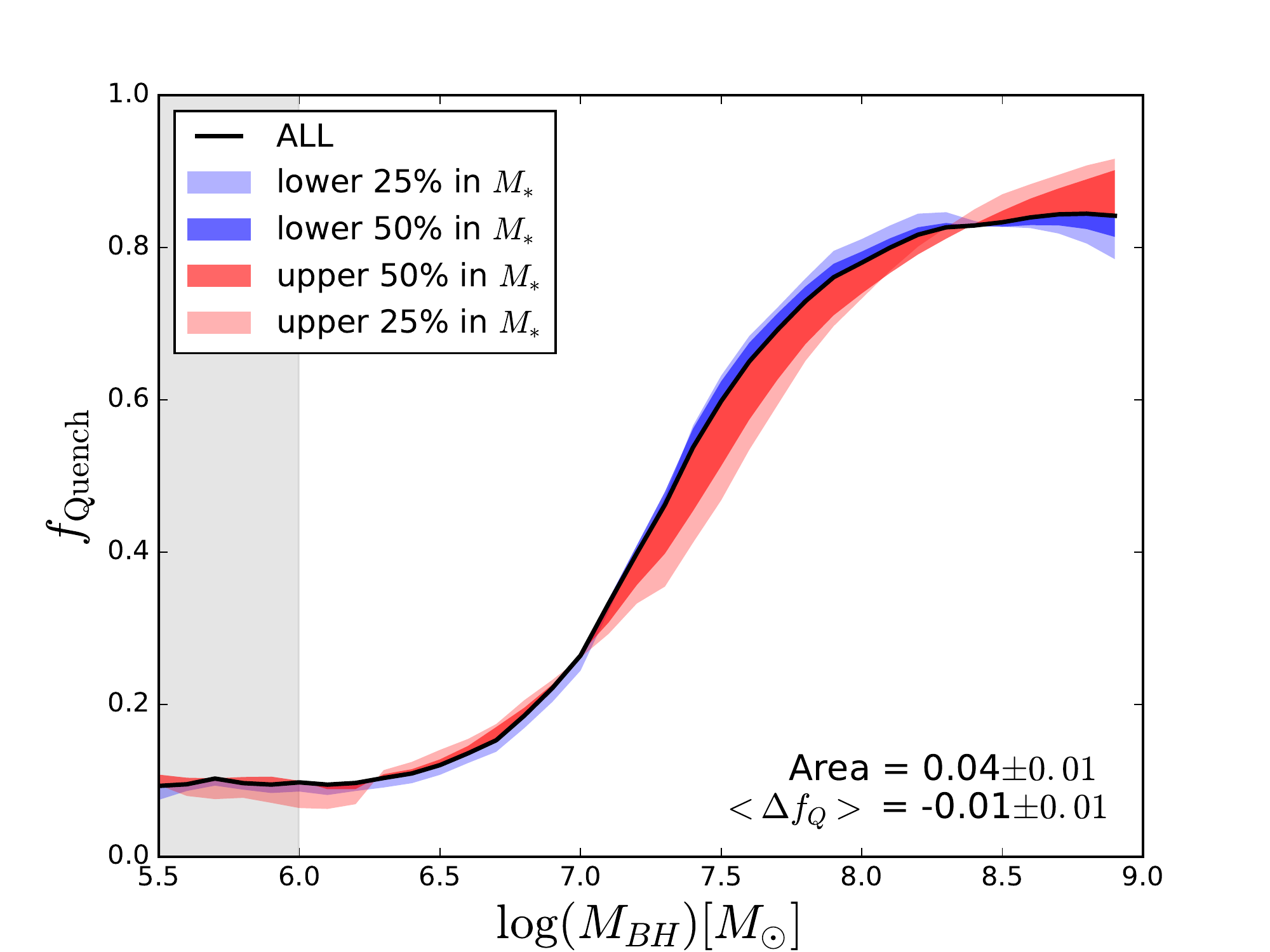}
\caption{Reproduction of Fig. 6 comparison of the relationship between quenched fraction and stellar, halo and black hole mass. For this figure black hole masses are estimated using separate scaling laws for classical and pseudo bulges (as in Saglia et al. 2016).}
\end{figure*}

In order to compare the dependence of central galaxy quenching on supermassive black hole mass in models to observations it is necessary to estimate black hole masses for observed galaxies. Since there are so few dynamically measured black hole masses in existence ($\sim$ 100, Saglia et al. 2016) indirect means must be used. Throughout the main body of the paper (particularly Section 6), we use the fiducial $M_{BH} - \sigma$ relationship for all morphological types presented in Saglia et al. (2016). This leads to our comparison with models being scaling law dependent. In this section we consider how significant a source of systematic bias this is for our analysis.

McConnell \& Ma (2013) find that galaxies with early-type (ETG) morphologies exhibit a different scaling law to those with late-type (LTG) morphologies. A part of the reason for this difference may come from the fact that galaxies in their sample are not restricted to being face-on. To test the potential impact on our results of adopting a different scaling law for ETGs and LTGs, here we relax our face-on criterion and weighting scheme (presented in Section 2) and fit galaxies with $(B/T)_{*} > 0.5$ and $(B/T)_{*} < 0.5$ with separate laws, assuming that galaxies with a dominant stellar fraction in a bulge component can be thought of as ETGs and those with a dominant stellar fraction in a disk component can be thought of as LTGs. Our results in this section are not particularly sensitive to the exact cut in $(B/T)_{*}$. Specifically, we compute (from McConnell \& Ma 2013):

\begin{equation}
{\rm ETGs:} \log(M_{BH}) = 5.20 \times \log(\sigma_{c} / 200 {\rm km/s}) + 8.39
\end{equation}

\begin{equation}
{\rm LTGs:} \log(M_{BH}) = 5.06 \times \log(\sigma_{c} / 200 {\rm km/s}) + 8.07
\end{equation}

\noindent In Fig. B1 we reproduce the results in Fig. 6 using these different scaling laws for different morphologies, and additionally allowing all disk inclinations to enter our sample. The results are almost identical between using a single scaling law for all galaxies (with a face-on restriction and weighting, Fig. 6) and using separate scaling laws for ETGs and LTGs (for the full sample of inclinations). It is clear that estimated black hole mass is a much tighter correlator with the quenched fraction than halo or stellar mass, as seen before. Even quantitatively the areas and mean difference statistics (shown on each plot) are very similar, within their respective errors. The only small difference is that the quenching threshold (black hole mass at which 50\% of central galaxies are quenched) increases slightly from $\sim 2 \times 10^{7} M_{\odot}$ (for a single scaling law) to $\sim 3 \times 10^{7} M_{\odot}$ (for separate ETG/ LTG laws). This small change leaves all of our conclusions unchanged.

In addition to the dependence of the scaling law on global morphology, Kormendy \& Ho (2013) and Saglia et al. (2016) both find strong evidence that the dependence of dynamically measured black hole mass on central velocity dispersion is much weaker for pseudo-bulges than for classical bulges. The most common way to identify a pseudo bulge is by its low S\'{e}rsic index. However, in the bulge - disk decompositions of Simard et al. (2011) and Mendel et al. (2014), which we use for our morphological measurements, a fixed $n_{S}$ = 4 bulge model is used. This is because a free $n_{S}$ bulge was found not to be supported by the data in most cases, making the fits ultimately degenerate. It is, however, possible to identify pseudo-bulges via other means.

In Fig. B2 we show the relationship between bulge mass (from SED fitting to the bulge - disk decompositions) and central velocity dispersion (from the width of absorption lines in the SDSS spectra). At high bulge mass and central velocity dispersion, there is a strong and steep correlation, as expected for classical bulges. At lower bulge masses and central velocity dispersions, the correlation between $M_{\rm bulge}$ and $\sigma_{c}$ weakens and eventually disappears. This lack of relationship between mass and velocity dispersion is exactly as expected for pseudo bulges. Additionally, in Fig. B2 we colour code each hexagonal bin in the relationship by the average S\'{e}rsic index of the galaxy (for a free S\'{e}rsic index fit in r-band, performed in Simard et al. 2011). Clearly, the classical bulges have higher values of $n_{S}$ than the pseudo bulges.

Using a cut of $n_{S}$ = 2 for the whole galaxy to separate likely pseudo bulges from the classical bulge sample, we recompute the black hole masses using the separate scaling laws for classical bulges and pseudo bulges in Saglia (2016). As with our $(B/T)_{*}$ cut above, the exact threshold in S\'{e}rsic index does not significantly affect the results. Specifically we compute:

\begin{equation}
{\rm Classical:} \log(M_{BH}{\rm [M_{\odot}]}) = 4.87 \times \log(\sigma_{c}{\rm [km/s]} ) - 2.83
\end{equation}

\begin{equation}
{\rm Pseudo:} \log(M_{BH}[M_{\odot}]) = 2.13 \times \log(\sigma_{c}{\rm [ km/s]} ) + 2.53
\end{equation}

\noindent In Fig. B3 we reproduce the results in Fig. 6 (and Fig. B1) this time for separate scaling laws for classical and pseudo bulges (as advocated in Kormendy \& Ho 2013; Saglia et al. 2016). Again our results are almost identical to the single scaling law implementation in Section 6. Black hole mass remains a much tighter correlator to the quenched fraction than either halo or stellar mass. In fact the difference in areas actually increases a little in the rendering with different scaling laws for different bulge types, compared to a single average scaling law (Section 6). The quenching threshold remains at $\sim 3 \times 10^{7} M_{\odot}$, identical to Fig. B1 and slightly higher than in Fig. 6.

In summary of this appendix, we find that our results and conclusions are highly insensitive to the exact rendering of the $M_{BH} - \sigma$ scaling law used for comparison to the models in Section 6. Specifically, using different laws for early- and late-types, relaxing the face-on inclination criterion, and fitting separately pseudo and classical bulges using the latest published results from the literature (e.g., McConnell \& Ma et al. 2013; Saglia et al. 2016) lead to no significant differences in any of our results. Hence, our comparison to Illustris and L-Galaxies (in Section 6.1) is largely free of systematic bias from our choice of scaling law parameterization.

\end{document}